\documentclass[amsmath,amssymb,twocolumn,showpacs,prb]{revtex4}
\usepackage{graphicx}
\graphicspath{{./}{./figures/}}
\usepackage{dcolumn}
\usepackage{bm}
\newcommand{\BEQ}{\begin{equation}}
\newcommand{\EEQ}{\end{equation}}
\newcommand{\beq}{\begin{equation}}
\newcommand{\eeq}{\end{equation}}
\newcommand{\BEA}{\begin{eqnarray}}
\newcommand{\EEA}{\end{eqnarray}}
\newcommand{\bea}{\begin{eqnarray}}
\newcommand{\eea}{\end{eqnarray}}

\newcommand{\ra}{\right\rangle}
\newcommand{\la}{\left\langle}

\newcommand{\p}{\partial}
\newcommand{\nn}{\nonumber }
\newcommand{\Tr}{{\rm Tr}}

\newcommand{\JJ}{{\mathbb J}}

\newcommand{\JU}{{1\hspace{-0.52ex}\mathrm{l}}}

\newcommand{\frgN}{\cite{LeDoussalWiese2001,LeDoussalWiese2003b} }
\newcommand{\ofrgN}{Refs.~\onlinecite{LeDoussalWiese2001,LeDoussalWiese2003b} }
\newcommand{\OfrgN}{Refs.~\onlinecite{LeDoussalWiese2001,LeDoussalWiese2003b}}
\newcommand{\oBBM}{\onlinecite{BalentsBouchaudMezard1996} }

\newcommand{\MP}{\cite{MezardParisi1991} }
\newcommand{\oMP}{\onlinecite{MezardParisi1991} }
\newcommand{\OMP}{\onlinecite{MezardParisi1991}}
\newcommand{\OBMP}{\onlinecite{BouchaudMezardParisi}}
\newcommand{\rme}{\mathrm{e}}
\newcommand{\tv}{{ \tilde  v}}
\newcommand{\tj}{{ \tilde j}}
\newcommand{\tu}{{ \tilde u}}
\newcommand{\1}{\mbox{1\hspace*{-0.55ex}l}}

\newcommand{\Fig}[1]{\includegraphics[width=\columnwidth]{#1}}

\begin{document}

\bibliographystyle{KAY}

\title{Cusps and shocks in the renormalized potential of glassy random manifolds: How Functional Renormalization Group and Replica Symmetry Breaking fit together}
\author{Pierre Le Doussal, Markus M\"uller and Kay J\"org Wiese} \affiliation{CNRS-Laboratoire
de Physique Th{\'e}orique de l'Ecole Normale Sup{\'e}rieure, 24 rue Lhomond,75231 Cedex 05, Paris, France}
\affiliation{Department of Physics, Harvard University, Lyman Laboratory, Cambridge MA 02138, USA.}
\date{\today}

\begin{abstract}
We compute the Functional Renormalization Group (FRG) disorder-correlator function $R(v)$ for $d$-dimensional elastic
manifolds pinned by a random potential in the limit of infinite embedding space dimension $N$. It measures the
equilibrium response of the manifold in a quadratic potential well as the center of the well is varied from $0$
to $v$. We find two distinct scaling regimes: (i) a ``single shock'' regime, $v^2 \sim L^{-d}$ where $L^d$ is
the system volume and (ii) a ``thermodynamic'' regime, $v^2 \sim N$. In regime (i) all the equivalent replica
symmetry breaking (RSB) saddle points within the Gaussian variational approximation contribute, while in regime
(ii) the effect of RSB enters only through a single anomaly. When the RSB is continuous (e.g., for short-range
disorder, in dimension $2 \leq d \leq 4$), we prove that regime (ii) yields the large-$N$ FRG function obtained
previously. In that case, the disorder correlator exhibits a cusp in both regimes, though with different
amplitudes and of different physical origin. When the RSB solution is 1-step and non-marginal (e.g., $d < 2$ for SR
disorder), the correlator $R(v)$ in regime (ii) is considerably reduced, and exhibits no cusp. Solutions of the FRG flow corresponding
to non-equilibrium states are discussed as well. In all cases the regime (i) exhibits a cusp non-analyticity at
$T=0$, whose form and thermal rounding at finite $T$ is obtained exactly and interpreted in terms of shocks. The results are compared with previous work, and consequences for manifolds at finite
$N$, as well as extensions to spin glasses and related
models are discussed.
\end{abstract}
\maketitle

\section{Introduction}

A major difficulty in devising analytical methods to handle glassy systems, such as systems with quenched
disorder, is to describe accurately the many metastable states which play a role both in the statics
(equilibrium) and the dynamics, as well as the barriers separating these states. Two main general methods have
been developed. The first is a mean-field theory, based on the Gaussian variational method (GVM), which, in the
statics, captures the many states by a multitude of saddle points exhibiting spontaneous replica symmetry
breaking (RSB) \cite{MezardParisiVirasoro}. Being highly versatile, the GVM has been applied to numerous
problems, notably in spin glasses, and has later been extended to the dynamics
\cite{CugliandoloKurchan1993,BouchaudCugliandoloKurchanMezardBookYoung}. The second method is the functional
renormalization group (FRG) which has been applied successfully to disordered elastic systems and random field
spin models, both in the statics
\cite{Fisher1985b,DSFisher1986,BalentsDSFisher1993,ChauveLeDoussalWiese2000a,ScheidlDincer2000,DincerDiplom,LeDoussalWiese2001,ChauveLeDoussal2001,LeDoussalWieseChauve2003,LeDoussalWiese2003b,LeDoussalWiese2004a,LeDoussalWiese2005a,Feldman2000,Feldman2001,Feldman2002,TarjusTissier2004,BalentsLeDoussal2002,BalentsLeDoussal2003,BalentsLeDoussal2004,LeDoussalWiese2005b,LeDoussalWiese2006b,TarjusTissier2005,TarjusTissier2006,FedorenkoLeDoussalWiese2006,FedorenkoLeDoussalWiese2006b}
as for the driven dynamics
\cite{NattermanStepanowTangLeschhorn1992,NarayanDSFisher1993a,ChauveGiamarchiLeDoussal1998,ChauveGiamarchiLeDoussal2000,LeDoussalWieseChauve2002,LeDoussalWiese2002a,LeDoussalWiese2003a,LeDoussalWieseRaphaelGolestanian2004}.
Although its range of applications is at present smaller, it is a powerful and promising method which allows to
compute
fluctuations not captured by mean-field theory.  The (relevant)
coupling constant in FRG is the disorder correlator $R(u)$. In contrast to standard field theories it is not a number but
a {\em function} of the field $u$, defined for the microscopic model in Eq.~(\ref{R0def}). Since the two
methods GVM and FRG  are rather different in spirit, and historically have developed along separate tracks, it is important
to compare them whenever possible. A further goal is to understand whether and how they may be extended to a
broader range of models.

In this paper we focus on disordered elastic systems, where both methods have been applied. As we discuss, some
of the conclusions and ideas may extend to other models with quenched disorder. Besides being of direct interest
for experiments, including vortex lattices, magnetic systems, density waves
\cite{DSFisher1986,BlatterFeigelmanGeshkenbeinLarkinVinokur1994,NattermannScheidl2000,GiamarchiLeDoussal1994,GiamarchiLeDoussal1995,GiamarchiLeDoussalBookYoung,NattermannBookYoung},
models of manifolds in a random potential provide the simplest example of a glass phase where numerous
metastable states occur beyond the so-called Larkin scale. As for random-field systems, the so-called
dimensional reduction phenomenon occurs, which renders conventional zero-temperature perturbation theory trivial
\cite{AharonyImryShangkeng1976, Grinstein1976,EfetovLarkin1977}, indicating the failure of the latter to capture
the complexity of the energy landscape. A big success of both methods,  GVM and  FRG, has been to circumvent
this problem. However, they achieve this in seemingly  rather different ways: The GVM approximates the Gibbs
measure by a hierarchical superposition of Gaussians, encoded in the Parisi Ansatz
\cite{MezardParisiVirasoro,MezardParisi1991}. In the FRG, the existence of a cusp (in the coupling function
$R''(u)$) beyond the Larkin scale is related to the existence of many metastable states. This was nicely
illustrated in an early paper by Balents, Bouchaud and M{\'e}zard \cite{BalentsBouchaudMezard1996}. The idea
that coarse graining leads to shocks in the force landscape, and the similarity of the pinning problem to the
Burgers equation, was introduced using a toy RG model with two degrees of freedom. However, the function $R(u)$
defined in that work does not coincide with the one usually studied in field theory, making a precise comparison
difficult.

In this article, we want to make this comparison quantitative. Let us denote $u(x)$, $x \in R^d$, the
$N$-component displacement (or ``height'') field which parameterizes the position of a manifold of internal dimension $d$ in
the $N$-dimensional embedding space. The GVM was applied to the problem by M{\'e}zard and Parisi (MP)
\cite{MezardParisi1991}, introducing replicated fields $u_a(x)$. For long-range disorder, or for short-range disorder and internal dimensions above
$d=2$, they found a solution  with continuous replica symmetry breaking (RSB) with a roughness exponent for the
manifold $u(x) \sim x^\zeta$ given by the Flory estimate. (The FRG allows one to go beyond this result and to compute deviations from Flory for $N<\infty$.)
For short-range disorder, with $d<2$, they found a 1-step RSB solution, analogous to the one in  infinite range
$p$-spin models\cite{CugliandoloKurchan1993,BouchaudCugliandoloKurchanMezardBookYoung}.

In the infinite-$N$ limit, the GVM becomes formally exact and hence can be compared with the FRG. Two of
us~\frgN obtained a self-consistent equation for the coupling function of the FRG, $R(u)$. It was defined from
the effective action of the replicated field theory (for a uniform mode $u(x)=u$), a standard field theory
definition, and computed in the large-$N$ limit, performing the usual rescaling $R(u)=N \tilde B(u^2/N)$.
Although the resulting self-consistent equation for $\tilde B(x)$ is formally valid only below the Larkin scale,
the corresponding FRG equation could be continued ``naturally'' to
scales beyond the Larkin scale (with a cusp present at $u=0$), extending the flow all the way to the RG fixed
point. For $2 \leq d \leq 4$, where the RSB is continuous, this FRG flow recovered~\frgN the MP result for small
overlap, i.e., it yielded only, even though exactly, the (non-trivial) contribution of the most distant states
to the correlation function. However, quite surprisingly, varying the IR cutoff $m$ in the confining potential
well finally allows for the reconstruction of the complete self-energy function obtained in the GVM, without
ever referring to ultrametric matrices~\frgN.

Despite this quantitative progress in understanding the connection between FRG and GVM, several questions
remained. It is natural that the FRG recovers the contribution of distant states, since it introduces an external field which
explicitly breaks replica symmetry and hence splits all the replicas: $u_a-u_b \neq 0$. However, one would
hope FRG to describe all states, not just the ones with smallest overlap. Further, no thermal rounding of the
cusp was found which is physically surprising in view of results relating finite-temperature droplets and FRG
\cite{BalentsLeDoussal2004,LeDoussal2006b,LeDoussalToBePublished}. Finally, one would like to describe better
the situation where the GVM yields a {\it non-marginal} 1-step RSB solution, namely the case $d < 2$ with short-range
disorder, which includes in particular the KPZ problem with $d=1$
\cite{KPZ,FreyTaeuber1994,Laessig1995,Wiese1997c,Wiese1998a}).

The first aim of this paper is to compute the FRG functions from first principles in the large-$N$ limit. We
take advantage of the recently obtained direct relation between the field theoretic definition of the FRG
function $R(u)$ and directly observable quantities \cite{LeDoussal2006b,LeDoussalToBePublished}. This has
allowed for a numerical determination of  $R(u)$ or the force correlator
\begin{equation}
\Delta(u)=-R''(u)
\end{equation}
for $N=1$ interfaces at $T=0$ in dimensions $d=0,1,2,3$ in Ref.~\onlinecite{MiddletonLeDoussalWiese2006}, as well as for
the depinning problem  in Refs.~\onlinecite{LeDoussalWiese2006a,RossoLeDoussalWiese2006a}. It was found that the
numerics compares remarkably well with the $\epsilon=4-d$ expansion,  to 1-loop, and even better to 2-loop
order. The idea, also implemented here, is to subject the manifold, in addition to the random potential, to an
external quadratic potential well $\frac{m^2}2 \int d^d x\, (u(x)-v)^2$, centered at $v$. The ``mass'' $m$ acts as
an infrared cutoff limiting the interface fluctuations. By measuring the free energy $\hat V(v)$ of the system
as a function of $v$ one obtains a random landscape whose second cumulant is the function $R(v)$. More
generally, if $v \to v(x)$ the whole second cumulant functional $R[v]$ is retrieved.

Here we compute this functional exactly. We find {\it two} distinct scaling regimes with non-trivial
infinite-$N$ limits, the first one where $v^2 \sim N^0\, L^{-d}$, the second for $v^2 \sim N$. The reason for this
peculiar property is that at $v=0$ there is spontaneous RSB, which implies the contribution of many  saddle
points which are equivalent under replica permutations. If the ``applied field'' $v$ remains ``small'' (first
regime) the spontaneous RSB saddle point is not significantly modified and all saddle points contribute to a
given observable (though, now, not all of them equivalently). This can be handled by a method introduced in
Ref.~\oBBM, and will be applied here with some improvements. In the second regime, which corresponds
to the more conventional scaling at large $N$, the applied field is stronger and the saddle point is modified.
We explicitly compute the $v$-dependence, and show how, for large $v$, a non-trivial FRG function emerges.
Specifically, in the case of a uniform field $v(x)=v$ we obtain:
\begin{equation}\label{summary}
R(v) - R(0) = \left\{\begin{array}{cccc} L^{-d} \tilde r(v^2 L^d) &\mbox{for}&  v^2 \sim L^{-d}, & (i)  \\
 N r(v^2/N) &\mbox{for}&   v^2 \sim N, & (ii)
\end{array}\right.
\end{equation}
with two different scaling functions which we compute. We check that the two regimes match, i.e., $\tilde r(z)
\sim A z$ at large $z$ and $r(z) \sim A z$ at small $z$. In the case where the RSB is continuous (e.g., for $2
\leq d \leq 4$) we prove that regime ({\it ii}) yields the large-$N$ FRG function obtained in our previous
study~\frgN. Hence the natural continuation (using the FRG flow-equation) performed there is correct, and one of
the main results of the present paper is to show this rigorously. Remarkably, RSB enters in this regime only
through a single number, an anomaly. In the case of continuous RSB (including the marginal 1-step solution in
$d=2$) we find that both regimes ({\it i}) and ({\it ii}) in (\ref{summary}) exhibit a cusp in $R''(v)$ at the origin,
though of different nature. Specifically, one finds $\tilde r(z) = B z + \tilde C z^{3/2}$ at small $z$ (at
$T=0$) and $r(z) \sim A z+C z^{3/2}$ at small $z$ (for all $T$ in the glass phase). These cusps are caused by jump discontinuities in $\hat V'(v)$, called shocks, as the center of the well $v$ is moved. In regime ({\it i}) the scaling function describes a
fluctuation of the energy, as measured by the connected correlator
\begin{equation} 
\overline{[\hat V(v)-\hat V(0)]^2}=2 L^d[R(v) - R(0)]
\end{equation}
 of order unity as $v$ is moved by a very small amount $\sim L^{-d/2}$. Hence we call this regime
``single shock regime''. By contrast, in regime ({\it ii}) the variance of the energy fluctuations scales with volume
and $N$.  Hence we call this regime ``thermodynamic'', since the scaling function $r(z)$ encodes the average
properties of many shocks. The two regimes also exhibit different properties with respect to temperature. In
regime ({\it ii}), as was found in \cite{LeDoussalWiese2001,LeDoussalWiese2003b}, the temperature dependence is weak
and the cusp survives even for $T>0$. Regime ({\it i}) exhibits a thermal rounding of the cusp, whose form is obtained
in an exact closed form. We compare its form with the predictions for the thermal boundary layer of the FRG
obtained previously \cite{LeDoussal2006b,LeDoussalToBePublished} in cases where it can be obtained using droplet
arguments (i.e., at finite $N$). For sufficiently short-ranged disorder in $d < 2$, the equilibrium 1-step RSB
solution yields a considerably reduced regime ({\it ii}) with no non-analyticity. A non-analyticity in this regime
would indicate the criticality of the system towards clustering of the states into an ultrametric superstructure
(higher-step RSB). This kind of criticality  is not present in typical samples in $d<2$, and only arise in
exponentially rare disorder realizations, as we confirm by studying the configurational entropy of
non-equilibrium states. Possible consequences of our results for manifolds at finite $N$ are discussed, as well
as extensions to spin glasses and related models.

The outline of the paper is as follows. In Section \ref{sec:model} we define the model and the observable to be
studied. In Section~\ref{s:small-v} we analyze regime ({\it i}) for both continuous and 1-step RSB, discussing various
subtleties of the phase diagram in the 1-step case. We study in particular the non-analytic cusp arising in the
force correlator at $T=0$ and its thermal rounding. Taking the limit of large $v\gg 1$ we establish the matching
with the regime $v^2\sim N$, which is analyzed in detail in Section~\ref{s:large-v}. We rigorously show how to
compute the FRG function $R(v)$ exactly by introducing two replica groups and derive from it the correct FRG
flow equations in all cases, including the anomaly arising from RSB. The physical significance of the presence
or absence of a cusp in this regime is discussed. The results are summarized in Section~\ref{s-discussion} and
possible applications are discussed.

\section{Model, observables and previous results}
\label{sec:model}

\subsection{Model}

We consider an elastic manifold parameterized by a $N$-component displacement field $u(x)$, also denoted $u_x$,
where $x$ belongs to the internal $d$-dimensional space. The manifold is exposed to a random potential, $V(x,u)$, which
lives in a $(D=d+N)$-dimensional space. Indices of the field $u^i_x$, $i=1,\dots,N$ are shown only when strictly
needed, and we use the notation $u \cdot v = \sum_{i=1}^N u^i v^i$. We study the (classical) equilibrium
statistical mechanics defined by the canonical partition sum $Z_V = \Tr\, e^{- \beta H_V}$ at temperature $T$,
and denote thermal averages by $\langle F[u] \rangle_V$ (or sometimes simply $\langle F[u] \rangle$) in a given
realization of the random potential. The model is defined by the total energy:
\begin{eqnarray}
 H_V[u] = \frac{1}{2} \int_k g^{-1}_k u_k \cdot u_{-k} + \int_x V(x,u_x)\,, \label{model}
\end{eqnarray}
where $u_k=\int_x u_x e^{i k x}$, $\int_x=\int d^d x$ and $u_x=\int_k u_k e^{- i k x}$, $\int_k =\int
\frac{d^dk}{(2 \pi)^d}$. To fix the average center-of-mass position
$\overline{u_{\mathrm{cm}}}:=\overline{u_{k=0}}/L^d$ we choose a non-zero value for $g_{k=0}^{-1}=c m^2$, which
takes the role of a mass, $c$ being the elastic constant. The mass provides a quadratic well for the manifold and thus serves as an IR cutoff to
limit the displacement fluctuations. One is often interested in the scale invariant limit, $m \to 0$. A UV
cutoff $\sim \Lambda^{-1}$ in $x$ space (e.g., due to a lattice) is implicit everywhere. For specific applications we consider
\footnote{Other choices of interest are, e.g., $g_k=\frac{1}{c (k^2 + m^{4/\alpha})^{\alpha/2}}$ for $\alpha<2$ which
models long-range elasticity.} :
\begin{eqnarray}
g_k=\frac{1}{c (k^2 + m^2)}\,,
\end{eqnarray}
even though most results apply to more general forms of $g_k$. The quenched disorder is chosen to possess
statistical translational invariance, with second cumulant
\begin{eqnarray}\label{R0def}
 \overline{V(x,u) V(x',u')} = \delta^{(d)}(x-x') R_0(u-u')\ .
\end{eqnarray}
This property entails a useful symmetry - see below - usually referred to as statistical tilt symmetry (STS). We
always assume $O(N)$ symmetry of the disorder, choosing the bare correlator to be~\footnote{The bare correlator $B$ is defined in the same way as in Refs.~\onlinecite{LeDoussalWiese2001,LeDoussalWiese2003b}, but corresponds to $-\hat{f}\equiv B$ defined in Ref.~\onlinecite{MezardParisi1991}.}:
\begin{eqnarray}
\label{bareR0}
R_0(u)=N B(u^2/N).
\end{eqnarray}
This scaling with $N$ yields a non-trivial large-$N$ limit. Among the variety of models parameterized by the function $B(z)$ one distinguishes short-range
(SR) disorder, one often studied example being:
\begin{equation} \label{modelI}
(\text{I}) \quad  B(z)=B_0 e^{-z/r_f^2} \quad [\textrm{SR}],
\end{equation}
and long-range (LR) disorder, often represented by the family of
power law force correlator \footnote{Corresponding to
$B(z)=\left[B(0) - \frac{\gamma B_0 }{\gamma-1}\right] \Theta(1-\gamma)
+ \frac{\gamma}{\gamma-1} B_0 \left(1 + \frac{z}{\gamma r_f^2}\right)^{1-\gamma}$
for $\gamma \neq 1$, and $B(z)=B(0)- B_0  \ln(1 + z/r_f^2)$ for $\gamma=1$.}:
\begin{equation} \label{modelII}
(\text{II}) \quad B'(z)=- \frac{B_0}{r_f^2 (1+ \frac{z}{\gamma r_f^2})^{\gamma}}
\quad [\textrm{LR}].
\end{equation}
Here, the parametrization of model II is chosen~\footnote{In \ofrgN the model I was parametrized with $g\equiv B_0$, $r_f\equiv 1$. The bare correlator of model II was taken in the form $B'(z)= - g \frac{1}{(a^2+z)^{\gamma}}$, corresponding to the reparametrization
$g a^{- 2 \gamma} = B_0/r_f^2$, $a^2= \gamma r_f^2$.} such that the limit ${\gamma\to \infty}$
corresponds to model I, and that the limits $\gamma \to 1$ and $\gamma \to 0$ are meaningful.
These two models  possess a special scale-invariance property
at infinite $N$: As discussed below in Section~\ref{s:large-v}, they arrive at their FRG fixed point after a finite renormalization time~\footnote{For finite $N$ there is a range of values $\gamma>\gamma^*(N)$ for which model II renormalizes to the SR disorder class I, with $\gamma^*(N) \to \infty$ as $N \to \infty$.}.

It is well known ~\cite{BlatterFeigelmanGeshkenbeinLarkinVinokur1994} that the effect of disorder in model (\ref{model})
for any $N$ becomes non-linear, and metastability appears when the mass is decreased beyond some characteristic scale. As easily seen from dimensional analysis,
the natural unit is the so-called Larkin scale
\footnote{One should distinguish this ''Larkin scale'', which is a simple combination
of dimensionful quantities meant only to provide an order of magnitude estimate, and a ''Larkin mass'' or a ''Larkin length'' which in some cases can be defined precisely
from an observable, e.g., for $N=\infty$ an exact phase transition to a regime with many pure states
occurs at a mass $m=m_c$, see below.}:
\BEQ
\mu_c \equiv \frac{1}{L_c}:=\left(\frac{B_0}{c^2r_f^4}\right)^{1/\epsilon},\\
\EEQ
where $\epsilon=4-d$. In finite dimensions $d>0$, $L_c=1/\mu_c$ has the loose interpretation of the smallest typical size of domains which may be trapped in different metastable states and thus exhibit glassiness at low temperature. The energy of such domains is naturally expressed in the unit of energy
\bea
E_c &:=& c \mu_c^{-d}\,(r_f \mu_c)^2,
\eea
while $r_f$ and $L_c=1/\mu_c$ are the natural scales for embedding space and internal space (i.e. inverse mass) lengths in the problem. We are free to choose units in which $E_c=L_c=r_f=1$, or equivalently, $c=B_0=r_f=1$, which we adopt throughout the paper. For completeness, we give the dimensions of all observables used in the present paper in App.~\ref{app:dimensions}, which allows to restore the full dependence on these parameters.

In order to study the model (\ref{model}) one introduces replicated fields $u^a(x)$, $a=1,\ldots, n$. Using standard methods,
all disorder-averaged correlation functions of the $u(x)$ field can be expressed as correlation functions of the
replicated fields $u^a(x)$ in the theory with partition sum $\overline{Z_V^n} = \Tr\, e^{- S[u]}$ and action
$S[u]$:
\begin{equation}
S[u] = \frac{1}{T} \sum_a \int_k g^{-1}_k u^a_k \cdot u^a_{-k} - \frac{1}{2 T^2} \sum_{ab} \int_x R_0(u^a_x -
u^b_x). \label{action}
\end{equation}

\subsection{Summary of previous studies}
\label{sec:summary}

\subsubsection{Gaussian Variational Method (GVM)}

Before defining the observable computed in the present work let us briefly review the quantities studied in
previous publications~\cite{MezardParisi1991,LeDoussalWiese2001,LeDoussalWiese2003b}, and
the main results obtained there (details are skipped and can be found in these original publications). The model
(\ref{action}) was studied in Ref.~\onlinecite{MezardParisi1991} using the Gaussian Variational Method (GVM),
which becomes exact in the limit $N=\infty$. The central observable calculated there is the two point
correlation function of the replicated field:
\begin{equation}
\langle u^{ai}_{k} u^{bj}_{k'} \rangle_S = G_{ab}(k) \delta_{ij} (2 \pi)^d \delta^{(d)}(k+k').
 \label{2point}
\end{equation}
Here we denote by $\langle . \rangle_S$ averages over the replicated action (we later drop the subscript $S$
when not strictly needed). The correlation (\ref{2point}) encodes \footnote{In the case of RSB (i.e., in the
glass phase and within the GVM) there are some subtleties in this correspondence. One can define the
intravalley connected average, given by $G_{aa}-G({\sf 1})$ (see below) - relevant for time scales in which no
transitions between valleys occur - as well as the full connected correlation $G_{aa}-\int_0^1 d{\sf u} G({\sf u})$.} the
averages $\overline{\langle u^i_k u^j_{k'} \rangle } = G_{aa}(k)\delta_{ij} (2 \pi)^d \delta^{(d)}(k+k')$
(diagonal part) and $\overline{\langle u^i_k u^j_{k'} \rangle^c } = \sum_b G_{a b}(k)\delta_{ij} (2 \pi)^d
\delta^{(d)}(k+k') $ (connected thermal average). The large distance behavior of the first one defines the
roughness exponent $\zeta$ of the manifold, i.e. $\overline{\langle (u_x-u_x')^2 \rangle } \sim |x-x'|^{2
\zeta}$, equivalently $G_{aa}(k) \sim k^{-(d+2 \zeta)}$ at small $k$. These hold at fixed scale in the limit $m
\to 0$, or at small but fixed $m$ for scales smaller than the IR cutoff, $k \gg 1/m$. Another important
exponent characterizes how the fluctuations of the ground state energy (or of the free energy) scale with system
size, $\Delta E \sim L^\theta$. Here, thanks to the statistical tilt symmetry, one has the relation
$\theta=d-2+2 \zeta$.

Quite generally (\ref{2point}) takes the form:
\begin{equation}
\label{GabMP}
T G^{-1}_{ab}(k) = \delta_{ab} g^{-1}_k - \sigma_{ab}(k)\ ,
\end{equation}
and within the GVM the self-energy is taken to be $k$-independent,  $\sigma_{ab}(k)\equiv \sigma_{ab}$. It obeys a self-consistent equation which arises as a large-$N$ saddle-point equation, reading:
\begin{equation}
\label{sigmaMP}
\sigma_{ab} = - \frac{2}{T} B'\left(2 \int_k \left[G_{aa}(k)-G_{ab}(k)\right] \right)\,.
\end{equation}
There are two types of saddle points: Either they respect the replica symmetry of the action (\ref{action}),
$\sigma_{a\neq b}=\sigma$, which happens in the high-temperature phase (where the roughness exponent assumes its thermal value, $\zeta=\zeta_{\rm th}={\rm max}(0,(2-d)/2$). Or, the saddle points
spontaneously break the symmetry (RSB), $\sigma_{ab} \to \sigma({\sf u})$, where ${\sf u} \in
[0,1]$ labels the distance of replicas in an ultrametric Parisi scheme describing the glassy, pinned phase.

Let us summarize the results for $N=\infty$ and $g^{-1}_k=k^2+m^2$, for which the dependence on the mass is
worked out in Ref.~\onlinecite{LeDoussalWiese2003b}.

We start with the case where the glass phase is described by continuous RSB
(also called infinite-step or full RSB). Within such a RSB scheme it is found that the roughness exponent equals
its ``Flory'' value:
\BEA
\zeta=\zeta_F=\left\{\begin{array}{cc} 0 & ({\textrm I}) \\ (4-d)/[2(1+\gamma)], & (\textrm{II})
\end{array} \right.
\EEA
and the continuous RSB solution is self-consistent if the corresponding energy exponent is positive, $\theta=\theta_F=d-2+2 \zeta_F >0$. This occurs in dimensions $4 \geq d \geq 2$ for both models and any $\gamma$, as well as in dimensions $d < 2$ for sufficiently long-ranged disorder in model II [$\gamma < \gamma_c(d) = 2/(2-d)$], including model $I$ in $d=2$.

\begin{figure}
\includegraphics[width=3.5in]{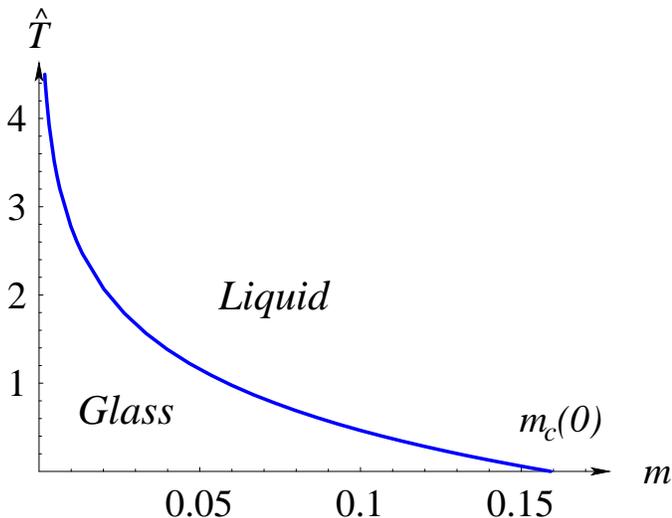}
\caption{Phase diagram for model I in $d=3$ for $\Lambda/m_c \gg 1$. The phase transition as given by (\ref{instability}) is everywhere continuous, and
the glass phase exhibits continuous RSB. The transition temperature $\hat{T}_c$ diverges as $m\to 0$. An
unconstrained system ($m=0$) is therefore always glassy. A similar phase diagram applies to $d>2$ for both models I and II, and to $d<2$ for model II with sufficiently long-ranged correlators, $\gamma<\gamma_c$ (the temperature scale being set, for $d<2$, by $I_1(m_c(T=0))$).}
\label{fig:phasediagram_d3}
\end{figure}

The replica symmetry is broken for small IR cut-off $m < m_c$, where $m_c=m_c(T)$
is the temperature dependent Larkin mass, which is determined by the instability of the replica-symmetric (RS) solution:
\begin{eqnarray} \label{instability}
1 &=& 4 B'' \big(2 T I_1(m_c) \big) I_2(m_c), \\
I_n(m)&=&\int_k \frac{1}{(k^2+m^2)^n},\label{In}
\end{eqnarray}
and decreases as a function of
$T$ from $m_c(T=0)=O(1)$
to zero as $T \to \infty$.
Hence, in that case, an unconstrained system ($m=0$) is always glassy,
while a strong confinement $m > m_c(T)$ leads to an ergodic (replica symmetric) high temperature phase. This is illustrated in Fig.~\ref{fig:phasediagram_d3} where we plot the phase diagram for model I in $d=3$. Since the temperature always enters in the combination $2 T I_1$, where $I_1$ is dominated by the UV cut-off for $d>2$, we have introduced the rescaled temperature:
\bea
\label{That}
\hat T= 2 T I_1(m=0)=\frac{4 T \Lambda^{d-2}}{(4\pi)^{d/2}(d-2)\Gamma(d/2)}.
\eea
for a circular UV cutoff in $k$ space.

The selfenergy function $\sigma({\sf u})$ of a continuous RSB solution generally interpolates continuously  between two
plateaux at small and large ${\sf u}$. For both models (I) and (II) (with $\gamma<\gamma_c$ in $d\leq 2$)
 $\sigma({\sf u})$ takes the form \footnote{In the limit of large UV cutoff
$\Lambda \gg m_c(T)$, corresponding to the weak-collective pinning regime. Note that the UV cutoff needs to be
retained only in the $T$ dependence of $m_c(T)$ via $\hat{T}$, cf., (\ref{That}).}:
\begin{equation} \label{sig}
\sigma({\sf u})=\frac{2}{2-\theta} \frac{m_c^2}{{\sf u}_c} \times\left\{ \begin{array}{cc} (m/m_c)^{2-\theta}, & {\sf u}<{\sf u}_m,\\
({\sf u}/{\sf u}_c)^{\frac{2}{\theta}-1}, & {\sf u}_m<{\sf u}<{\sf u}_c,\\
1, & {\sf u}_c<{\sf u}<1,
\end{array}\right.
\end{equation}
Here $m_c=m_c(T)$, and~\footnote{According to  (\ref{That}), in dimensions $d>2$ the temperature scales as $T\sim
\hat{T}\Lambda^{-(d-2)}$ with $\hat{T}\leq O(1)$ at the transition and in the glassy regime. Thus, in the weak collective pinning regime where $\Lambda\gg \mu_c=1$, one always has ${\sf u}_c\sim T\sim
\Lambda^{-(d-2)}\ll 1$. In long-range models $d<2$ with $\gamma<\gamma_c$, one finds that on the transition line the fluctuation dissipation ratio ${\sf u}_c=(1+\gamma)(2-d)/(4-d)\leq (1+\gamma_c(d))(2-d)/(4-d)=1$ is always smaller than 1.}
\bea
\label{21}
{\sf u}_c&=&A T m_c^\theta,\\
{\sf u}_m&=&A T m^\theta,
\label{22}
\eea
where, with $I_{n}$ defined in (\ref{In})
\BEA
\label{A}
A&=&\frac{I_2(m_c)^2}{m_c^{d-2} I_3(m_c)}= \frac{4A_d}{\epsilon^2}, \quad \text{(model I)}, \\
A & =&  \frac{4 A_d}{\epsilon^2} \left(1+\frac{1}{\gamma}\right)
\left( \frac{\epsilon}{4 A_d}\right)^{\frac{1}{1+\gamma}} , \quad \text{(model II)},\qquad
\EEA
and $A_d:=\epsilon \int_k (1+k^2)^{-2}=2\Gamma(3-d/2)/(4 \pi)^{d/2}$.

In the sequel the value of $\sigma({\sf 0})$ will play a central role, and we give an explicit expression for later reference:
\bea
\label{sigma0}
\sigma({\sf 0})=\sigma({\sf u}_m)=\frac{2}{2-\theta}\frac{m^{2-\theta}}{AT}.
\eea
In the case of sufficiently short-ranged disorder in $d\leq 2$ (model I or model II with $\gamma>\gamma_c(d)$) the glass phase is described by a 1-step RSB solution which is fully characterized by three numbers:
the break point ${\sf u}_c$, and the self-energy parameters
\BEA
\sigma_0&\equiv& \sigma({\sf u}<{\sf u}_c)=\sigma({\sf 0}), \label{sigma1step}\\
\sigma_1&\equiv& \sigma({\sf u}>{\sf u}_c)=\sigma({\sf 1}).\nn
\EEA

\begin{figure}
\includegraphics[width=3.5in]{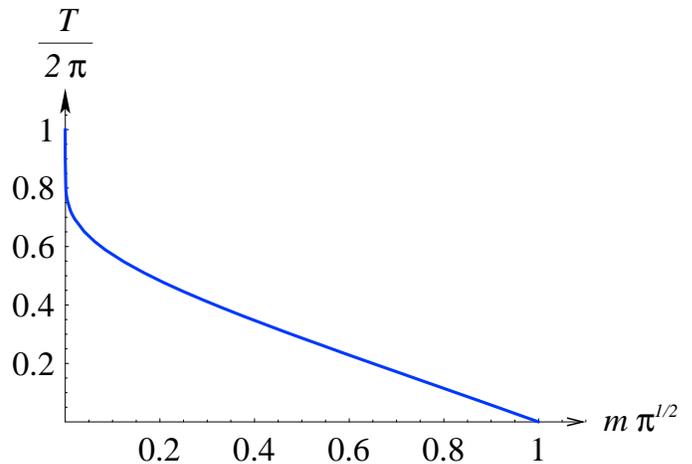}
\caption{Phase diagram for model I in $d=2$ (with a UV cut-off $\Lambda=10$). The phase transition is everywhere continuous, and
the glass phase exhibits continuous one-step RSB. The transition temperature
has a finite limit $T_c\to 2\pi$ as $m\to 0$. A similar phase diagram
applies to model II with critical long-range disorder $\gamma=\gamma_c$ in $d<2$. The transition line has a
cusp-like behavior as $m\to 0$, $[T_c(0)-T_c(m)]/T_c(0)\sim m^{2-d}$ (and $\sim 1/\log(1/m)$ in $d=2$).}
\label{fig:phasediagram_d2}
\end{figure}

The borderline between 1-step and full RSB is characterized by $\theta=\theta_F=0$: $\gamma=\gamma_c$ for model II in $d<2$, or model I in $d=2$. In this case the 1-step solution can equivalently be obtained as the limit of a continuous RSB solution, which entails that the 1-step scheme is only marginally stable. In this limiting case of continuous RSB, $m_c(T)$ remains a decreasing function of $T$, However, it vanishes at a finite $T=T_c$ 
signalling a continuous glass transition for the unconstrained system ($m=0$) at $T_c$. The phase diagram for the marginal case of SR disorder in $d=2$ is shown in Fig.~\ref{fig:phasediagram_d2}.

Away from the borderline in the $(d,\gamma)$ plane, the 1-step solution is genuinely stable. This includes in
particular the case of the directed polymer and the KPZ problem ($d=1$). The phase diagram in the ($T$,$m$)
plane is more complicated: (cf., Fig.~\ref{fig:phasediagram1step} on page \pageref{fig:phasediagram1step}) and will be discussed together with the 1-step solution within the GVM in Section~\ref{sec:phasediagram1step}. The essential difference with the cases discussed above is the nature of the temperature-driven glass transition at small mass which becomes a discontinuous random first-order transition. Similarly to ordinary first-order transitions, two locally stable solutions coexist  at least close enough to the transition: Here these are the  RS solution and
the 1-step solution.
The roughness exponent of all 1-step solutions equals the thermal exponent, $\zeta=\zeta_{\rm th}=(2-d)/2$.

\subsubsection{Effective action and functional RG}

In an effort to connect the results described above to the functional RG approach,
two of us performed a calculation of
the effective action $\Gamma[u]$ for the model (\ref{model}), cf.\ Refs.\ \onlinecite{LeDoussalWiese2001,LeDoussalWiese2003b}.
One starts by defining the
standard generating functional $W[j]$ for connected correlations in the replica theory with, in general, $j=(j^1(x),\ldots,j^n(x))$:
\begin{eqnarray} \label{defW1}
&& e^{W[j]} = Z[j] = \overline{\prod_{a=1}^n Z_V[j^a]}, \\
&& Z_V[j^a] = \int {\cal D}[u]\, e^{- \beta H_V[u] + \int_x j^a_x \cdot u_x },
\end{eqnarray}
and the effective action functional is defined via the Legendre transform:
\begin{eqnarray} \label{legendre}
\Gamma[u] + W[j] = \sum_a \int_x j^a_x \cdot  u^a_x \quad , \quad \frac{\delta W}{\delta j_x^a}[j]=u_x^a.
\end{eqnarray}
In \ofrgN  the effective action for a uniform configuration $\Gamma(u)=\frac{1}{L^d} \Gamma[\{u_x^a=u^a\}]$ was
studied in the large-$N$ limit. It has a non-trivial limit in the regime
\BEQ \label{Nregime1}
u^2 \sim N,\quad \Gamma \sim N.
\EEQ
It was further computed in an expansion in the ``number of replica sums''
(which is effectively a cumulant expansion) as follows:
 \BEA \label{expGamma}
\Gamma(u) &=& \sum_a\frac{m^2 \left(u^a\right)^2}{2 T } - \frac{1}{2 T^2} \sum_{ab} R(u^{ab}) \nn\\
&& \quad- \frac{1}{6 T^3} \sum_{abc} S^{(3)}(u^{abc}) + \dots \EEA up to a constant. Here and below $u^{ab}=u^a-u^b$,
$u^{abc}=u^a,u^b,u^c$. This defines unambiguously (for any $N$) the second cumulant of the renormalized disorder
potential $R(u)=R_m(u)$, where we will usually keep the $m$-dependence implicit. At large $N$, it
obeys the scaling form
\begin{equation} \label{relationRB0}
R(u)=N \tilde B(u^2/N).
\end{equation}
Remarkably, the scaling function $\tilde B(z)$ satisfies a closed equation:
\begin{equation} \label{Brecursion}
\tilde B'(z) = B'\left(z + 2 T I_1 + 4 I_2 \left[\tilde B'(z)-\tilde B'(0)\right]\right),
\end{equation}
where $z=u^2/N$, and $I_n=I_n(m)$ was given in (\ref{In}). As we will show later, this equation is only valid in the non-glassy regime since it was derived under the implicit assumption that the replica symmetry is not spontaneously broken as $u\to 0$.
Taking a derivative with respect to the mass $m$ and
introducing a scaled function $\tilde b(x)=4 A_d m^{4 \zeta-\epsilon} \tilde B(x m^{-2 \zeta})$ a FRG equation
was derived for the renormalized and scaled potential correlator $\tilde b(x)$ (denoted $b(x)$ in Refs.~\onlinecite{LeDoussalWiese2001,LeDoussalWiese2003b}). Remarkably, this FRG equation admits a natural continuation to the glassy regime. In that continuation a linear cusp exists for all $m \leq m_c$. The resulting large-$N$ equation, valid to all orders in $\epsilon=4-d$, exactly matches the equation previously derived to first
order in $\epsilon$ (1 loop) but for arbitrary $N$ by Balents and Fisher~\cite{BalentsDSFisher1993} by a quite different
method. In addition, for all universality classes characterized by ($d$,$\gamma$), such that $\theta>0$
(ensuring continuous RSB in the GVM), the fixed points $\tilde b^*(x)$ could be related to the solution of the
GVM. All the above strongly suggested that the derived flow equation was the correct large-$N$ limit of the FRG.
\begin{figure}
\Fig{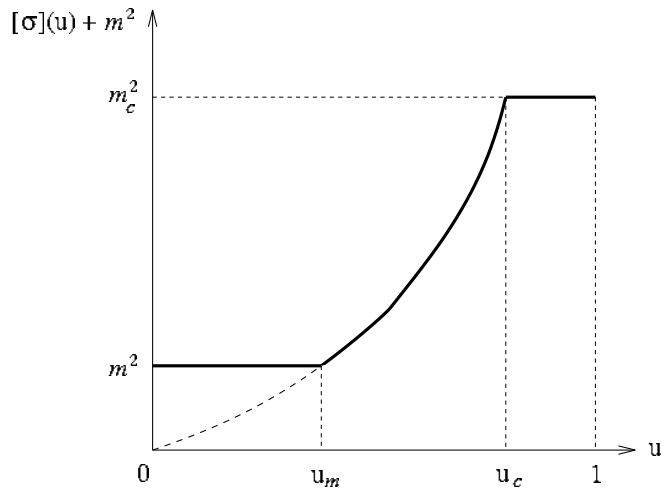}
\caption{The MP function $[\sigma](u)+m^{2}$ for  $m>0$. Changing $m^{2}$, only the lower plateau will move, while the remainder of the function (for ${\sf u}>{\sf u}_{m}$) remains unchanged.}
\label{f:moving-parabola}
\end{figure}

\subsubsection{Relation between GVM and FRG in the case of continuous RSB}

However, establishing a link between the GVM results and the FRG is rather subtle. To a considerable extent,
this has been achieved for the case of continuous RSB in ~\ofrgN, while the same task for
situations exhibiting 1-step RSB has remained an open problem, which is addressed in the present paper. In order to
review the results for the continuous case, we recall the multiplication formulae for two ultrametric matrices
$AB=C$:
\BEA
A^cB^c &=& C^{c},\label{inv1}\\
(A^c-[A]({\sf u}))(B^c-[B]({\sf u}))&=&C^c-[C]({\sf u}),\label{inv2}\\
 A({\sf 0})B^c+B({\sf 0})A^c&=&C(0),\label{inv3}
\EEA
where:
\BEA
A^c&:=&A_{aa}-\int_0^1 A({\sf u})d{\sf u},\\
\ [A]({\sf u})&:=&{\sf u} A({\sf u})-\int_0^{\sf u}  A({\sf u}') d{\sf u}'.
\EEA
We also introduce the notation for the diagonal element $\tilde A:=A_{aa}$.

From (\ref{sig}) above one finds:
\begin{equation} \label{sig2}
m^2+[\sigma]({\sf u})=m_c^2 \times\left\{ \begin{array}{cc} (m/m_c)^{2}, & {\sf u}<{\sf u}_m,\\
({\sf u}/{\sf u}_c)^{\frac{2}{\theta}}, & {\sf u}_m<{\sf u}<{\sf u}_c,\\
1, & {\sf u}_c<{\sf u}<1,
\end{array}\right.
\end{equation}
and from (\ref{GabMP}) we obtain
\BEA
&&G^c(k)=T g_k,\label{Gc}\\
&&G^c(k) - [G](k,{\sf u}) = \frac{T}{g_k^{-1} + [\sigma]({\sf u})},\label{[G]}\\
&&G(k,{\sf 0})=G(k,{\sf u}_m)=T \sigma({\sf u}_m) g_k^2.\label{G0}
\EEA
Further, using $\frac{d}{d{\sf u}} [\sigma]({\sf u}) = {\sf u} \frac{d}{d{\sf u}} \sigma({\sf u})$, we
find:
\begin{eqnarray}
 G(k,{\sf u})= T \sigma({\sf u}_m) g_k^2 + T \int_{{\sf u}_m}^{\sf u} \frac{ \dot \sigma({\sf u})\,d{\sf
u}}{[g_k^{-1}+[\sigma]({\sf u})]^2}\,,\label{GofuMP}
\end{eqnarray}
(we often denote $\dot \sigma \equiv \frac{d}{d{\sf u}} \sigma$ here and below) which relates $G(k,{\sf u})$ to
the self-energy $\sigma({\sf u})$, which for $g_k^{-1}=k^2+m^2$ is given by (\ref{sig}) and (\ref{sig2}).

In the GVM the 2-point correlation function is thus given by contributions from states at all distances ${\sf
u}$ as:
\begin{equation} \label{diagonalG}
\frac{\tilde G(k)}{T}= g_k + \sigma({\sf u}_m) g_k^2 + g_k \int_{{\sf u}_m}^1 \frac{d{\sf u}}{{\sf u}^2}
\frac{  [\sigma]({\sf u})}{g_k^{-1}+[\sigma]({\sf u})}
\end{equation}
Using (\ref{sig2}) and integrating over ${\sf u}$ yields the final result:
\begin{eqnarray} \label{mp}
&& \tilde G(k) \sim C m^{- (d + 2 \zeta)} f(k/m),
\end{eqnarray}
in the limit $m \to 0$, with $f(0)=1$ and $f(z) \sim z^{-(d+ 2 \zeta)}$ with $\zeta=\zeta_F$, exhibiting the
anticipated scaling.

On the other hand the FRG allows one to compute the two point function using (\ref{expGamma}) through the exact
relation:
\begin{eqnarray} \label{exact}
 (2 \pi)^d \delta^{(d)}(k+k') G^{-1}_{ab}(k)=\frac{\delta^2 \Gamma[u]}{\delta u^a_{k} \delta u^b_{k'}} |_{u=0}\, ,
\end{eqnarray}
which, upon matrix inversion yields the zero-momentum correlation function:
\begin{eqnarray}
\label{FRGcorrelator}
 \tilde G(k=0)= -\frac{ 2 \tilde B'(0)}{m^4} + \frac{T}{m^2}.
\end{eqnarray}
It turns out that this {\it does not} yield the full MP result (\ref{diagonalG},\ref{mp}). Indeed, the contribution from the integral over ${\sf u}$
in (\ref{diagonalG}) is missing. However, Eq.~(\ref{FRGcorrelator}) exactly reproduces the first two
terms in (\ref{diagonalG}), i.e., the contribution from the most distant states corresponding to $0<{\sf u}<{\sf
u}_m$, as well as the trivial term $T/m^2$ describing the connected correlations $G^c(k=0)$. Using the natural continuation to the glassy regime of the RG flow associated to the solution of (\ref{Brecursion}), it was indeed inferred that
\begin{eqnarray} \label{resfrg}
&& - 2 \tilde B'(0) = T \sigma({\sf u}_m),
\end{eqnarray}
as will be derived rigorously in Section~\ref{s:large-v}.
This is already a non-trivial result: obtaining it within the GVM does require
the RSB Ansatz, while the FRG considered in ~\ofrgN
does not make reference to ultrametric matrices and their properties.

The fact that the FRG calculation of \ofrgN reproduced only the contribution of the most distant states within
the Parisi hierarchy was argued to arise because $\Gamma$ and $W$ were computed by
introducing sources which split all replicas, i.e., $u^{ab} \neq 0$ (and of order $\sqrt{N}$) for all $a\neq b$.
However, it remained to be understood how the full result (\ref{mp}) can be recovered and how the crossover to
spontaneous RSB takes place as $u^{ab} \to 0$.

This is the aim of the present paper. We will show the existence of a regime whose large-$N$ limit is different
from (\ref{Nregime1}):
\BEA  \label{Nregime2} u^2 \sim N^{0} L^{-d},\quad \Gamma \sim  N^{0},
\EEA
and where RSB cannot be
neglected. This will allow us to recover the complete GVM result, i.e., to correctly perform the (non-trivial) zero-field limit $u^a_k \to 0$ in (\ref{exact}). We first show, in the following
section, that it is equivalent, but physically more transparent, to study the generating functional $W$
rather than $\Gamma$. Further, we will revisit the attempt by Balents, Bouchaud, and
M\'ezard~\cite{BalentsBouchaudMezard1996} to connect RSB and FRG. Although we choose a different observable than
those authors (so as to connect more directly to the FRG) we employ techniques similar to theirs, at least in 
the regime $v^2=O(1)$.

\subsection{Observable studied here}
\label{sec:observable}
The observable studied here is the free energy $\hat V[v]$ of the elastic manifold
in a quadratic potential well centered at position $v_x$, and defined by~\footnote{The definition of $\hat V[v]$ can alternatively be interpreted
as the free energy of the elastic manifold (with elastic properties described by  $g_k$),
where the quenched disorder has been displaced by $-v(x)$.}:
\begin{eqnarray} \label{defhatv}
&& e^{- \hat V[v]/T} = \int {\cal D}[u]\, e^{- H_V[u;v]/T}\, , \\
&& H_V[u;v] = \frac{1}{2} \int_k g^{-1}_k |u_k-v_k|^2 + \int_x V(x,u_x)\, , \nonumber
\end{eqnarray}
with $g_k=k^2+m^2$. For a uniform $v_x=v$, $v_k= v (2 \pi)^d \delta^{(d)}(k)$
one has \footnote{As usual formulae for discrete sums are obtained replacing
$\int_x \equiv \sum_x$, $\int_k \equiv L^{-d} \sum_k$ and for a uniform $v_x=v$ one has $v_k=L^d v \delta_{k,0}$.}:
\begin{eqnarray}
H_V[u;v] = H_V[u] - m^2 v \int_x u_x + \frac{1}{2} L^d m^2 v^2.
\end{eqnarray}
It is then clear from the definitions (\ref{defW1},\ref{defhatv}) that the
statistics of $\hat V[v]$ can be obtained from the functional $W[j]$ with:
\begin{eqnarray} \label{changejtov}
 j_k = \frac{g_k^{-1} v_k}{T}.
\end{eqnarray}
Hence we denote $\hat W[v] = W[j=(g^{-1} v)/T]$, such that:
\begin{eqnarray}
 e^{\hat W[v]} = e^{\frac{1}{2 T } \int_k g^{-1}_k \sum_a v^a_k v^a_{-k}}
\overline{\prod_a e^{- \hat V[v_a]/T}}\,.
\end{eqnarray}
It is thus clear that the functional can be expanded in a cumulant expansion, i.e., in the number of replica sums:
\BEA
\label{Wrepsum}
 \hat W[v] &=& \frac{1}{2 T } \sum_a\int_k g^{-1}_k v^a_k v^a_{-k} +
\frac{1}{2 T^2} \sum_{ab} \hat R[v^{ab}] \nn\\
&&\quad + \frac{1}{6 T^3} \sum_{abc} \hat S^{(3)}[v^{abc}] + ...,
\EEA
up to a constant, where again $v^{ab}_x=v_x^a-v_x^b$, etc. Each term in the expansion is a functional,
$\hat R[v^{ab}] \equiv \hat R[\{v_x^{ab}\}]$ with fixed $a,b$, etc., and
represents the free-energy cumulants:
\begin{eqnarray}
\label{correlator}
&& \overline{\hat V[v] \hat V[v']}^c = \hat R[v-v'],
\end{eqnarray}
and similarly for higher cumulants (the overbar denoting the disorder average over the random
potential $V$). For a uniform configuration $v_x=v$ one has
\BEQ
\hat R[v]=L^d \hat R(v)
\EEQ
which
defines $\hat R(v)$. Note that with arguments in brackets $[...]$ we denote functionals, while $(...)$ is reserved for functions. When discussing uniform $v$ we usually switch to the function $R(v)$, separating out the volume factor.

As shown in Refs.~\onlinecite{LeDoussal2006b,LeDoussalToBePublished} by performing the Legendre transform
(\ref{legendre}), this observable is directly related to the function $R(u)$ of the FRG:
\begin{eqnarray}
 \hat R(v) = R(v),\label{Rhat=R}
\end{eqnarray}
i.e., the two functions are the same, and this holds for the functionals, too. Therefore, by computing $\hat
R[v]$, which we do here for large $N$, we simultaneously compute $R[u]$ as defined from the effective action
(\ref{expGamma}). The fact that (\ref{correlator}) defines an observable which is easy to measure has allowed for a numerical
determination\cite{MiddletonLeDoussalWiese2006} of $R(u)$ or, more precisely, of the force correlator
$\Delta(u)=-R''(u)$, for $N=1$ interfaces at $T=0$.

The derivative of $R(u)$ at $T=0$ contains information about the
shocks. Indeed, if one computes the ground state $u(x;v)$ for a fixed
well position and defines the center of mass displacement $\bar u(v):=L^{-d} \int u(x;v)$,
one finds that the latter exhibits jumps as $v$ is varied. The statistics of these jumps is encoded in the functions $\hat R$, $\hat S$ etc., for instance one has:
\begin{equation}
 \overline{[\bar u^i(v) - v^i][\bar u^j(v') - {v'}^j]} = m^{-4} L^{-d} \Delta_{ij}(v-v'),
\end{equation}
and the cusp of $\Delta_{ij}(v)=- \partial_i \partial_j \hat R(v)$, i.e.\ the derivative of $\Delta$ at argument $0$,
is proportional to the second cumulant of the jump sizes.

We now turn to the calculation of this observable in the large-$N$ limit.
We will perform separate calculations
for the two scaling regimes $v^2_{x} \sim O(1)$ and $v^2_{x} \sim O(N)$ in the next Sections and check that they match.

\section{Regime of small $v$, $v^2_{x} = O(1)$}
\label{s:small-v}

We want to compute the generating functional (\ref{defW1}) in the form
\begin{eqnarray} \label{defW}
&& e^{\hat W[v]} =  \overline{\prod_a Z_V[j^a]}  \quad , \quad j^a_k = \frac{g_k^{-1} v^a_k}{T}\,.
\end{eqnarray}
This can be carried out at large $N$ through the saddle-point method as in Ref.~\onlinecite{MezardParisi1991},
introducing the term $i \lambda_{ab} (N \chi_{ab} - u_a \cdot u_b)$ and integrating over the field $u$.
This is detailed in the next Section where we study the case where $v_{ab}^2\sim O(N)$ which
distorts the saddle point away from (\ref{GabMP},\ref{sigmaMP}). In this section, we study $v_{ab}^2 = O(1)$ and
the above saddle points are unchanged (more precisely,
they are shifted only by terms at most of order $O(1/N)$ which are discarded). Since there is spontaneous RSB,
there are in fact many saddle points equivalent under replica permutations. Hence the
above average must be written as a sum over all equivalent saddle points:
\begin{eqnarray}
&& \overline{\prod_a Z_V[j^a]} = C_n \sum_{\pi} \langle e^{\sum_a \int_x j^a_x u_x} \rangle_{\pi},
\end{eqnarray}
where $\pi \in S_n$ belongs to the group of permutation of $n$ replica which is used to label the saddle points.
Around each saddle point the measure is Gaussian with correlator $\langle u^a_x u^b_y \rangle_{\pi}
= G^\pi_{ab}(x-y):= G_{\pi(a) \pi(b)}(x-y)$ where $G_{ab}(x)$ is the ultrametric matrix given by (\ref{GabMP}). Hence we obtain:
\begin{equation}
 e^{\hat W[v] - \hat W[0] } =  \tilde \sum_\pi \exp\left[\frac{1}{2 T^2} \sum_{ab}
 \int_k g_k^{-2} G^\pi_{ab}(k) v^a_{-k} \cdot v^b_k\right]\, ,
 \label{start}
\end{equation}
where $\tilde \sum_\pi$ denotes a normalized average over permutations, i.e., $\tilde \sum_{\pi} 1=1$.
The remainder of this Section is devoted to the analysis of this formula.

In principle, by expanding this formula in powers of $v$ to all orders, and regrouping terms one could check
that it is indeed possible to put it into the form (\ref{Wrepsum}),
and to compute all the cumulants (we denote by $\hat S^{(n)}$ the $n$-th cumulant). This is a formidable task, however, and we
will focus here only on the second cumulant function $\hat R$. Before computing it directly,
let us give a flavor of the direct expansion in powers of $v$.

\subsection{Direct expansion in powers of $v$}
\label{sec:directexpansion}
The expansion of (\ref{start}) in powers of $v$ starts as:
\begin{eqnarray}\label{expRSB}
 \hat W[v] - \hat W[0]
&=& \tilde \sum_\pi \frac{1}{2 T^2} \sum_{ab} \int_k g_k^{-2} G^\pi_{ab}(k) v^a_{-k} \cdot v^b_k \nn\\
&& \quad\quad + O(v^4)\,.
\end{eqnarray}
Let us recall the structure and parametrization of a hierarchical Parisi matrix.  Dropping temporarily the
$k$-dependence one has \cite{footnoteRSB,footnote:order_of_u's}
\begin{eqnarray}\label{hierarchy}
 G_{ab} &=& [\tilde G - G(n)] \delta_{ab} + G(n) (\1_n)_{ab}\,,  \quad [\text{RS}], \\
 G_{ab} &=& [\tilde G - G(1)] \delta_{ab} + [G(1)-G(n)] (\1_{{\sf u}_c}^{(n)})_{ab} \nonumber \\
&& + G(n) (1_n)_{ab}\,, \quad [\text{1-step RSB}], \\
 G_{ab} &=& [\tilde G - G(1)] \delta_{ab} + \int_n^1 d{\sf u} \,\frac{dG({\sf u})}{d{\sf u}} (\1_{\sf u}^{(n)})_{ab} \nonumber \\
&& + G(n) (\1_n)_{ab}\,, \quad  [\text{continuous RSB}].\label{hierarchyRSB}
\end{eqnarray}
We have defined $\1^{(n)}_m$ as the matrix made of $n/m$ identical blocks along the diagonal, each block being
the $m$ by $m$ matrix with all entries equal to one, and $\1^{(n)}_n=\1_n$. As an example,
\begin{equation}
\1_{3}^{(12)}= \left(\begin{array}{cccccccccccc}
1&1&1&0&0&0&0&0&0&0&0&0\\
1&1&1&0&0&0&0&0&0&0&0&0\\
1&1&1&0&0&0&0&0&0&0&0&0\\
0&0&0&1&1&1&0&0&0&0&0&0\\
0&0&0&1&1&1&0&0&0&0&0&0\\
0&0&0&1&1&1&0&0&0&0&0&0\\
0&0&0&0&0&0&1&1&1&0&0&0\\
0&0&0&0&0&0&1&1&1&0&0&0\\
0&0&0&0&0&0&1&1&1&0&0&0\\
0&0&0&0&0&0&0&0&0&1&1&1\\
0&0&0&0&0&0&0&0&0&1&1&1\\
0&0&0&0&0&0&0&0&0&1&1&1
\end{array} \right)
\ .
\end{equation}
With these formulae one easily checks that
\begin{eqnarray}\label{38-1883213724}
 G^c &:=& \sum_b G_{ab}=\tilde G - \int_n^1 G({\sf u}) d{\sf u} + n G(n)\, , \nn\\
&\stackrel{n \to 0}{=}& \tilde G - \int_0^1 G({\sf u}) d{\sf u}\,.
\end{eqnarray}
For symmetry reasons, the permutation average must yield a replica symmetric matrix
\begin{eqnarray}\label{Gres}
 \tilde \sum_\pi G^\pi_{ab} = \alpha \delta_{ab}+\beta\,.
\end{eqnarray}
Setting $a\neq b$ one finds $\beta=(n-1)^{-1}\sum_{b \neq 1} G_{1b}=(G^c-\tilde G)/(n-1)$, while multiplying with $\delta_{ab}$ shows that $\alpha+\beta=\tilde G$. Thus, one finds:
\begin{eqnarray}
 \tilde \sum_\pi G^\pi_{a b} &=& \frac{1}{1-n} \left[ (G^c - n \tilde G) \delta_{ab} + \tilde G - G^c\right] \nonumber \\
&\stackrel{n\to 0}{\to}& \left[G^c+n(\tilde G - G^c)\right]\delta_{ab} + \tilde G - G^c, \label{resG}\qquad
\end{eqnarray}
which is a replica symmetric matrix by construction, but with non-trivial
entries in the case of RSB. Terms of higher order in $n$ have been neglected, except for the linear term in the replica diagonal part which we retain for later use.

Inserting in (\ref{expRSB}) we finally have
\begin{eqnarray}\label{expRSB1}
\!\!&& \hat W[v] - \hat W[0]\\
\!\!&&\quad = \frac{1}{2 T^2} \sum_{a}\int_k g_k^{-2} \left[G^c(k)+n(\tilde G(k) - G^c(k))\right] |v^a_{k}|^2\nn\\
\!\!&& \quad\quad + \frac{1}{2 T^2} \sum_{ab} \int_k g_k^{-2}(\tilde G(k) - G^c(k))  v^a_{k} \cdot v^b_{-k} + O(v^4)\,.\nn
\end{eqnarray}
On the other hand, expanding (\ref{Wrepsum}) we obtain, for $n=0$:
\BEA\label{expW}
\hat W[v] - \hat W[0]
&=& \frac{1}{2} \sum_{ab} \int_k \left( \frac{\delta_{ab}}{Tg_k}   - \frac{1}{T^2} \hat R''_k[0] \right) v^a_k \cdot v^b_{-k}\nn\\
&&\quad\quad\quad + O(v^4), \EEA since the fourth cumulant must start as $v^4$ and the third as $v^6$ (as one
cannot construct STS invariant combinations of smaller degree - see Appendix B of Ref.
\onlinecite{BalentsLeDoussal2004} for an argument in the case $N=1$). Here and below we adopt the notation
\begin{eqnarray} \label{matrder}
 \hat R''_{xx'}[v]&=&\frac{\delta^2 \hat R[v]}{\delta v_x^i \delta v_{x'}^i}\, , \\
 \int_{xx'} e^{ik x+ik'x'} \hat R''_{xx'}[v] &=& (2 \pi)^d \delta(k+k') \hat R''_k[v]\,,\qquad
\end{eqnarray}
where the last line uses translational invariance and one has set $v_x=v$ after taking the derivatives. Further, we have assumed
$O(N)$ symmetry so that the second derivative of $R$ is diagonal and independent of the $O(N)$ index $i$ in
(\ref{matrder}) (where no index summation is assumed).

Identifying (\ref{expRSB1}) and (\ref{expW}) 
shows that for $n=0$:
\begin{eqnarray}
\tilde G(k) = T g_k - g^2_k \hat R''_k[0]
\quad , \quad G^c(k) = T g_k, \label{secder}
\end{eqnarray}
the second identity being a simple consequence of the STS symmetry. Hence, once the functional $\hat R=R$ of the
FRG is known, the correlation function computed via the GVM can be retrieved as:
\begin{equation}
\tilde \sum_\pi \langle u^a_{-k} u^b_{k} \rangle_\pi = \tilde \sum_\pi G^\pi_{ab}(k)
= T g_k \delta_{ab} - g_k^2 R''_k[0].
\end{equation}
This property of $R$, here guaranteed by its definition (\ref{start}) via the effective action, does not hold for the observable
defined in Ref.~\onlinecite{BalentsBouchaudMezard1996} (having a similar form but with $G \to G^{-1}$).
This made the comparison of their results with the FRG problematic.

To evaluate the identity (\ref{secder}) at $k=0$, we use that for a uniform $v$, $d/dv = \int dx \frac{\delta}{\delta v_x}$, and thus:
\begin{eqnarray}
&& \hat R''_{k=0}[v]=\frac{1}{L^d} \int_{xx'} \hat R''_{xx'}[v] = \hat R''(v),
\end{eqnarray}
which yields:
\begin{eqnarray}
\label{FRGcorrelator2}
\tilde G(k=0) = \frac{T}{m^2} - \frac{\hat R''(0)}{m^4}.
\end{eqnarray}
This relation is exact, and the task is hence to evaluate $\hat R''(0)$. There is however a crucial subtlety in
evaluating this derivative at $v \to 0$. If one uses that $\hat R(v)=R(v)$, together with the result of \OfrgN:
\begin{equation} \label{relationRB}
\hat R(v)=N \tilde B(v^2/N) \quad , \quad v^2 \sim N
\end{equation}
a relation also obtained directly in Section~\ref{sec:uniform v}, one finds that (\ref{FRGcorrelator2})
coincides with formula (\ref{FRGcorrelator}). However, this result is valid only in the region $v^2 \sim N$ and,
as pointed out above, it does {\it not} reproduce the full MP result for $\tilde G$. To obtain the latter, as
we show below, one needs to be more careful in the $v \to 0$ limit and compute $\hat R''(0)$ in the region where
$v^2 \sim L^{-d} \ll N$. One could say that the $v \to 0$ and $N \to \infty$ limit do not commute, or more
accurately, that to obtain contributions of all ultrametric states (and not just the most distant ones) one must
take the limit $v \to 0$ with great care.

The next order $O(v^4)$ is obtained in App.~\ref{app:v4} by the direct expansion method and in the next Section by a more
powerful method which can handle all orders, and to which we now turn.

\subsection{Second cumulant from a two-group analysis}
\label{sec:twogroups}

While the results of the previous section were completely general and independent of the RSB scheme, we now
focus on a specific choice of external sources. In order to compute the
second cumulant function, one best uses two sets of replica, which we denote by $v_x^a=v^1_x$ for $a=1,\ldots,n/2$
and $v_x^a=v^2_x$ for $a=1+n/2,\ldots,n$. Inserting into (\ref{Wrepsum}) we find
\begin{eqnarray}
\hat W[v]-\hat W[0] &=& \frac{1}{2 T} \frac{n}{2} \int_k g_k^{-1} (v^1_k \cdot v^1_{-k} + v^2_k \cdot v^2_{-k}) \qquad \label{expn} \\
&&+ \frac{1}{2 T^2} \frac{n^2}{2} (\hat R[v^{21}] - \hat R[0]) + O(n^3),\nn
\end{eqnarray}
where
\begin{equation}
v^{21}:=v^2-v^1,
\end{equation}
since all higher cumulants yield higher powers of $n$. We compare this
with expression (\ref{start}), slightly rewritten as:
\bea
\label{start2}
&& e^{\hat W[v] - \hat W[0] } =\\
&& \quad \tilde \sum_\pi \exp\left[\frac{1}{2 T^2} \sum_{ab} \int_k g_k^{-2} G_{ab}(k) v^{\pi(a)}_{-k} \cdot v^{\pi(b)}_k\right]\,.\nn
\eea
In order to perform the sum over permutations, we introduce Ising spins $\tau^a$ as in Ref.~\onlinecite{BalentsBouchaudMezard1996}:
\begin{equation}
v^a = \frac{v^1+v^2}{2} + \tau^a \frac{v^{21}}{2},\label{param}
\end{equation}
where $\tau^a=-1$ if $\pi(a) \in [1,n/2]$, $\tau^a=+1$ otherwise. Each configuration
$\{ \tau_a \}$ is left invariant by $[(n/2)!]^2$ permutations, the number of distinct
spin configurations being $C_n^{n/2}=n!/[(n/2)!]^2$. They all correspond to vanishing total magnetization, i.e.,
$\sum_{a=1}^n \tau^a=0$. Inserting (\ref{param}) into (\ref{start2}) the term linear in $\tau$ vanishes, and one finds:
\begin{eqnarray}
&& \!\!\!e^{\hat W[v] - \hat W[0] } = \exp \left[\frac{n}{4 T^2} \int_k g_k^{-2} G^c(k) (|v^1_k|^2 + |v^2_k|^{2}) \right]
\nonumber  \\
&&  \times
\frac{1}{C_n^{n/2}} \sum'_{\{\tau\}} \exp \left[\frac{1}{2 T^2} \int_k g_k^{-2}{ \frac{|v_k^{21}|^2}4 }\sum_{ab}
\hat G_{ab}(k) \tau^a \tau^b \right]\,,  \nn
 \label{start2p}\\
 && \!\!\! \hat G_{ab} := G_{ab} - G^c \delta_{ab}
\end{eqnarray}
The first factor equals the term proportional to $n$ in (\ref{expn}), as seen upon using (\ref{secder}). The prime on the sum indicates that the sum extends over all
Ising spin configurations subject to the global constraint $\sum_{a=1}^n \tau^a=0$. Identifying the above expression with
(\ref{expn}),  we arrive at the formula:
\begin{eqnarray}
 \hat R[v] - \hat R[0] \label{Rising} & =& \lim_{n \to 0} \frac{4 T^2}{n^2}
 \frac{1}{C_n^{n/2}} \\
&& \hspace{-30pt}\times \sum'_{\{\tau\}} \left[ \exp \left(\int_k \frac{g_k^{-2}}{8 T^2} |v_k|^2 \sum_{ab}
\hat G_{ab}(k) \tau^a \tau^b \right) - 1 \right]\,. \nonumber
\end{eqnarray}
Note that we have simplified notations by renaming $v^{21} \to v$.
At this stage we can check the small-$v$ expansion again:
\begin{eqnarray}
\label{v2term}
&& \frac{1}{2} \int_k \hat R''_k[0] |v_k|^2 = \\
&&\quad\quad \lim_{n \to 0} \frac{T^2}{2 n^2}
 \int_q \frac{g_k^{-2}}{2T^2} |v_k|^2 \tilde\sum'_{\{\tau\}}
 \hat G_{ab}(k) \tau^a \tau^b . \nonumber
\end{eqnarray}
If we denote the normalized average
$A_{ab}:=\tilde \sum'_{\{\tau\}}
\tau_a \tau_b = \frac{1}{C_n^{n/2}} \sum'_{\{\tau\}}
\tau_a \tau_b$ then one has $A_{aa}=1$. Further, the identity $0=\sum_b A_{ab}=A_{aa} + (n-1) A_{a \neq b}$ implies
$A_{a \neq b}=1/(1-n)$ and one can thus write $A_{ab}=(1-n\delta_{ab})/(1-n)$ and
$\sum_{ab} \hat G_{ab} A_{ab} = - n^2/(1-n) [\tilde G(k)-G^c(k)]$.
Using this in (\ref{v2term}) we recover
the result (\ref{secder}) of the previous Section. Note that the same result is obtained from the piece $\sim n^2$ in the replica-diagonal part of (\ref{expRSB1}), while the two-replica sum vanishes.

To evaluate the restricted spin sum in (\ref{Rising}) we use the same method as in Ref.~\OBMP, leading to Parisi's nonlinear diffusion equation in the form discussed by Duplantier~\cite{Duplantier1981}. We first eliminate the constraint of zero magnetization and rewrite (\ref{Rising}) as:
\begin{widetext}
\begin{equation}
\label{completeR}
\hat R[v] - \hat R[0]  = \lim_{n \to 0} \frac{4 T^2}{n^2} \frac{2^{n+1}\Gamma(-n)}{[\Gamma(-n/2)]^2} \int_{-\infty}^\infty
dy \sum_{\{\tau\}} \left[ \exp \left(\int_k \frac{g_k^{-2}|v_k|^2}{8 T^2}  \sum_{ab=1}^n
\hat G_{ab}(k) \tau^a \tau^b + y \sum_{a=1}^n \tau^a \right) - \exp \left(y \sum_{a=1}^n \tau^a \right) \right]
\end{equation}
\end{widetext}
which makes use of an identity derived in Ref.~\OBMP, valid for $n <0$ only.

The evaluation of a spin sum such as it appears in (\ref{completeR}) is standard in the mean-field theory of
spin glasses. Here we summarize the main steps following Duplantier~\cite{Duplantier1981}. We write the matrix which couples the spins as
\BEA
q_{ab} &=& \frac{1}{4 T^2} \int_k g_k^{-2} |v_k|^2 \hat G_{ab}(k),\nn\\
q({\sf u}) &=& \frac{1}{4 T^2} \int_k g_k^{-2} |v_k|^2  G(k,{\sf u}),\label{qdeu}
\EEA
with $q^c=\sum_{b} q_{ab}=0$.
Using (\ref{GofuMP}) this becomes
\begin{equation}
q({\sf u}) = \frac1{4T}\int_{k} |v_{k}|^{2 }\left[ \sigma({\sf u}_{m}) +\int_{{\sf u}_{m}}^{{\sf u}} \frac{\dot \sigma({\sf u}) d {\sf u}}{\left[g_{k}^{-1}+[\sigma]({\sf u})\right]^{2}}\right]
\end{equation}
Let us now assume that $q_{ab}$ has a $K$-step ultrametric structure with breakpoints at
\BEA
\label{Kstep}
n\equiv {\sf u}_0 \prec {\sf u}_1\prec \dots\prec {\sf u}_K\prec {\sf u}_{K+1}\equiv 1,
\EEA
and entries parametrized by $q({\sf u})$ with
\BEA
q({\sf u})=q_\ell,\quad \text{for}\quad  {\sf u}_\ell\preccurlyeq {\sf u} \prec {\sf u}_{\ell+1}.
\EEA
Further we define $q_{-1}\equiv 0$.
Let us introduce the ``partial partition sums'':
\BEA
&&\!\!\!\!\!g_\ell(y)\equiv e^{{\sf u}_\ell \psi_\ell(y)} := \\
&& \!\!\sum_{\{\tau^a\}, a=1,..,{\sf u}_\ell}  \exp \left[ \frac{1}{2} \sum_{a,b=1}^{{\sf u}_\ell} \tau^a \left(q_{ab}-q_{\ell-1}\right) \tau^b +
\sum_{a=1}^{{\sf u}_\ell} \tau^a y \right], \nonumber
\end{eqnarray}
which defines $g_\ell(y)$ and $\psi_\ell(y)$ for $\ell=0,\dots K+1$. Obviously,
$g_{K+1}(y)=2\cosh(y)\exp[\tilde q-q({\sf 1})]$.

The expression (\ref{completeR}) can then be rewritten as:
\begin{eqnarray}
\hat R[v] - \hat R[0]  &=& \lim_{n \to 0} \frac{-2 T^2}{n}  \int_{-\infty}^\infty
dy \left[g_0(y) - (2 \cosh y)^n\right] ~~\nonumber \\
& =& - 2 T^2 \int_{-\infty}^\infty dy \left[\psi_0(y) - \ln(2 \cosh y)\right]\,, \label{integral}
\end{eqnarray}
where we used that $[\Gamma(-n/2)]^2/\Gamma(-n) = -4/n+O(n^0)$.

As shown in Ref.~\onlinecite{Duplantier1981}, the functions $g_\ell(y)$ satisfy a recursion relation, the idea being as follows.
Consider one of the $n/{\sf u}_\ell$ equivalent groups of size ${\sf u}_\ell$, for instance the first one
$a=1,\ldots,{\sf u}_\ell$. It contains $\frac{{\sf u}_\ell}{{\sf u}_{\ell+1}}$ subgroups of size ${\sf u}_{\ell+1}$\cite{footnote:order_of_u's}. The only coupling between the subgroups is through
the matrix $\Delta q_\ell 1^n_{{\sf u}_\ell}$ with $\Delta q_\ell =q_{\ell}-q_{\ell-1}$ \footnote{The term
$-q_{\ell-1}$ naturally arises in the construction~\cite{footnoteRSB} for a complete Parisi matrix, see also
(\ref{hierarchyRSB}). Here it arises from the extra term $-q_{\ell-1}$ in the definition of $g_\ell$ which
involves an incomplete Parisi matrix, (the replica sum is interrupted at ${\sf u}_{\ell}$).}, which couples
uniformly all spins in the group. They can be decoupled by a Hubbard-Stratonovich transformation, adding a term
$z \sqrt{\Delta q_\ell} \sum_{a=1}^{{\sf u}_\ell} \tau_a$, where $z$ is a Gaussian random field of variance
$\Delta q_\ell$, acting on all spins in the group. This allows to perform configurational sums independently
within each subgroup and yields the recursion relation:
\begin{equation}
\label{recursion}
g_{\ell}(y) = \left\langle g_{\ell+1}\left(y+z \sqrt{\Delta q_\ell}\right)^{\frac{{\sf u}_\ell}{{\sf u}_{\ell+1}}} \right\rangle_z\,,
\end{equation}
where here and below $\langle \ldots  \rangle_z := \int_{- \infty}^\infty \frac{dz}{2 \pi}  \ldots e^{-z^2/2} $ denotes the
average over a unit Gaussian.

\subsection{Continuous RSB: Second cumulant from an evolution equation}
\label{sec:evolution}
In the case of continuous RSB, we have to take the continuum limit $K\to \infty$ in the above: $\Delta {\sf u}_\ell={\sf u}_{\ell+1}-{\sf u}_{\ell} \to 0$, $q({\sf u}_\ell)\equiv q_\ell\to q({\sf u})$ and $\Delta q_\ell\to 0$, while $\Delta q_\ell/\Delta {\sf u}_\ell\rightarrow dq({\sf u})/d{\sf u}$. For the function $\psi({\sf u}_\ell,y) \equiv \psi_\ell(y)\rightarrow \psi({\sf u},y)$, Eq.~(\ref{recursion}) yields a differential equation:
\begin{eqnarray}
 \partial_{\sf u} \psi = - \frac{1}{2} \frac{d q({\sf u})}{d{\sf u}} \left[\partial_y^2 \psi + {\sf u} (\partial_y \psi)^2\right]\,,
\label{evolution}
\end{eqnarray}
with initial condition:
\BEA
\psi({\sf u}_c,y)=\psi({\sf 1},y)=\frac{\tilde q - q({\sf 1})}{2} + \ln[2 \cosh(y)]\,, \label{initial}
\EEA
where we assume $q({\sf u})$ to be constant for ${\sf u}_c<{\sf u}< 1$, as is generally the case, cf., (\ref{sig}).

Eq.~(\ref{evolution}) must be integrated from ${\sf u}_c$ to ${\sf u}_m$, where $q({\sf u})$ reaches its lower plateau $q_0=q({\sf u}_m)$. If $q_0\neq 0$, the recursion (\ref{recursion}) for $\ell=0$ shows that we have to perform a last convolution to obtain
\begin{equation}
 \psi({\sf 0},y) 
 = \langle \psi({\sf u}_m , y + z \sqrt{q({\sf 0})}) \rangle_z\,.
\end{equation}
The solution of (\ref{evolution}) with initial condition (\ref{initial})
behaves at large $|y|$ like:
\begin{equation}
 \psi({\bf {\sf u}},y) \stackrel{|y|\rightarrow \infty}{\approx} |y| + \frac{1}{2} \left(\tilde q - \left[\int_{\sf u}^1 d{\sf u}'\,q({\sf u}') \right]- {\sf u} q({\sf u})\right)\,, \label{asymptot}
\end{equation}
as can be seen by substituting (\ref{asymptot}) into (\ref{evolution}).
This is also confirmed by noting that for
 $|y|\gg 1$ the sum in (\ref{completeR}) is dominated by the configuration with all $\tau^a= \text{sign}(y)$, which leads to the simple approximation ({\ref{asymptot}) for $\psi({\bf {\sf u}},y)$. This behavior
 implies $\psi({\sf u}_m,y) \approx |y| + q^c = |y|$, since $q^c=0$ from
the condition $\hat G^c=0$. Hence the integral (\ref{integral}) converges, the subleading terms of $\psi$ decaying exponentially at large $|y|$.

We can now state the main result of this section. In the regime $v^2 = O(1)$, the
second cumulant {\it functional} can be obtained from the saddle point of the GVM
as follows:
\begin{eqnarray}
 \hat R[v] - \hat R[0] = 2 T^2 \int_{-\infty}^\infty dy\, y \left[M({\sf 0},y) - \tanh (y)\right]\,, \label{intM}
\end{eqnarray}
where the function $M({\sf u},y)=\partial_y \psi({\sf u},y)$ is the solution of
\begin{eqnarray}
&& \partial_{\sf u} M =  - \frac{1}{2} \frac{d q({\sf u})}{d{\sf u}} \left(\partial_y^2 M + 2 {\sf u} M \partial_y M\right), \label{evolM} \\
&& M({\sf u}_c,y)=\tanh(y),
\end{eqnarray}
in the interval ${\sf u} \in [{\sf u}_m,{\sf u}_c]$, and \BEA M({\sf 0},y) = \left\langle M({\sf u}_m , y + z
\sqrt{q({\sf 0})}) \right\rangle_z\,. \label{M0} \EEA Eq.~(\ref{intM}) follows from Eq.~(\ref{integral}) by
integration by parts, using that $\lim_{y \to \pm \infty} y \left\{\psi({\sf 0},y)-\ln[2 \cosh(y)]\right\} = 0$
as discussed above.

From the above and (\ref{qdeu}) we see that the dependence of the functional $\hat R[v]$ on the field $v_x$ occurs only through the combination
$q({\sf u}) = \frac{1}{4 T^2} \int_{xy} h(x-y,{\sf u}) v(x) \cdot v(y)$, where $h(k,{\sf u})=g_k^{-2} G(k,{\sf u})$. Hence for
a uniform $v_x=v$ one easily sees that $q({\sf u}) \sim L^d v^2$ and we can expect that (\ref{intM}) will assume a scaling form with scaling variable $v L^{d/2}$. This dependence on the system size is very different from the one in the regime $v^2 = O(N)$
and will be commented on further below.

The convolution (\ref{M0}) only results in an additive contribution to the potential correlator. This can be
seen using the identity: \BEA \label{identity}
&&\int_{-\infty}^\infty dy \,y \left[\langle \Phi( y + \sqrt{Q} z ) \rangle_z - \Phi(y)\right] =\nn\\
&&\quad\quad\quad\quad - \frac{Q}{2}   \left[\Phi(\infty)-\Phi(-\infty)\right],
\EEA
proven in App.~\ref{app:identity} for functions $\Phi$ whose derivative decreases faster than $1/|y|$ at infinity. Applying (\ref{identity}) to $\Phi(y)=M({\sf u}_m,y)$, we can rewrite (\ref{intM}) as:
\begin{eqnarray}
 \hat R[v]-\hat R[0] &=& - \frac{1}{2} \int_k g_k^{-2} G(k,{\sf u}_m) |v_k|^2 \label{intM2} \\
&& + 2 T^2 \int_{-\infty}^\infty dy\, y [M({\sf u}_m,y) - \tanh (y)]. \nn
\end{eqnarray}
This result can also be derived via an alternative route, providing a useful check of the above formalism:
In (\ref{completeR}) we can directly separate out the contribution form the most distant states by writing $\hat G_{ab}(k)=\overline{G}_{ab}(k)+\hat{G}(k,{\sf 0})(1-n\delta_{ab})$, with $\overline{G}^c(k)=0$, and $\overline{G}(k,{\sf 0})=0$. One can easily check that the piece $\hat{G}(k,{\sf 0})(1-n\delta_{ab})$ produces the first term in (\ref{intM2}). The remaining part leads to the same expression as (\ref{completeR}), with the replacement $\hat G\to\overline{G}$. The only difference in the subsequent evaluation is that there is no need for a convolution in the end, since $\overline{G}^c(k)=0$. This establishes (\ref{intM2}).

Note that the first term in (\ref{intM2}) has the expected form for the contribution from the plateau $0\leq {\sf u} \leq{\sf u}_m$, cf., Eqs.~(\ref{diagonalG},\ref{resfrg}), with a coefficient of $|v_k|^2$
\begin{eqnarray}
-\frac{1}{2}g_k^{-2} G(k,{\sf u}_m)= -\frac{1}{2}T \sigma({\sf u}_m)=\tilde B'(0),
\end{eqnarray}
independent of $k$.
Clearly, this term is the only contribution in the case of a replica
symmetric solution. We now show that it also gives the dominant contribution
at large $v^2$ of order $O(N^0)$, but $v^2\gg L^{-d}$.

\subsubsection{Limit of $L^d v^2\gg 1$ }
\label{sec:largevFRSB}
In this case, Eq.~(\ref{evolM}) can be rewritten as
\begin{eqnarray}\label{104}
 -\frac{1}{2}\left(M'' + 2 {\sf u} M' M\right) =  \frac{dM}{dq} \to 0
\end{eqnarray}
since $q$ is large (from now on we denote $\partial_y$ by a prime). One can integrate this equation to $M'+{\sf
u}(M^2-1)=0$, where the integration constant (w.r.t.\ $y$) is fixed by the fact that $M({\sf u}_{c},y)\to 1$ for large $y$, which cannot be changed by the evolution (\ref{evolM}). The solution of (\ref{104}) is then
$M({\sf u},y)=\tanh({\sf u} y)$. One can check that the flow of $M$
is attracted to this simple ``fixed point'' as ${\sf u}\to {\sf u}_m$, if $v\gg v_*$, with $v_*$ defined below. In that case,
the second term in (\ref{intM2}) becomes:
\begin{eqnarray}
&& 2 T^2 \int_{-\infty}^\infty dy\, y \left[\tanh({\sf u}_m y) - \tanh(y)\right] \nn\\
&& \quad =  4 T^2 \left(\frac{1}{{\sf u}_m^2}-1\right) \int_{0}^\infty dy\, y \left[\tanh(y) - 1\right],\qquad
\end{eqnarray}
and the large-$v$ behavior of the second cumulant in this $v^2=O(1)$
regime is thus:
\begin{equation}
\hat R[0]-\hat R[v] \stackrel{v \to \infty}{\approx} \frac{T\sigma({\sf u}_m)}{2}  \int_x v_x^2
+ \frac{\pi^2}{6} T^2 \left(\frac{1}{{\sf u}_m^2}-1\right).  \label{largev}
\end{equation}
The leading behavior is quadratic in $v$ and corresponds to the contribution of the most distant states to the
full (inverse) correlation. Hence it should match the result obtained in the FRG for the regime $v^2\sim N$, cf.\ Eqs.~(\ref{FRGcorrelator},\ref{resfrg}) or
equivalently, Eqs.~(\ref{FRGcorrelator2},\ref{relationRB}). Indeed, it does!

For a uniform $v_x=v$ and in the limit $T\to 0$, one finds for both models 
(\ref{modelI},\ref{modelII}), using (\ref{sig},\ref{22}):
\begin{equation}
\hat R[0]-\hat R[v] \stackrel{v \to \infty}{\approx} \frac{m^{2-\theta}v^2 L^d}{A(2-\theta)}
+ \frac{\pi^2}{6 A^2} m^{-2 \theta}.  \label{largevmod}
\end{equation}
This suggests a crossover around $v \sim v_*$ defined as:
\begin{equation}
\label{vstar}
m^2 L^d v_*^2 =  m^{-\theta} \quad \leftrightarrow \quad v_* = (m L)^{- d/2} m^{-\zeta}.
\end{equation}

The physical meaning of this crossover scale will be discussed in more detail in the context of the models with one-step RSB below, cf. Sec.~\ref{ss:InterpretationofTBL}.

\subsubsection{Perturbation expansion for $L^d v^2\ll 1$}
\label{sec:PTfullFRG}

Since for $q({\sf u})=0$, one has $M({\sf u},y)=\tanh(y)=:m_0(y)$ there is a uniquely defined expansion in
powers of $q$, which since $q\sim v^{2}$ is equivalent to the direct expansion in powers of $v^2$ of Sec.~\ref{sec:directexpansion}:
\begin{equation}
M({\sf u},y) = m_0(y) + m_1({\sf u},y) + m_2({\sf u},y) + \dots\,.
\end{equation}
Each $m_n({\sf u},y)$ contains only terms of degree $q^n$. They satisfy the recursion:
\begin{eqnarray} \label{eqs1}
 \dot m_n = - \frac{1}{2} \dot q({\sf u}) \left[m_{n-1}'' + {\sf u} \left(\sum_{p=0}^{n-1} m_p \label{eqsn} m_{n-p-1}\right)'\right]\,,\qquad
\end{eqnarray}
where here and below dots denote $\p_{\sf u}$. 
The initial conditions are $m_n({\sf u}_c,y)=0$. The final result for
$\hat R[v]$ is:
\begin{eqnarray} \label{sumR}
 \hat R[v]-\hat R[0] &=& - 2 T^2 q({\sf 0}) + 2 T^2 \sum_{n \geq 1} \int_{-\infty}^\infty dy\, y ~ m_n({\sf u}_m,y) \nonumber \\
& =:& \sum_{n \geq 1} R_n[v],
\end{eqnarray}
where we recall that $4 T^2 q({\sf u}) = \int_k g_k^{-2} |v_k|^2 G(k,{\sf u})$. Hence $R_n$
contains all terms of degree $q^n \sim v^{2 n}$, and we are thus
 effectively computing the derivatives at the origin, $\hat R^{(2 n)}[v=0]$.

This calculation is performed in Appendix \ref{app:PTfullFRG}, and the results are
indeed consistent with the direct expansion of Sec.~\ref{sec:directexpansion} although the method is quite different.
The lowest order term reads:
\begin{eqnarray}
\label{smallvFRSB1}
R_1[v] &=& - 2 T^2 \int_0^1 q({\sf u}) d{\sf u} \nonumber \\
&=& - \frac{1}{2} \int_k g_k^{-2} \left[\tilde G(k)-G^c(k)\right] |v_k|^2\,,
\end{eqnarray}
using that $\int_0^1 G({\sf u})=\tilde G - G^c$,
recovering the result (\ref{expRSB},\ref{resG}) which
yields, at small $v$, the full result of M\'ezard-Parisi for the correlation function.

The next-order term is:
\begin{equation} \label{finalR2}
R_2[v] = \frac{2}{3} T^2 \left(
\int_0^1 d{\sf u}\, q^2({\sf u}) - \left[\int_0^1 d{\sf u}\, q({\sf u})\right]^2 \right)\,.
\end{equation}

As discussed in Refs.~\onlinecite{NattermannScheidl2000,ChauveLeDoussal2001,BalentsLeDoussal2004,LeDoussal2006b,LeDoussalToBePublished}
the fourth derivative at zero of the FRG function $R^{(4)}[0]$ is a direct measure of susceptibility
fluctuations. Indeed we find that our present result has the general form of susceptibility fluctuations within
a Parisi Ansatz, defined and derived in Ref.~\oMP.

\subsubsection{Thermal boundary layer: General formula}
\label{sec:TBLgeneral}

It is possible to resum the derivatives, order by order in temperature and
derive the thermal boundary layer (TBL) form:
\begin{eqnarray}
\hat R[v] - \hat R[0] &=& - \frac{1}{2} \int_k g_k^{-2} \tilde G_{T=0}(k)  |v_k|^2 \nonumber
\\
&& + T^3 \hat r_1[\hat v] + T^4 \hat r_2[\hat v] + \dots, \label{tblexp}
\end{eqnarray}
where the first term is the ``zero-temperature limit''
of the leading small-$v$ quadratic
term \footnote{When comparing (\ref{tblexp}) with (\ref{smallvFRSB1}), note that $G^c_{T=0}=0$ because of (\ref{FRGcorrelator2}).}, and the higher-order terms in the expansion in $T$ are scaling functions of the
boundary-layer variable:
\begin{eqnarray} \label{TBLvariable}
 \hat v_k &=& v_k/T,\\
 q({\sf u}) &=& \frac{1}{4} \int_k g_k^{-2} |\hat v_k|^2 G(k,{\sf u}).
\end{eqnarray}
This structure, which appears already in the 1-loop FRG, can
be computed exactly in the large-$N$ limit here.

This structure appears already in the 1-loop FRG where one finds that at $T>0$ the non-analyticity of $R[v]$ is thermally rounded.~\cite{ChauveGiamarchiLeDoussal1998,ChauveGiamarchiLeDoussal2000,BalentsLeDoussal2004}
Since temperature is irrelevant in the RG sense
when $\theta>0$ this rounding occurs only in a layer determined by $v m^\zeta \sim T
m^\theta$, which becomes smaller and smaller as $T\to 0$, hence the name thermal boundary layer. Here, for the
first time we compute its exact expression in the large-$N$ limit.

We now perform a (slightly more formal) expansion which allows to
obtain (\ref{tblexp}). The idea, looking at (\ref{evolM}), is that the nonlinear
term contains an extra factor of ${\sf u}$ and that for ${\sf u}_m \leq {\sf u} \leq {\sf u}_c$ one has
${\sf u} \sim T$. Hence it is natural to expand in the nonlinearity to
generate a low-temperature expansion. Of course this must be checked
a posteriori. Further, we find it more convenient to use $q({\sf u})$ instead of ${\sf u}$ to parameterize the ultrametric distance. Thus rewriting (\ref{evolM}) as $\partial_q{M}=-\frac12 \left[M''+{\sf u} (M^{2})'\right]$, and expanding $ M = M_0 + M_1 + \dots, $ formally in powers of $\sf u$, we
have to solve the hierarchy:
\begin{eqnarray}
&& \partial_q M_0 = - \frac{1}{2} M_0'', \\
&& \partial_q M_1 = - \frac{1}{2}\left( M_1'' + 2 {\sf u} M_0 M_0'\right), \label{nextorder}\\
&& \partial_q M_n = - \frac{1}{2}\left( M_n'' + 2 {\sf u} \sum_{p=0}^{n-1}M_p M_{n-1-p}'\right).\qquad
\end{eqnarray}
Note that this hierarchy is formally similar to (\ref{eqsn}), apart from a shift in the index of the linear term on the RHS which indicates that we actually perform a different resummation.

The initial condition is $M_0({\sf u}_c,y)=\tanh(y)$, $M_{n\geq 1}({\sf u}_c,y)=0$. Again, primes stand for
$\partial_y$. We define:
\begin{eqnarray}
\hat R[v]-\hat R[0] &=& {\cal R}_0[v]+{\cal R}_1[v]+\dots, \label{TBLexpansionFRSB}\\
{\cal R}_0[v] &=& - 2 T^2 q({\sf 0}) \label{calR0}\\
&& + 2 T^2 \int_{-\infty}^\infty dy\, y \left[M_0(q({\sf u}_m),y)-\tanh(y)\right], \nonumber \\
 {\cal R}_{n\geq 1}[v]&=& 2 T^2 \int_{-\infty}^\infty dy\, y M_n(q({\sf u}_m),y),
\label{calm1}
\end{eqnarray}
where we use the notation ${\cal R}_n$ to distinguish from
the expansion in powers of $q$ discussed in the previous Section, and different from the $\hat r_{i}[v]$ defined in (\ref{tblexp}).

The equation for $M_0$ is a simple diffusion equation with solution:
\begin{eqnarray}
M_0(q({\sf u}),y) &=& \langle \tanh(y + \sqrt{q({\sf u}_c,{\sf u})} z ) \rangle_z\,, \label{soluM0} \\
  q({\sf u}_c,{\sf u}) &:=& q({\sf u}_c)-q({\sf u}).
\end{eqnarray}
Thanks to the identity (\ref{identity}), when $M_0(q({\sf u}_m),y)$ from (\ref{soluM0}) is substituted into (\ref{calR0}), we obtain
\begin{eqnarray}
&& {\cal R}_0[v] = - 2 T^2 q({\sf u}_c),\label{R0FRSB}
\end{eqnarray}
whereby the terms proportional to $q({\sf 0})=q({\sf u}_m)$ cancel. In the limit $T \to 0$ (\ref{R0FRSB}) becomes
the same as the first term in (\ref{tblexp}), since:
\begin{eqnarray}
\tilde G(k) &=& G^c(k) + (1-{\sf u}_c) G(k,{\sf u}_c) + \int_0^{{\sf u}_c} d{\sf u}\,G(k,{\sf u}) \nonumber
\\
& \stackrel {T \to 0}{\approx} & G(k,{\sf u}_c).
\end{eqnarray}
Here, we used $G^c,{\sf u}_c \sim T$ and the fact that $G(k,{\sf u})$ has a finite limit as $T \to 0$.

To obtain the next order term, ${\cal R}_1[v]$, we solve Eq.~(\ref{nextorder}):
\bea
\label{solM1}
 M_1(q,y) &=& - \int_q^{q({\sf u}_c)} dq' dy'\, {\cal D}(q, y ; q', y') \\
 &&\quad\quad\quad \times {\sf u}(q') \left[(M_0M_0')(q',y')\right]\,,\nn
\eea
where:
\begin{equation}
 {\cal D}(q, y ; q', y') = -\frac{\theta(q'-q)}{\sqrt{2 \pi (q'-q)}} \exp\left[ - \frac{(y-y')^2}{2(q'-q)}\right]
\end{equation}
is the (reverse) diffusion kernel satisfying $(\partial_q + \frac{1}{2} \partial_y^2){\cal D}=\delta(q-q') \delta(y-y')$,
and ${\sf u}(q)$ is the inverse function of $q({\sf u})$ for ${\sf u}_m < {\sf u} < {\sf u}_c$. Eq.~(\ref{solM1}) can
be rewritten as:
\BEA
 M_1(q,y) &=&  \int_{q}^{q({\sf u}_c)} dq' {\sf u}(q') \times\\
 &&\quad\quad\quad \left\langle (M_0 M_0')(q',y + z \sqrt{q'-q}) \right\rangle_{z}\,,\nn
\EEA
which inserted into (\ref{calm1}) gives:
\begin{eqnarray}
{\cal R}_1[v] &=&  2 T^2 \int_{q({\sf 0})}^{q({\sf u}_c)} dq' {\sf u}(q')\times\\
&& \quad\times
 \int_{-\infty}^\infty dy\, y  \langle (M_0 M_0')(q',y + z \sqrt{q'-q({\sf 0})}) \rangle_{z}\,.
\nonumber
\end{eqnarray}
We can now use the identity (\ref{identity}) for $\Phi=M_0 M_0'$ which has rapidly decreasing derivatives and satisfies $\Phi(\pm\infty)=0$. This yields:
\begin{eqnarray}
 {\cal R}_1[v] =  2 T^2
\int_{q({\sf 0})}^{q({\sf u}_c)} dq\, {\sf u}(q) \int_{-\infty}^\infty dy\, y  (M_0 M_0')(q,y)\,. \nonumber
\end{eqnarray}
Integrating by parts over $y$ and using the solution (\ref{soluM0}) for $M_0$ one finally obtains:
\begin{widetext}
\begin{eqnarray} \label{w1}
&& {\cal R}_1[v]
=  T^2
\int_{q({\sf 0})}^{q({\sf u}_c)} dq {\sf u}(q) \int_{-\infty}^\infty dy \left[1 - \langle \tanh(y+z_1 \sqrt{q({\sf u}_c)-q})
\tanh(y+z_2 \sqrt{q({\sf u}_c)-q}) \rangle_{z_1,z_2}\right]\,.
\end{eqnarray}
\end{widetext}
Note that we have used $(M_{0}^{2}-1)$ as primitive of $2 M_{0} M_{0}'$, since we need it to vanish for large argument.
Using
\begin{eqnarray} \label{phiw2}
\psi(a-b)&:=&\int_{- \infty}^\infty dy \left[1 - \tanh(y+b) \tanh(y+a)\right] \nn\\
&=& 2 (a-b) \coth(a-b),
\end{eqnarray}
we obtain the exact scaling functional for the thermal boundary layer
\footnote{Note that the function $T^3 r_1[\hat v]$ in (\ref{TBL}) differs from $T^3 \hat r_1[\hat v]$ in the systematic expansion (\ref{tblexp}) by a trivial
quadratic term $-2T^3 dq({\sf u}_c)/dT|_{T=0} \propto T^3\hat v^2 = T v^2$.}:
\begin{eqnarray}
 {\cal R}_1[v] & =& T^3 r_1[\hat v], \label{TBL}\\
 r_1[\hat v] &=&  2  \int_{q_m}^{q_c} dq\, \hat {\sf u}(q) \times\nn\\
 &&\quad \quad \left\langle z \sqrt{2(q_c-q)} \coth(z \sqrt{2(q_c-q)})
\right\rangle_{z}\,,\nn\\
\end{eqnarray}
where:
\BEQ
\hat {\sf u}(q):=\frac{{\sf u}(q)}T
\EEQ
has a finite limit as $T \to 0$, where we introduced the definitions:
\begin{eqnarray}
 q_c&:=&q({\sf u}_c)=q({\sf 1})=\frac{1}{4} \int_k g_k^{-2} |\hat v_k|^2 G(k,{\sf u}_c), \qquad\\
 q_m &:=& q({\sf u}_m)=q({\sf 0})=\frac{T \sigma({\sf 0})}{4}  \int_k |\hat v_k|^2,\\
\tilde q &=& \frac{1}{4} \int_k g_k^{-2} |\hat v_k|^2 \tilde G(k).
\end{eqnarray}

\subsubsection{Thermal boundary layer: Uniform $v$ and scaling function
for $m \to 0$}
\label{sec:TBL}
Let us consider here a uniform $v_x=v$ for
which one has:  \BEA \label{quniform} q({\sf u})= \frac{m^4 \hat v^2
L^d}{4}\, G(k=0,{\sf u}). \EEA Here and below we define: \BEA G(\hat{\sf
u})&=&G(k=0,{\sf u}=T\hat{\sf u}). \EEA
One can rewrite $r_1[\hat v]\equiv L^d
r_1(\hat v)$ with:
\begin{widetext}
\begin{eqnarray} \label{double}
r_1[\hat v] &=&  \frac{m^4 \hat v^2 L^d}{2 } \int_{G_m}^{G_c} dG\, \hat
{\sf u}(G)  \left\langle
z \frac{m^2 |\hat v| L^{d/2}}{2 }  \sqrt{2(G_c-G)} \coth\left[z \frac{m^2
|\hat v| L^{d/2}}{2 }
\sqrt{2(G_c-G  )}\right]
\right\rangle_{z}\,,
\end{eqnarray}
\end{widetext}

where here we denote $G_m:=G({\sf u}_m)$ and $G_c:=G({\sf u}_c)$ (not to
be confused with the connected
correlator $G^c$).
Let us now specify to models I and II. One finds:
\BEA \label{GG}
&& G(\hat{\sf u})= \frac{8}{A(4-\theta^2)} \frac{1}{m^{2+\theta}} -
\frac{2}{2+\theta}
\frac{A^{\frac{2}{\theta}}}{\hat{u}^{1+\frac{2}{\theta}}},
\EEA
where $A$ was given in (\ref{A}). Hence:
\begin{eqnarray} \label{uhat}
&& \hat {\sf u}(G)= \frac{A m^\theta (4-2
\theta)^{\frac{\theta}{2+\theta}}}{
\left[8 -  (4-\theta^2) A m^{2+\theta}
G\right]^{\frac{\theta}{2+\theta}}}\,,
\end{eqnarray}
with $\hat {\sf u}_c=A m_c^\theta$, $A m^{2+\theta}
G_m=\frac{2}{2-\theta}$ and
$A m^{2+\theta} G_c= \frac{8}{4-\theta^2} -  \frac{2}{2+\theta}
(\frac{m}{m_c})^{2+\theta}$.
Using (\ref{uhat}) we obtain the TBL function as a double integral in
(\ref{double}).
In the limit $m \ll m_c$ this simplifies to:
\begin{eqnarray} \label{tblexact}
 r_1[\hat v] &=& m^2 \hat v^2 L^d H_\theta\left( \frac{1}{\sqrt{2 A}}
\frac{\hat v}{v_* m^\theta} \right), \\
H_\theta(x) &=& \frac{1}{2}
\left(\frac{2}{2+\theta}\right)^{\frac{\theta}{2+\theta}}
\int_{0}^{\frac{2}{2+\theta}} \frac{dt}{t^{\frac{\theta}{2+\theta}}}
\left \langle z x \sqrt{t} \coth(z x \sqrt{t})
\right \rangle_{z}, \nn
\end{eqnarray}
where $v_*=L^{-d/2} m^{-1-\frac{\theta}{2}}$ is the scale obtained in
(\ref{vstar}). Further, from
(\ref{A}), $A$ depends only on $d$ and $\gamma$. The scaling
function $H_\theta$ depends only on $\theta$, while the argument is
scaled by
the characteristic displacement $v_*$ multiplied by the scaled
temperature $T m^\theta$. This indicates that
the scaling function of the thermal boundary layer exhibits universality
for $\theta>0$
since only scales of order $1/m$ contribute to the final scaling function,
all
features of $q({\sf u})$ with ${\sf u} \gg {\sf u}_m$ being subdominant.

\subsubsection{Non-analytic cusp from the $T\to 0$ limit of the TBL}
\label{sec:cusp}
The TBL functional (\ref{TBL}) exhibits a non-trivial large-argument
limit, $\hat v=v/T \to
\infty$ (or, equivalently, $q_{c} \to \infty$):
\begin{equation}
r_1[\hat v] \approx  \frac{4}{\sqrt{\pi}}
\int_{q_m}^{q_c} dq\, \hat {\sf u}(q) \sqrt{q_c-q}\,,
\end{equation}
using that $\langle  |z| \rangle_{z} = \sqrt{2/\pi}$. This limit must
match the $T=0$ limit
of the functional $R[v]$, denoted $R_{T=0}[v]$. More precisely, as we will show in the next subsection, $r_1$ must match the
cubic term $\sim v^3$
of $R_{T=0}[v]$, i.e., the cusp non-analyticity.

We now perform an
explicit calculation
for a uniform $v_x=v$.
From (\ref{double}) one finds in the limit of large $\hat v L^{d/2}$:
\begin{eqnarray}
L^d {\cal R}_1(v) &=& {\cal R}_1[v] \approx  \rho (m^4 v^2 L^{d}/4)^{3/2},
\label{TBLRSB}\\
\rho &=& \frac{4}{\sqrt{\pi}} \int_{G_m}^{G_c} dG \,\hat {\sf u}(G)
\sqrt{G({\sf u}_c)-G}\,,\qquad\label{Bcoefficient}
\end{eqnarray}
which produces a cusp non-analyticity at zero temperature.
The coefficient $\rho$ can be computed explicitly for
models I and II using Eq. (\ref{GG},\ref{uhat}) and one
finds:
\begin{eqnarray}
\rho &=& \frac{16}{3} \sqrt{\frac{{2}}{\pi A}} m_c^\theta
\left(\frac{m^{-2-\theta}-m_c^{-2-\theta}}{ 2+\theta}\right)^{3/2}
\label{B}\\
&& \quad\times
{}_2F_1\left[\frac{3}{2},\frac{\theta}{2+\theta},\frac{5}{2},1
- \left(\frac{m_c}{m}\right)^{2+\theta}\right]\,.\nn
\end{eqnarray}

One can check that in the limit $m \ll m_c$, one has $ \rho \sim
m^{-(3+\theta/2)}$ and that the
prefactor agrees, using
(\ref{TBLRSB}), with the one obtained from the large $\hat v$ limit of
(\ref{tblexact}).
We have thus obtained the exact leading non-analyticity at zero temperature: it consists in a linear cusp in the force correlator, $-N^{-1}\sum_i \p_{v_i}^2 R[v]\sim |v|$. This non-analyticity is of the same
kind (i.e., proportional to $|v|$) as the one found in the standard FRG (e.g., to one loop for any $N$), and
in the large $N$ FRG in the regime $v^2 \sim N$, which will be
discussed in detail in the next Section. However,
there are differences in the dependence of the amplitude of the
cusp on $m$, $L$ and $N$. Anticipating the results of Sect.~\ref{s:large-v}, they can be summarized as follows.
In the regime $m \ll m_c$ one finds:
\bea
&& \hat R_{T=0}[v] - \hat R_{T=0}[0] = L^d \left[\hat R_{T=0}(v) - \hat
R_{T=0}(v)\right] \label{firsts} \\
&& = \frac{C_1}{ m^{2 \theta}} \left(\frac{v}{v_*}\right)^2 \left[1 - b
\frac{|v|}{v_*} + O\left(\frac{v^2}{v_*^2}\right)\right] ,\,\,  v^2 =
O\left(\frac{ N^0}{L^{\frac{d}{2}}}\right), \nn \\
&& = \frac{C_N}{ m^{2 \theta}} \left(\frac{v}{v_*}\right)^2 \left[1 - a
\frac{|v|}{v_c} + O\left(\frac{v^2}{v_c^2}\right)\right] ,\,\,  v^2 = O(N L^{0}), \nn
\eea
where $a, b$, as well as $C_1,C_N$ are numerical prefactors (with $C_1/C_N= G_c/G_m>1$). Note that the two scales
$v_* = (m L)^{-d/2} m^{-\zeta}$
and $v_c=\sqrt{N} m^{-\zeta}$ are very different.
The non-analytic corrections in the regime $v^2=O(L^{-d/2}N^0)$ are
related to the occurrence of shocks
in the system, as the quadratic well is
moved. Their contribution starts dominating once $v>v_*$.

From (\ref{B}) one finds that the amplitude of the cusp is
zero at the Larkin mass $m_c$, and then
grows linearly as function of $m_{c}-m$. This is in contrast to the
large-$v$ regime, where the
amplitude jumps to a finite value at $m_c$, cf. Eq.~(\ref{O(N)cusp})
below.

Let us finally point out that a cusp non-analyticity proportional to $|v|$ in the regime
$v^2 = O(L^{-d/2} N^0)$ was found in
Ref.~\oBBM, but with an amplitude scaling differently with $m$. This resulted from a
calculation of a rather different observable which is reviewed in
Appendix~\ref{app:BBMrevisited}. A closely related cusp singularity was also found in the study of shocks
in Burger's turbulence.~\cite{BouchaudMezardParisi} (for a recent discussion of their relation to FRG see Ref.
\cite{LeDoussal2006b}).

It is important to note that the non-analyticity found here at $T=0$ in
the regime $v=O(1)$ is
a robust feature that occurs irrespectively of the type of the ultrametric RSB
scheme
(whether continuous or 1-step, marginally or fully stable). As will be
discussed
below it reflects the switches in the minimum energy, and shock-like jumps
in position, which
occur as the energies of two states cross upon moving the harmonic well.
It is, not surprisingly,
rounded by temperature. In that sense it has some similarities with shocks
discussed for
interfaces $N=1$, and for Burgers turbulence. In contrast, in Section~\ref{s:large-v} the cusp in the
regime $v^2=O(N)$ will
be seen to rely on the
marginality of the RSB-scheme with respect to clustering fluctuations on
the largest scales. That type of marginality only occurs naturally for systems exhibiting continuous RSB. As we
will see, in that case
a cusp occurs even at finite $T$, in contrast to the above discussed
non-analyticity which forms
only in the limit $T=0$.
As we will discuss later, it is related to a more global transformation of the
set of states as $v$ is
varied.

\subsubsection{$T\to 0$ limit of $R[v]$}

The above described perturbation expansion is also perfectly suited to
analyze the limit $T=0$,
where it turns into a rigorous
expansion in $|v|$. A similar expansion was pointed out in Ref.~\oBBM. To
exhibit the dependence on
$T$
and $v$, we define $q=v^2 L^d \gamma/T^2$ by introducing the reduced
coupling function:
\bea
\gamma(\hat{\sf
u})=\frac{1}{4}\int_k g_k^{-2}\left[ \frac{|v_k|^2}{v^2 L^d} G(k,{\sf
u}=T\hat{\sf u})\right],
\eea where in
this paragraph we denote by $v^2$ a suitable average of $v_k$, such as
$v^2 L^{d} = \int_x v_x^2=
\int_k v_k^2$,
such that $v_k/|v| L^{d/2}$ is just a form factor. As mentioned before,
$\gamma(\hat{\sf u})$ and
its inverse
$\hat{\sf u}(\gamma)$ have finite $T=0$ limits, taking values between
$\gamma(\hat{\sf
u}_{c,m})=\gamma_{c,m}$
and $\hat{\sf u}_{c,m}$, respectively. After rescaling $y\equiv |v|
L^{d/2} \hat y/T$, the
evolution equation
(\ref{evolM}) for $M(q,y) \equiv \hat M(\gamma,\hat y)$ becomes:
\begin{eqnarray}
&& \partial_\gamma \hat M =  - \frac{1}{2} \left[\partial_{\hat{y}}^2 \hat
M + |v|\, L^{d/2}
\hat{\sf u}(\gamma) \partial_{\hat{y}} (\hat M^2) \right],
\label{evolM_T0} \\
&& \hat M(\hat {\sf u}_c,\hat y)=\tanh\left(\frac{L^{d/2}|v|\hat
y}{T}\right)\stackrel{T\to
0}{\to}{\rm sign}(\hat y),
\end{eqnarray}
and the second cumulant takes the form;
\begin{eqnarray}
\hat R[v]-\hat R[0] &\stackrel{T=0}{=}& - \frac{1}{2} \int_k g_k^{-2}
G_{T=0}(k,\hat{\sf u}_m)
|v_k|^2 \label{intM2_T0} \\
&& + 2 v^2 L^d \int_{-\infty}^\infty d\hat y\, \hat y [\hat M(\hat {\sf
u}_m,\hat y) - {\rm
sign}(\hat y)]. \nn
\end{eqnarray}
From (\ref{evolM_T0}) it is easy to see that the above described procedure
corresponds
to a power series expansion in $|v|$, the term corresponding to ${\cal
R}^{T=0}_n[v]$ being
proportional to $|v|^{2+n}$.
We will show below that in the case of a one step RSB the zero temperature
correlator can be
obtained
in closed form.

\subsubsection{Comparison with the TBL from droplet arguments}
\label{sec:shocks}
It is interesting to note certain analogies
with formulae obtained from droplet
arguments, where one assumes rare quasi-degeneracies
of the ground state. In $d=0$ it was found in
Refs.~\onlinecite{LeDoussal2006b,LeDoussalToBePublished}
that:
\begin{eqnarray}
&&\hat R_{ij}''(v) :=  \p^2_{v^{i}v^j}\hat R(v) = \hat R_{ij}''(0) + \label{droplet}\\
&& \quad\quad+ m^4 T \left\langle y^i y^j F\left(m^2 y \cdot
\, \frac{v}{T}\right)
\right\rangle_y + O(T^2), \nonumber \\
\label{F(z)}
&& F(z) = \frac{z}{4} \coth(z/2)-1/2= \frac{1}{4}\left[\psi\left(z/2\right)-2\right],
\end{eqnarray}
where $y \equiv u^{21}=u^2-u^1$ is the difference between the center of
mass displacement
of the ground state ($u^1$) and of the excitation ($u^2$). They are
characterized by
an (unnormalized) distribution $D(y)=P(y,E=0)$ where $P(y,E) dy dE$ is the
(normalized)
joint distribution
of position and energy differences between the two states. Here we denote
$\langle ... \rangle_y := \int dy ... D(y)$, which can be normalized using
the STS identity $\langle y_i y_j \rangle_y = 2 \delta_{ij} /m^2$.
In some simple cases (e.g., the Sinai model, corresponding to $N=1$ with
random field
disorder)
$D(y)$ is known analytically. The above formula can be generalized~\cite{LeDoussal2006b,LeDoussalToBePublished}
to any $d$:
\begin{eqnarray}
\hat R[v] &=& \frac{1}{2} \int_{xy} \frac{\delta \hat{R}[v] }{\delta v_x^i \delta v_y^j}|_{v=0} v^i_x v^j_y \nn\\
&& +
T^3  \left\langle {\cal F} \left( \int_{xx'} g^{-1}_{xx'} v_x \cdot
u^{12}_{x'}/T\right)
\right\rangle_{u^{12}},
\end{eqnarray}
with ${\cal F}''(z)=F(z)$, and a formula for all cumulants was also obtained.

Given the similarities between the formulae (\ref{droplet}) and (\ref{TBL}), it is tempting to interpret the latter in terms of a droplet size
density, summed over all ultrametric distances ${\sf u}$.
To this end, we rewrite (\ref{TBL}) for uniform $\hat v=v/T$ as:
\bea
L^d{\cal R}_1(v)&=& \hat{v}^2T^3 \int_{\hat q_m}^{\hat q_c} d\hat q\, \hat{\sf u}(\hat q) \langle
\psi(|\hat{v}| \sqrt{2(\hat q_c-\hat q)} z) \rangle_z,\quad\,\,
\eea
where we note that $\hat{q}(\hat{{\sf u}}):=q(\hat{{\sf u}})/\hat{v}^2$ and its inverse function are independent of $\hat v$. 


As shown in App.~\ref{app:droplet}, the force correlator splits into a longitudinal and transverse part, describing forces parallel and orthogonal to the displacement vector $\vec v$. Note that due to the $O(N)$ symmetry the correlator is only a function of $|v|$.
The TBL contribution to the force correlators is $L^d \Delta^{(1)}_{L}(v)=-L^d{\cal R}_1''(|v|)$ and $L^d \Delta^{(1)}_{T}(v)=-L^d{\cal R}_1'(|v|)/|v|$, respectively. They can be cast in the form (for details see App.~\ref{app:droplet}):
\bea
\label{TBLforcelong}
-L^d \Delta^{(1)}_{L,T}(v)
&=& T \int_0^\infty  db \,\rho_{L,T}^{\rm RSB}(b)\, \psi(b |\hat v|),
\eea
with the densities 
\bea
\label{PRSB}
&&\rho_L^{\rm RSB}(b) = \\
&&\quad\quad\int_{\hat q_m}^{\hat q_c} d\hat q\, \hat{\sf u}(\hat q) \left\langle  (z^4-z^2) \delta(b-z \sqrt{2(\hat q_c-\hat q)})\right\rangle_z\,,\nn\\
&&\rho_T^{\rm RSB}(b) = \\
&&\quad\quad\int_{\hat q_m}^{\hat q_c} d\hat q\, \hat{\sf u}(\hat q) \left\langle  (1+z^2) \delta(b-z \sqrt{2(\hat q_c-\hat q)})\right\rangle_z\,.\nn
\eea
This has precisely the same form as the droplet expressions for the force correlators, which from (\ref{droplet}) are found as:
\bea
\label{Pdrop}
&&-\Delta_{L,T}^{\rm drop}(v)= \\
&&\quad\quad  \hat R''(0) -m^2 T+T\int_0^\infty db\, \rho^{\rm drop}_{L,T}(b)\, \psi(b |\hat v|)+ O(T^2),\nn
\eea
where $R''(0)\equiv \p_{v_i}^2 R(v=0)$ (for any fixed $i$), and the densities $\rho^{\rm drop}_{L,T}(b)$ are given in (\ref{PdropL}).
Note also that formulae (\ref{w1},\ref{phiw2}), and the appearance of the
function $\psi$, bear similarities to expressions obtained
in the case $N=1$ for averages over a uniform density of shocks rounded by
temperature~\cite{LeDoussal2006b,LeDoussalToBePublished}.

In order to go further in the comparison and extract observables such as
shock and droplet density and their size distribution, a more detailed
description of higher moments is needed for the present model. Work in this direction is in progress. 

\subsection{The case of 1-step RSB ($d\leq 2$, $\gamma \geq \gamma_c$)}
\label{1stepO(1)}

As mentioned before, the GVM saddle point for a 1-step situation is characterized by the ``break point'' ${\sf
u}_c$ [$\equiv {\sf u}_1$ in the notation of (\ref{Kstep})] and the two self-energy parameters $\sigma_{0,1}$
(\ref{sigma1step}). Similarly, the correlation function $G(k,{\sf u})$ and the coupling $q({\sf u})$ assume only
two off-diagonal values $G_{0,1}$ and $q_{0,1}$. The former has the following interpretation:
The measure on configuration space describing fluctuations both due to disorder and
thermal noise can be
constructed independently for each mode $k$, following
Ref.~\onlinecite{MezardParisi1991}.

For the displacement in each environment one picks a set of
``state centers'' $u^\alpha_k $ as
\footnote{Independently for each of the $N$ components $u^{i\alpha}_k$.
This is left implicit here in order not to burden the notations.}
\begin{eqnarray}
\label{states1}
&& u^\alpha_k = u^0_k + \sqrt{G_1(k)-G_0(k)}\, \xi^\alpha_k, \\
\label{states2}
&& u^0_k = \sqrt{G_0(k)}\, \xi^0_k,
\end{eqnarray}
where the $\xi^0_k$, $\xi^\alpha_k$ are chosen from
independent univariate Gaussian distributions. Note that
in a given environment the (infinite) set of $u^\alpha_k$ are globally
displaced
by the same (random) $u^0_k$. Each state corresponds to a partial Gibbs measure
which is assigned a weight $W_\alpha$ (with $\sum_\alpha W_\alpha=1$), such that
the total Gibbs measure in this environment
is the weighted superposition of the partial Gibbs measures, i.e., in one
thermal realization one picks a state $\alpha$ with probability $W_\alpha$
and the mode $u_k$ takes the value
\begin{equation}
\label{thermalnoise}
u_k = u^\alpha_k + \sqrt{\tilde G(k) - G_1(k)}\, \eta^\alpha_k\, ,
\end{equation}
where $\eta^\alpha_k$ are chosen from
independent univariate Gaussian distributions and
account for
thermal noise inside a given state. In each environment the weights
$W_\alpha$
are independent random variables chosen from the distribution $P(W)
\sim_{W \to \infty} W^{-(1+{\sf u}_c)}$,
the glass transition
corresponding to the divergence of its first moment at large $W$, which implies that
only a few states dominate the
total Gibbs measure.~\footnote{Equivalently, one can write $W_\alpha =
\exp(-\beta F_\alpha)$ with a tail $P(F) \sim \exp( {\sf u}_c \beta F)$ for large
negative $F$.} An analogous construction applies in the case of continuous RSB, where the obvious generalization of the above to $K$-step RSB has to be taken to the limit $K\to \infty$.

With a general
1-step Ansatz of the form (\ref{GabMP}) we obtain the correlation functions:
\BEA  \label{G01}
G_0(k)&=& T \sigma_0 g_k^2, \\
G_1(k) &=& 
T \sigma_0 g_k^2 + \frac{T}{{\sf u}_c} \left(g_k - \frac{1}{g_k^{-1}+\Sigma_1}\right)\,, \label{G01_2}
\end{eqnarray}
where $\Sigma_1:= [\sigma]({\sf u}_c)={\sf u}_c(\sigma_1-\sigma_0)$
As always, on the saddle point, STS implies $G^c(k)=Tg_k$, and we note:
\BEA
\label{tildeG}
 \tilde G(k)=G^c(k)+{\sf u}_c G_0(k)+(1-{\sf u}_c)G_1(k).
\EEA
With such a 1-step Ansatz the GVM free-energy density takes the form:~\MP
\begin{eqnarray}
\lefteqn{\phi({\sf u}_{c})\equiv\frac{{\cal F}({\sf u}_c)}{N L^d}} \nn\\
&=& f_0-\frac{1}{2T}\sum_b B\left(2\int_k \tilde G(k)-G_{ab}(k) \right)
\nn\\
&& +\frac{1}{2}\int_k \left[g_k^{-1}\tilde G(k)-\frac{T}{n}{\rm Tr} \log G(k)\right]\label{fenergygen} \\
&=& \tilde f_0 +  \frac{1}{2 T} \left[{\sf u}_c B(\chi_0) + (1- {\sf u}_c) B(\chi_1)\right]\label{1stepaction}\\
&&+\frac{T}{2} \frac{1-{\sf u}_c}{{\sf u}_c} \int_k \left[
\frac{\Sigma_1}{k^2 + m^2 + \Sigma_1} - \ln\! \left(
\frac{k^2 + m^2+\Sigma_1}{k^2 + m^2}\right)\right]\,,\nn
\EEA
where we denote the square of transverse intra/inter-state fluctuations as:
\BEA
\chi_{0,1} &=& 2 \int_k \left[\tilde{G}(k)-G_{0,1}(k)\right], \label{chi01}
\EEA
and we have absorbed quantities that depend only on $T$ and $m$, but not on ${\sf u}_c$ and $\Sigma_1$, into the constants $f_0$ and $\tilde f_0$.
The free-energy density (\ref{fenergygen}) is minimized with respect to $G$ by the saddle point  satisfying $\delta {\cal F}/\delta G_{ab}=0$ [Eq.~(\ref{sigmaMP})], taking the 1-step form:
\BEQ\label{sigma01}
\sigma_{0,1} = - \frac{2}{T} B'(\chi_{0,1}).
\EEQ
In order to describe equilibrium thermodynamics, the breakpoint ${\sf u}_c$ has to be chosen so as to extremize ({\em maximize}) $\cal F$, as usual in replica treatments. However, other choices of ${\sf u}_c$ are of physical interest as well, as discussed in detail below.

\subsubsection{Instability of the RS solution}
Let us derive the phase diagram in the 1-step case. For simplicity we
restrict to $d<2$ and $\Lambda/m_c = \infty$.
We recall that we use the natural units introduced in Section \ref{sec:model}.
Performing the integrals, the free-energy density reads (dropping the constant $\tilde f$):
\bea
\phi({\sf u}_c)&\equiv& \frac{{\cal F}({\sf u}_c)}{N L^d} =
\frac{1}{2 T} \left[{\sf u}_c B(\chi_0) + (1- {\sf u}_c) B(\chi_1)\right]\nn
 \\
 \label{phionestepexpl}
&&  +\frac{A_d T}{\epsilon(2-d)} \frac{1-{\sf u}_c}{{\sf u}_c}  \left\{\frac{{\Sigma_1}} {(m^2 +
{\Sigma_1})^{\frac{2-d}{2}}}\right.\\
&&\quad\quad\quad\quad\quad\quad\left. -\frac{2}{d}\left[ (m^2+{\Sigma_1})^{\frac{d}{2}}-m^{d}\right]
\right\},\nn
\eea
where we have used
\bea
\int_q \frac{1}{1+q^2}=\frac{\Gamma(1-d/2)}{(4\pi)^{d/2}}=\frac{2A_d}{\epsilon(2-d)}.
\eea
In $d=0$ (a particle) the last line of (\ref{phionestepexpl}) becomes $-2 \ln(1 + \Sigma_1/m^2)$.
In $\phi({\sf u}_c)$ above, $\Sigma_1$ and ${\sf u}_c$ can be considered as two independent
variational parameters. The variation with respect to $\Sigma_1$ (at fixed ${\sf u}_c$) yields back the saddle-point equation:
\bea
{\Sigma_1}&=&-\frac{2 {\sf u}_c}{T}\left[B'(\chi_1)-B'(\chi_0)\right]\,,\label{SPeq_onestep}
\eea
where from (\ref{chi01}) one has:
\bea
\chi_0&=&
\frac{4A_d}{\epsilon(2-d)}\frac{ T}{{\sf u}_c}\left[\frac{1}{m^{2-d}}-\frac{1-{\sf u}_c}{(m^2+{\Sigma_1})^{1-d/2}}\right],\nn\\
\chi_1&=&\frac{4A_d}{\epsilon(2-d)}\frac{T}{(m^2+{\Sigma_1})^{1-d/2}}.
\label{chisp}
\eea
The replica-symmetric solution (${\sf u}_c=1$, $\chi_1=\chi_0=\chi^{\text{RS}}$, ${\Sigma_1}=0$) is valid at high temperature/large mass, but becomes unstable when the condition
\BEQ
\label{tc}
\frac{4A_d}{\epsilon m^{\epsilon}} B''\left(\chi^{\text{RS}}[m,T_c]\right)= 1
\EEQ
is met, with $\chi^{\text{RS}}=\frac{4A_d}{\epsilon(2-d)}T m^{d-2}$.
This defines a unique function $T_c(m)$ 
for masses $m\leq m_c$ smaller than the zero temperature critical mass:~\footnote{We make the natural assumption that $B''(x)$ is monotonously decreasing with $x$.}
\bea
m_c=m_c(T=0)=\left[\frac{4A_d}{\epsilon} B''(0)\right]^{1/\epsilon}.
\eea
The function $T_c(m)$ describes the location of a continuous glass transition towards a 1-step RSB phase for $m_c\geq m\geq m_*$. Here, $m_*$ denotes the mass where $T_c(m)$ attains its maximum. For $m<m_*$, the line $T_c(m)$ has little physical significance, since in that regime the glass transition occurs in a discontinuous manner at $T>T_c(m)$, as will be discussed below.

It is interesting to note~\cite{MezardParisi1991} from  (\ref{states1}-\ref{thermalnoise}), using (\ref{G01}-\ref{tildeG}), that as $m \to 0$ the thermal fluctuations of the displacement field $u$
within a state, $\tilde G-G_1 = T/(k^2 +m^{2} +\Sigma_1)$ remain bounded and massive and occur only at the Larkin scale.

By contrast, the r.m.s. difference in displacement between two states in a given sample, $u^\alpha - u^{\alpha'}$, scales as
$ G_1-G_0 =  \frac T{{\sf u}_c} [\frac1{k^2+m^2}-\frac1{k^{2}+m^{2} + \Sigma_{1}}]$. Hence, the thermal fluctuations between
states, which are active at any $T>0$ in a given sample,
occur at all scales up to $1/m$.  Note that these r.m.s. differences in displacement
have a finite $T=0^+$ limit, if $T/{\sf u}_c$ goes to a constant in that
limit (which is generally the case as shown below). These fluctuations
are responsible for the roughness of the manifold $\zeta=(2-d)/2$ with
a non-zero amplitude even as $T\to 0$.

Finally,  sample-to-sample fluctuations include in addition the term  $G_0 \sim T \sigma_0/ (k^2+m^2)^2$ whose mass dependence is controlled by $\sigma_0$. These fluctuations also occur at all scales smaller than $1/m$.

\subsubsection{Metastable states and configurational entropy}
\label{sec:1step_u_choice}
The break point $0<{\sf u}_c<1$ is a priori a free parameter of the 1-step Ansatz. In order to describe the thermodynamic equilibrium, one should choose ${\sf u}_c={\sf u}_c^{\text{eq}}$ which extremizes the free energy:
\bea
\label{equilibrium}
\frac{d \phi}{d {\sf u}_c}({\sf u}_c^{\text{eq}})=0, \quad\quad \text{(Equilibrium)}
\eea
(at fixed $\Sigma_1$) as was done in Ref.~\OMP.
As in most glassy problems with a 1-step structure, at low temperatures $T\ll T_c$ the equilibrium break point is proportional to the temperature~\footnote{We correct a misprint of the 1-step solution given in Ref.~\oMP, which suggested that ${\sf u}(T\to 0)\sim T^{(2-d)/(4-d)}$.} ${\sf u}_c^{\rm eq}\approx T\omega^{\rm eq}$. This is shown in App.~\ref{app:lowTonestep}, where $\omega^{\rm eq}$ is computed as well.

However, the function $\phi({\sf u}_c)$ encodes much more information than just the equilibrium physics. It can be interpreted~\cite{Monasson95} as the quenched ``replicated free energy'', i.e.,
the free-energy density per replica of a set of ${\sf u}_c$ clones (replica) which are bound to remain in the same metastable state of the energy landscape.
With this interpretation one can alternatively write the total free energy as a sum over states. Their number ${\cal N}(f)$ at fixed free-energy density $f$ 
increases exponentially with the volume and  $N$. One thus introduces the configurational entropy, or complexity, $\Sigma(f)=\log[{\cal N}(f)]/N L^d$. The total Boltzmann weight of ${\sf u}_c$ clones can then be written as:
\bea
&&\exp\left[-N L^d \beta {\sf u}_c \phi({\sf u}_c)\right]\\
&&\quad\quad\approx \int df \exp\left\{-N L^d \left[\beta {\sf u}_c f - \Sigma(f)\right]\right\}\,,\nn
\eea
where $\beta=1/T$.
In the large-$N$ limit, the configurational entropy becomes the Legendre transform of the replicated free energy:
\bea
\beta {\sf u}_c \phi({\sf u}_c)= \beta {\sf u}_c f-\Sigma(f),\quad\quad
\Sigma'(f)=\beta {\sf u}_c.
\eea
Knowing $\phi({\sf u}_c)$, one easily obtains an implicit parametrization of the configurational entropy as a function of the free-energy density:
\bea
f&=&\frac{d[{\sf u}_c\phi({\sf u}_c)]}{d{\sf u}_c}=\phi({\sf u}_c)+{\sf u}_c \phi'({\sf u}_c),\\
\Sigma&=&\beta {\sf u}_c^2 \phi'({\sf u}_c).
\eea
We see in particular that the equilibrium prescription (\ref{equilibrium}) corresponds to choosing the states with vanishing configurational entropy. Since the configurational entropy is an increasing function of $f$, the so selected states have the lowest free-energy density in typical samples and thus describe indeed the quenched equilibrium free energy.

However, the equilibrium may not be possible to reach dynamically, which suggests a different choice for ${\sf u}_{c}$.
The choice is constrained by the requirement that the 1-step solution be stable (as was the case for ${\sf u}_{c}={\sf u}_{c}^{\mathrm{eq}}$).

Metastable states of the free-energy density above the equilibrium  value $f^{\mathrm{eq}}$ are described by breakpoints in the range ${\sf u}_c^{\rm th}<{\sf u}_c<{\sf u}_c^{\rm eq}$, where the lower boundary is determined by the so-called replicon instability of the 1-step scheme, i.e., the condition: \cite{footnoteinstability} \bea
\label{replicon}
\frac{4A_d}{\epsilon (m^2+{\Sigma_1})^{\frac{\epsilon}{2}}} B''(\chi_1)= 1. \quad \text{(Threshold)}
\eea
A comparison with (\ref{tc}) shows that at fixed temperature $T<T_c(m_*)$, this implies $m^2+{\Sigma_1}=m_c^2(T)$, where $m_c(T)>m_*$ is uniquely defined as the solution of $T_c(m_c(T))=T$. One can prove that solutions of the 1-step saddle-point equations with condition (\ref{replicon}) exist for all $T<T_c(m_*)$ and $m<m_c(T)$ (but nowhere outside this parameter regime).~\footnote{However, ${\sf u}_c^{\rm th}$ will exceed 1 at small $m$ and $T$ close enough to $T_c(m_*)$.}
The states described by ${\sf u}_c^{\text{th}}$ are merely marginally stable and are often referred to as ``threshold states''. Since they are usually the most abundant metastable states of the system
they are most likely to trap the dynamics after a fast temperature quench for $m<m_*$.

Another choice for ${\sf u}_c$ of interest is the value ${\sf u}_c={\sf u}_c^{\text{cp}}> {\sf u}_{c}^{\mathrm{eq}}$ where the 1-step scheme becomes unstable with respect to a clustering fluctuation among the existing states~\cite{footnoteinstability}. The latter is equivalent to the condition:
\bea
\label{cuspmarginality}
\frac{4A_d}{\epsilon m^{\epsilon}} B''(\chi_0)= 1. \quad \text{(Cusp for FRG at large $v$)}
\eea
This condition ensures that the second cumulant in the regime $v^2\sim N$ is non-analytic at $v=0$, as we will discuss in Sec.~\ref{sec: General FRG}.

We emphasize that the kind of marginal stability imposed by (\ref{cuspmarginality}) is clearly distinct from the marginality (\ref{replicon}) due to the replicon mode, which is usually imposed to select dynamical threshold states in 1-step systems~\footnote{Even though distinct
they may occur at the same value of parameters in the case of the marginal
one step solution.}. The latter is also the marginality property of continuous RSB systems that ensures the presence of collective soft modes in classical~\cite{BrayMoore79} and quantum mean-field spin, elastic or electron glasses~\cite{CugliandoloGiamarchiLeDoussal2006,GiamarchiLeDoussal1996b,Schehr2005, MuellerIoffe07}, and leads to a universal, saturated pseudogap in the local field distribution of spin and electron glasses~\cite{MuellerIoffe04,Pankov06,MuellerPankov07}.

While the replicon instability (\ref{replicon}) indicates the fragility
towards fragmentation of existing states into substates, the condition
(\ref{cuspmarginality}) signals the instability towards the formation of
clusters among previously equivalent states.
When $m \ll m_c$ these two instabilities involve rather different length
scales. This can be seen as follows: The replicon instability corresponds
to each state $u_k^\alpha$ in (\ref{states1}) giving birth to a new cluster of
substates labelled by $\beta$, $u_k^{\alpha}\to
u_k^{\alpha,\beta}=u_k^{\alpha}+\delta u_k^{\beta}$, where each $\delta u_k^{\beta}$ occurs with a probability $W'_\beta$ (with $\sum_\beta W'_\beta=1$). This
rearrangment implied by the additional
substructure involves scales of order $1/m_c$ since
$dG({\sf u}) = T d\sigma({\sf u})/(g_k^{-1}+[\sigma]({\sf u}))^2$ with
$[\sigma]({\sf u}\geq {\sf u}_c) = \Sigma_1 \sim m_c^2$. On the contrary
(\ref{cuspmarginality}) signals the instability
towards $u_k^0$ splitting into several clusters $u_k^0\to u_k^0+\delta
u_k^\beta$, generating substates $u_k^{\beta,\alpha}$.
This rearranges the original $u_k^{\alpha}$ into clusters labelled by
$\beta$, and obviously involves scales up to $1/m$.

The above two kinds of instabilities have been analyzed in detail in the
context of the spherical p-spin model in
Ref.~\onlinecite{CrisantiSommers1992}.
More recently they have been
discussed in the context of lattice-glass models~\cite{Rivoire03} and
optimization
problems~\cite{MontanariRicci-TersenghiParisi,KrzakalaPagnaniWeigt,ZdeborovaKrzakala07},
where they are sometimes referred to as instabilities of the first
(clustering) and second (fragmentation) kind~\footnote{In the case of
lattice glasses one finds the same regions of stability for ${\sf u}_c$ as
here: The 1-step solution is unstable to replicon fluctuations for ${\sf
u}_c<{\sf u}^{\rm th}$, while it is unstable to the clustering instability
for ${\sf u}_c>{\sf u}^{\rm cp}$.}. They also appear as the instabilities
driving the transitions between the two types of one-step phases adjacent
to an intermediate two-step RSB phase in certain mixed spherical spin
models.~\cite{CrisantiLeuzzi07}

For model I in $d=1$, i.e., the directed polymer problem, we have checked explicitly that ${\sf u}_c^{\rm th}<{\sf u}_c^{\rm eq}<{\sf u}_c^{\rm cp}$, and we expect this to be generally true for 1-step solutions.
This is illustrated in Fig.~\ref{fig:us}.
For the associated configurational entropies this implies $\Sigma^{\rm th}>\Sigma^{\rm eq}=0>\Sigma^{\rm cp}$, and thus the states selected by the clustering instability (\ref{cuspmarginality}) have negative configurational entropy. Anticipating the analysis of Sec.~\ref{sec: General FRG},
we conclude that samples exhibiting non-analytic shock-like behavior
in the regime $v^2\sim N$ where an FRG equation was previously derived,
correspond to exponentially
rare realizations
of the disorder occurring with probability $P\sim \exp\left[ -N L^d|\Sigma({\sf u}_c^{\rm cp})|\right]$, as derived explicitly in App.~\ref{app:lowTonestep}.

Note that upon approaching the limit of a marginal one-step
solution, i.e., for $d\leq 2$ and $\gamma\downarrow \gamma_c$, the three values ${\sf u}_c^{\rm th}<{\sf u}_c^{\rm eq}<{\sf u}_c^{\rm cp}$ merge and the one-step saddle point becomes simultaneously marginal with respect to both instabilities discussed above.

\subsubsection{Phase diagram}
\label{sec:phasediagram1step}

\begin{figure}
\includegraphics[width=3.5in]{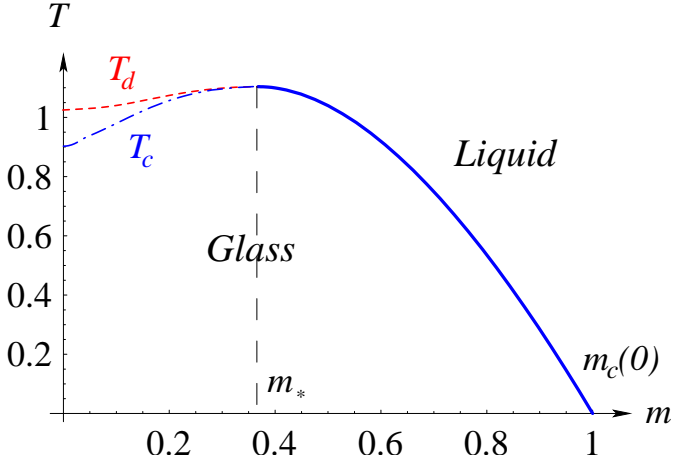}
\caption{Phase diagram for model I in $d=1$. The solid line for $m>m_*$ indicates a
continuous glass transition, where the renormalized force correlator $\tilde{B}'(x)$ displays a cusp at the origin. For
smaller mass, $m<m_*$, the glass transition towards the one-step RSB phase (as a function of $T$) is
discontinuous and takes place dynamically as a freezing transition at $T_d$, or, if equilibrium
can be attained, as a genuine thermodynamic transition at the lower temperature $T_c$. A similar phase diagram applies in
$d<2$ for model I and model II with $\gamma>\gamma_c$.} \label{fig:phasediagram1step}
\end{figure}

In this section we establish the phase diagram and discuss some of the subtleties associated with the choice of ${\sf u}_c$. The situation is closely analogous to a particle in a random environment ($d=0$) analyzed in Ref.~\onlinecite{CugliandoloPLD}. The phase diagram of a typical case, model I in $d=1$, is shown in Fig.~\ref{fig:phasediagram1step}, and explicit values given in the analysis below refer to this specific case.

\begin{figure}
\includegraphics[width=3.5in]{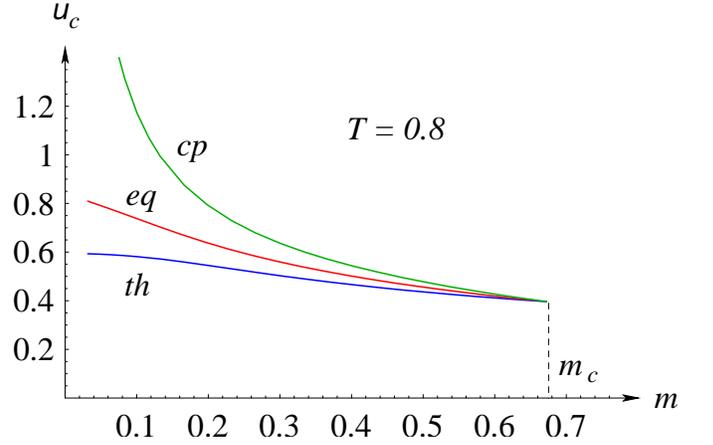}
\caption{The breakpoint ${\sf u}_c$ as a function of $m$ at constant temperature $T=0.8$ in the glass phase ($m<m_c(T)=0.673$) for model I in $d=1$. The labels ${\rm cp}$, ${\rm eq}$, and ${\rm th}$ indicate threshold, equilibrium 
and cusp states. Notice that ${\sf u}^{\rm cp}$ exceeds 1 at small enough $m$. The same would be true for ${\sf u}^{\rm eq}$ and ${\sf u}^{\rm th}$ for $T^{\rm max}>T>T_c(0)$ and $T^{\rm max}>T>T_d(0)$, respectively.}
\label{fig:us}
\end{figure}

The glass phase exhibits everywhere 1-step RSB.
Right on the instability line $T_c(m)$ where (\ref{tc}) is obeyed, there is only
one admissible value for ${\sf u}_c$:
\bea
{\sf u}_c^{\rm crit}(m)
&=&-\frac{T_c(m) I_2^2(m)}{I_3(m)}
\frac{B'''(\chi^{\rm RS})}{B''(\chi^{\rm RS})}\nn\\
&=& -\frac{4A_d}{\epsilon^2}\frac{T_c(m)}{m^{2-d}}\frac{B'''(\chi^{\rm RS})}{B''(\chi^{\rm RS})},
\eea
which yields the fluctuation-dissipation ratio relating response and correlations at the glass transition.~\cite{CugliandoloPLD}
It increases monotonously with decreasing $m$, and reaches  ${\sf u}^{\rm crit}_c=1$ at $m=m_*$ ($=1/e=0.3678$, with $T_c^{\rm max}=T_c(m_*)=3/e=1.1036$ in $d=1$ for model I where $A_1=3/4$).
For states to be dynamically or thermodynamically of significance, ${\sf u}$ must always be smaller than $1$, and thus the line $T_c(m)$ looses its significance below $m_*$.

For $m>m_*$ the glass transition is continuous in the sense that $G_1-G_0$ smoothly goes to $0$ as $T\uparrow T_c(m)$. For $m<m_*$ the temperature-driven transition is discontinuous with a sudden jump between intervalley and intravalley correlations occurring at some $T\in [T_c(m),T_c^{\rm max}]$. The location of the transition depends on the point of view. From a thermodynamic ({\em static}) standpoint the glass transition takes place at the line given by ${\sf u}^{\rm eq}_c(m,T)=1$, indicating that the replica symmetry
must be broken spontaneously  to extremize the free energy,
which gives rise to a non-analyticity in the free energy.
For model I, one finds the explicit result
\bea \label{Tcd}
T_c(m=0)=\left(\frac{2A_d}{d\epsilon} \,(2e)^{d/2}\right)^{-\frac{2}{\epsilon}} \stackrel{d=1}{=}0.9028.
\eea
However, metastable states exist already at higher temperature. In the mean-field limit $N\to \infty$ they induce a {\em dynamical} freezing transition at the line $T_d(m)$ defined by ${\sf u}^{\rm th}_c(m,T_d)=1$.
For model I one finds
\bea
\label{Td}
T_d(m=0)= \left(\frac{e 2^{d/2} A_d}{\epsilon}\right)^{-2/\epsilon}\stackrel{d=1}{=}1.0268.
\eea
The resulting phase diagram is naturally very similar to the one for a particle in a random environment (the limit $d=0$ of a random manifold)~\cite{CugliandoloPLD}, and it also strongly resembles the one of the spherical $p$-spin model, whereby the mass $m$ takes the role of the external field $h$~\cite{CrisantiSommers1992}. In the latter the glass phase is known to be everywhere of 1-step nature, the transition (at fixed $h$) being continuous for $|h|>h_*$ and discontinuous for $|h|<h_*$.

\subsubsection{Second cumulant}
To evaluate the second cumulant of the renormalized disorder potential $\hat V$,
we introduce as before the coupling $q_{0,1}$, connected to $G_{0,1}$ via:
\begin{eqnarray}
q_{0,1}=\frac{1}{4T^2} \int_k g_k^{-2} G_{0,1}(k) |v_k|^2\,.
\EEA
We need to apply the recursion relations (\ref{recursion}) only once (since here $K=1$) to calculate:
\begin{eqnarray}
g_1(y)&\equiv& \exp[{{\sf u}_c\psi_1(y)}]\label{psi1}\\
&=& e^{\frac{\tilde q - q({\sf 1})}{2}{\sf u}_c}
\left\langle \left[2 \cosh (y+z\sqrt{q_1-q_0})\right]^{{\sf u}_c} \right\rangle_z\,, \nonumber
\end{eqnarray}
and using similar steps as in the derivation of (\ref{completeR}) and (\ref{intM2}) we obtain:
\begin{eqnarray}
&&\hat R[v]-\hat R[0] = - 2 T^2 q_0 \label{integral1} \\
&& \quad\quad - 2 T^2 \int_{-\infty}^\infty dy\, \left\{\psi_1(y)-\ln[2 \cosh(y)]\right\}\,.\nn
\end{eqnarray}
Recalling that $q^c=\tilde q - \left[{\sf u}_c q_0 + (1-{\sf u}_c) q_1\right]=0$, one can see
that $\psi_1(y) - \ln[2 \cosh(y)] \sim  q^c/2 = 0$ at large $|y|$, and thus the
integral in (\ref{integral1}) indeed converges.
In the following we examine various limits of this formula.

\subsubsection{Large $L^d v^2$}
\label{sec:onestepO(1)largev}
Let us introduce the notation $Q=q_1-q_0$.
At large $v$ (large $Q$) one can use that
$\langle (2 \cosh(y+z\sqrt{Q}))^{{\sf u}_c} \rangle_z \stackrel{Q \to \infty}{\sim} 2 \cosh({\sf u}_c y) e^{{\sf u}_c^2 Q/2}$,
as can be shown using a saddle-point method. This yields the
large-$v$ limit:
\begin{eqnarray}
&&\hat R[v]-\hat R[0] \stackrel{v \to \infty}{\sim}  - 2 T^2 q_0 \\
&& \quad- 2 T^2 \frac{1}{{\sf u}_c} \int_{-\infty}^\infty dy \left\{\ln [2 \cosh(y {\sf u}_c)] - {\sf u}_c \ln[2 \cosh(y)]\right\}\,. \nn
\end{eqnarray}
Evaluating the integral we find the final result:
\begin{equation}
\hat R[0]-\hat R[v] \stackrel{v \to \infty}{\sim} \frac{T}{2} \sigma({\sf 0}) \int_x v_x^2
+ \frac{\pi^2 T^2}{6} \left(\frac{1}{{\sf u}_c^2}-1\right)\,,  \label{largev1}
\end{equation}
which matches the full RSB result (\ref{largev}), if we formally replace ${\sf u}_m$ by ${\sf u}_c$.
These two expressions are indeed identical in the case of a marginal 1-step solution, as it occurs in $d=2$. In that case, $\theta=0$, and the analysis of the crossover to large $v$ in
Section (\ref{sec:largevFRSB}) remains unchanged.

\subsubsection{Small $L^d v^2$}
\label{sec:1stepO(1)smallv}
For small $L^dv^2$ one can expand in $Q$
as follows:
\begin{eqnarray}
&& \psi_1(y)- \ln[2 \cosh(y)] \label{psionestep}\\
&& \quad = \frac{1}{{\sf u}_c} \ln \langle e^{{\sf u}_c [\ln \cosh(y+z \sqrt{Q}) - \ln \cosh(y)]} \rangle_z +
\frac{\tilde q - q({\sf 1})}{2} \nonumber \\
&&
\quad = \frac{1}{2} (1-{\sf u}_c) [1- \tanh^2(y)] Q + \dots\, ,\nn
\end{eqnarray}
where $q^c=0$ was used. Performing the $y$-integrals, one eventually finds:
\begin{eqnarray}
&&\hat R[v]-\hat R[0] = - 2 T^2 q_0 - 2 T^2  (1-{\sf u}_c)\times \label{onestepresfull} \\
 &&\quad \left[ Q  - \frac{{\sf u}_c}{3}  Q^2 + \frac{4{\sf u}_c}{45}  Q^3
 + \frac{4 {\sf u}_c}{315}  (3 u_c -5) Q^4 + O(Q^5)\right]
 .\nn
\end{eqnarray}
As it must be, the leading term is:
\begin{eqnarray}
\label{onestepres1}
&& \hat R[v]-\hat R[0] = - 2 T^2 \int_0^1 q({\sf u}) d{\sf u},
\end{eqnarray}
which yields the GVM result for two-point correlations. Indeed, (\ref{onestepres1}) matches the expression
(\ref{smallvFRSB1}), which should not depend on the RSB scheme. Similarly, inserting the one-step form for
$q({\sf u})$ in (\ref{finalR2}) one checks that it reproduces the second-order term from (\ref{onestepresfull}). In
fact, one can check that (\ref{onestepresfull},\ref{finalR2}) agree with:
\begin{eqnarray}
&&\hat R[v]-\hat R[0] = - 2 T^2 \left[ {\rm tr} (q) + \frac{1}{3} {\rm tr}(q^2)  \right.\\
&&\quad +\frac{4}{45} \left( \sum_b q_{ab}^3 +2 {\rm tr}(q^3)- 3{\rm tr}(q^2){\rm tr} (q) \right) \nn \\
&& \quad +\frac{4}{315}\left( -3 {\rm tr}(q^4) - 12 [{\rm tr}(q^2)]^2 + 30 {\rm tr}(q^2) [{\rm tr}(q)]^2 \right.\nn\\
&& \quad\quad\quad \left.\left.+ 5 \sum_b q_{ab}^4 - 20 {\rm tr}(q) \sum_b q_{ab}^3  + O(q^5) \right)\right].\nn
\end{eqnarray}
where we have used $q^c=0$ and defined ${\rm tr}(A)=\lim_{n \to 0} \frac{1}{n} {\rm Tr}(A)$. Note that
$\sum_{b} q_{ab}^{3} = {\rm tr}((q\cdot q) q)$, where the dot is the Hadamard-product.

\subsubsection{Low-$T$ expansion}
\label{sec:onesteplowT} Let us now consider the low $T$ expansion, i.e., the thermal boundary layer, using
similar notations as in (\ref{sec:TBLgeneral}). In the $T \to 0$ limit ${\sf u}_c\to 0$, but $\hat{\sf u}_c \equiv {\sf u}_c/T \to \omega^{\rm eq} $ has a non-zero limit, and so do $q_0$ and $q_1$ when expressed in terms of fixed $\hat v=v/T$, as in (\ref{TBLvariable}). Hence it is convenient
to define, for the present 1-step case, an expansion in ${\sf u}_c$\footnote{As in Sec.~\ref{sec:TBLgeneral}, it does
not exactly coincide with the expansion in powers of $T$ since some extra (weak) $T$ dependence is hidden in
other parameters such as, e.g., $\Sigma_1$. However, as before the series can easily be rearranged later.},
analogous to (\ref{TBLexpansionFRSB}):
\begin{eqnarray}
 \hat R[v]-\hat R[0]= \sum_{p=0}^\infty {\cal R}_p[v],
\end{eqnarray}
obtained by expanding (\ref{psionestep}) in powers of ${\sf u}_c$ at fixed $Q$. This yields:
\begin{eqnarray}
&&{\cal R}_0[v] = - 2 T^2 q_1, \label{1stepTBL0}\\
&&{\cal R}_{1}[v] = - T^2 {\sf u}_c  \int_{-\infty}^\infty dy
\left\{\left\langle \left[\ln \cosh(y+z \sqrt{Q})\right]^2\right\rangle_z \right.\nn\\
&&\quad\quad\quad\quad\quad\quad\left.- \left\langle\ln \cosh(y+z \sqrt{Q})\right\rangle_z^2-Q\right\}\,, \label{1stepTBL1}\\
&&{\cal R}_{p \geq 2}[v] = - 2 T^2 \frac{{\sf u}_c^{p}}{(p+1)!} \label{1stepTBLp}\\
&& \quad \times \int_{-\infty}^\infty dy
\left\langle \left[\ln \cosh(y+z \sqrt{Q}) - \ln \cosh(y)\right]^{p+1} \right\rangle^c_z\,. \nonumber
\end{eqnarray}
The expression for ${\cal R}_0$ has been simplified using
an integration by parts and
the identity (\ref{identity}), and is found to match the result (\ref{R0FRSB}) for the continuous case. The term $\tilde q - q_1 = - {\sf u}_c Q$ has been included in ${\cal R}_1$. The term multiplied by ${\sf u}_c^p$ in(\ref{1stepTBLp}) is the $(p+1)^{\text{th}}$ connected average (cumulant), as indicated by the superscript $c$.
After a calculation summarized in App.~\ref{app:TBL1step}
we finally obtain the thermal boundary layer:
\begin{equation}
\label{1stepR1}
{\cal R}_1[v] =  2 T^2 {\sf u}_c \int_{0}^{Q} dq \left\langle z \sqrt{2 q} \coth(z \sqrt{2 q})
\right\rangle_{z}\,,
\end{equation}
which agrees with the result for continuous RSB (\ref{TBL}) as expected, since both formulae should apply for
the borderline case of a marginal 1-step solution.

\subsubsection{Cusp and full correlator at $T=0$ }

The case of one-step RSB is sufficiently simple to allow for a complete calculation of the non-analytic $T=0$ limit of the full functional $\hat R[v]$ in the regime $v^2 \sim 1$.
Let us define the variable:
\begin{eqnarray}
\label{w2}
 w &:=&  {\sf u}_c \sqrt{Q} =
   \frac{\hat {\sf u}_c}{2}  \left[\int_k g_k^{-2} \left[G_1(k)-G_0(k)\right] |v_k|^2 \right]^{1/2}\,,
\end{eqnarray}
which will be shown to be of order $O(1)$ when the crossover to the shock-dominated regime occurs, see (\ref{R''1step}) below.
It tends to a finite limit as $T\to 0$.
For a uniform $v_x=v$ this becomes:
\bea
\label{w}
w &=& \frac{1}{2} \left[L^{d} \hat {\sf u}_c^2 v^2 m^4 (G_{1}-G_0) \right]^{1/2} \nn\\
&=&
\frac{v}{2 v_*}\left[ \frac{\hat {\sf u}_c}{m^\theta}\left(\frac{\Sigma_1}{m^2+\Sigma_1}\right) \right]^{1/2}\,,
\eea
with $G_{0,1}\equiv G_{0,1}(k=0)$ as given by (\ref{G01},\ref{G01_2}). Note that the characteristic scale
for $v$ (determined by $w\approx 1$) is not exactly the scale $v_*=L^{-d/2} m^{-1-\frac{\theta}{2}}$ for shocks
found for continuous RSB in (\ref{vstar}),
but rather (for $m \ll \Sigma_1$) $\sim v^{\rm 1step}_* = v_* m^{\frac{\theta}{2}} \sim L^{-d/2} m^{-1}$.
However, the two scales become the same in the case of marginal one step RSB since
there $\theta=0$.

Let us first obtain the $T=0$ cusp for a uniform $v_x=v$ by taking the limit of large $v/T$ of the thermal boundary layer. In that limit the expression (\ref{1stepR1}) tends to
\footnote{Agreement with (\ref{TBLRSB}) and (\ref{B}) can be checked for the
marginal one step case using $\theta=0$, $\hat{\sf u}_c=A$ from (\ref{A}),
and $m^2+\Sigma_1=m_c^2$.}
\begin{eqnarray}
&& L^d {\cal R}^{T=0}_1(v) =  2\hat{\sf u}_c \int_{0}^{(\frac{w}{\hat {\sf u}_c})^2} dq' \langle |z| \sqrt{2 q'}  \rangle_{z}
= \frac{1}{\hat{\sf u}_c^2}\frac{8 w^{3}}{3\sqrt{\pi}}  \nn \\
&& =  \frac{\left(m L^{d/2} |v| \right)^{3}}{3\left(\pi\hat{\sf u}_c\right)^{1/2}}
\left(\frac{\Sigma_1}{m^2+\Sigma_1}\right)^{3/2}, \label{cusp_O1_1step}
\end{eqnarray}
using $\langle |z| \rangle_z=\sqrt{2/\pi}$. This finite non-zero limit as
$T\to 0$ (cf., App.~\ref{app:lowTonestep}) is thus
the exact expression for the leading non-analyticity of $\hat R[v]-\hat R[0]$ at $T=0$,  which  is again cubic $ \sim |v|^3$ (corresponding to a ``linear
cusp'' of the force correlator $-\hat R''[v]$). Note that in this regime, $v^2 \sim L^{-d}$, the cusp exists irrespective of the choice of ${\sf u}_c$, including the equilibrium solution. This is to be contrasted to
the regime $v^2 \sim N $ where, in the case of a non-marginal one-step solution, a cusp exists only for the choice ${\sf u}_c={\sf u}^{\rm cp}$, see the discussion in Sec.~\ref{sec: General FRG}.

We now obtain the full functional $\hat R_{T=0}[v]$ for arbitrary $v_x$, using the variable $w$
defined in (\ref{w2}).
By using (\ref{psi1}) in (\ref{integral1}) and taking the limit of small ${\sf u}_{c}$, we find at $T=0$:
\begin{eqnarray}
\hat R[v]-\hat R[0]|_{T=0} &=&- \frac{1}{2} \int_k g_k^{-2}
G_0^{T=0}(k) |v_k|^2 \nn\\
&&\quad \quad\quad +{\cal R}_{T=0}(w),
\label{intM2_T01step}
\eea
where:
\bea
\label{RT=0}
{\cal R}_{T=0}(w)&=& -\frac{2}{\hat{\sf u}_c^{2}} \int_{-\infty}^\infty d\hat y\,  \left[ - \frac{w^{2}}{2} +\ln \langle e^{
|\hat y + z w|} \rangle_z - |\hat y|\right] \nn\\
&=&- \frac{w^2}{\hat{\sf u}_c^{2}} \int_0^\infty \frac{8\hat y\, d\hat y}{1+\frac{\exp(2 w \hat y)\left(1+{\rm erf}[(w+\hat y)/\sqrt{2}]\right)}{1+{\rm erf}[(w-\hat y)/\sqrt{2}]}},\nn\\
&=&  \left\{
\begin{array}{ll} -\frac{w^2}{\hat{\sf u}_c^{2}}\left[2-\frac{8 |w|}{3\sqrt{\pi}}+O(w^2) \right], & w\ll 1, \\
-\frac{\pi^2}{6\hat{\sf u}_c^{2}}, & w\gg 1.
\end{array}
\right.
\eea
From the expansion for $w\ll 1$ we again recover (\ref{cusp_O1_1step}) with, in the uniform-$v$
case ${\cal R}_{T=0}(w) \to L^d {\cal R}_1^{T=0}(v)$. The asymptotics for $w\gg 1$ is in agreement with the result (\ref{largev1}) for continuous RSB (with ${\hat {\sf u}}_m\to {\hat {\sf u}}_c$).

From this expression one can compute the force correlator, which, from $O(N)$ symmetry, splits in
transverse and longitudinal parts as defined in App.~\ref{app:droplet}. 

The non-analytic parts of the longitudinal and transverse force correlators in the regime $v^2 \sim L^{-d/2}$ are proportional to $-{\cal R}_{T=0}''(w)$ and $-\frac{{\cal R}_{T=0}'(w)}{w}$, respectively> They are plotted in Fig.~\ref{fig:Rcusp1step}, both exhibiting a linear cusp at the origin.
Note that the full correlator of the force in addition contains a constant which derives from the $v^2$ term in (\ref{intM2_T01step}). Further decay from
this constant to zero at infinity occurs (for any $\gamma>0$) in the regime $v^2 \sim N$.


\begin{figure}
\Fig{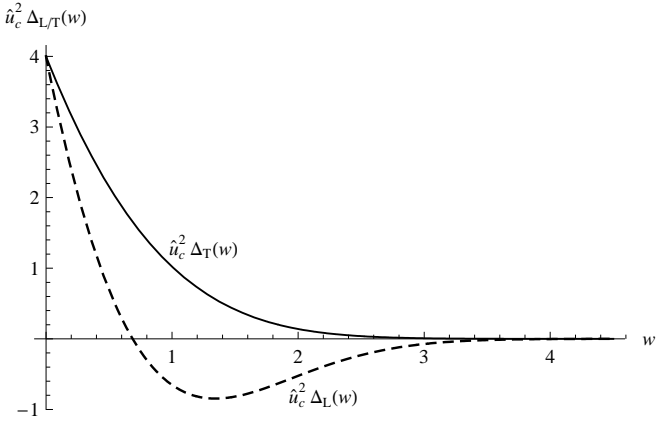}
\caption{Nonanalytic contribution to the force correlator in longitudinal $\Delta_{L}(w)$ and transversal $\Delta_{T}(w)$ direction, at $T=0$ in the case of 1-step RSB. The rescaled displacement $w$ is defined in (\ref{w}).}
\label{fig:Rcusp1step}
\end{figure}%

\subsection{Interpretation of $v_*$ and the thermal boundary layer}
\label{ss:InterpretationofTBL}

What happens to the manifold as the center of the harmonic well is moved? Let us first focus on the case of a one step RSB
for which the construction of the statistical measure is significantly simpler and was recalled in (\ref{states1},\ref{states2}). From this construction
we know that at low temperatures a certain state center $\xi^\alpha_k$ dominates, but there is at least one excited state which differs in energy by a typical amount $T/{\sf u}_c$. Changing the position of the harmonic well by an amount $v$ shifts the relative energies of these two low lying configurations (states), noted $u^{1,2}$, by an amount:
\bea
\label{extraenergy}
\Delta E = \int_k g_k^{-1} v_k\cdot [u^1(k)-u^2(k)].
\eea
One can expect a switch from one state to the other, and thus a macroscopic jump, as the energies of two states cross.
This occurs typically when $|\Delta E| \sim T/{\sf u}_c=\hat{{\sf u}}_c^{-1}$. Recalling that
$\overline{[u^1(k)-u^2(k)]^2}$ is of the order of $G_1(k)-G_0(k)$, one finds:
\bea
\label{jump}
\overline{\Delta E^2} = 2 \int_k g_k^{-2} |v_k|^2 [G_1(k)-G_0(k)].
\eea
Hence, comparing with Eq. (\ref{w2}) the jump occurs typically when:
\bea
w \sim O(1).
\eea
We thus understand the scale of variation of the functional $\hat R[u]$ given by (\ref{intM2_T01step},\ref{RT=0}).
The argument is even simpler for a uniform center $v$ ($v_k=L^{d/2}v\delta_{k,0}$)
which couples only to the $k=0$ mode of the excitation, the energy shift being
$m^2 L^{d/2}v [u^1(0)-u^2(0)]$. Since $\overline{[u^1(0)-u^2(0)]^2}=2(G_1-G_0)$, we see that the energy shift (\ref{extraenergy}) competes with the typical excitation energy $\hat{{\sf u}}_c^{-1}$ when $v$ is of the order of $v_*^{\rm 1step}$ as defined below Eq. (\ref{w}). Since the energy difference is finite, the jump is thermally smoothed at finite temperature, but turns into a sharp shock for $T\to 0$, which is at the root of the non-analyticity in (\ref{RT=0}).

We can obtain better insight into the physics for shifts of order $v\sim v^{\rm 1step}_*$ by examining the force-correlator.
We consider uniform $v$ for simplicity. To be specific we define $-\hat R''(v) := - \partial^2_{|v|} R(v)$ which corresponds to the longitudinal force correlator.
From (\ref{intM2_T01step}), we see that for $v\gg v^{\rm 1step}_*$ the correlator is simply constant and equal to: 
\bea
\label{constantforcecorrelator}
- \hat R''(v\gg v^{\rm 1step}_*)=m^4 G_0(k=0).
\eea
This has a simple interpretation in terms of the hierarchical construction (\ref{states1},\ref{states2}): One can imagine the states (characterized by $\xi^\alpha$) as parabolic potential valleys within a big parabolic potential valley as determined by the global center $u^0_k$, and accordingly write schematically $V(u)=V_0(u^0)+V_\alpha(\xi^\alpha)$ for the total potential. For $v\gg v^{\rm 1step}_*$, the manifold has jumped to a different state center, and therefore the part of the force arising from $V_{\alpha}^{\prime}(\xi^{\alpha})$ does not contribute to the correlator. However, $u^0$ is very robust under shifts of $v\ll N^{1/2}$, and jumps only once $v^2\sim N$, as we will see in the next section. The force corresponding to the big valley, $V_0^{\prime}(u^0)$, thus remains constant. Since the disorder potential competes against the quadratic well $L^d m^2 (u^0)^2/2$, that part of the force is of order $V_0^{\prime}(u) \sim L^{d/2} m^2 u^0_{k=0}$, which leads to (\ref{constantforcecorrelator}).

For $v\ll v_*^{\rm 1step}$ we see from (\ref{cusp_O1_1step}) that the force correlator behaves as
\bea
&&-\hat R''(v\ll v_*^{\rm 1step}) =
\label{R''1step}\\
&&=m^2\left[ m^2 G_1(k=0) -\frac{|v|}{v^{\rm 1step}_*}\frac{2}{\left(\pi\hat{\sf u}_c\right)^{1/2}}
\left(\frac{\Sigma_1}{m^2+\Sigma_1}\right)^{3/2}\right].\nn
\eea
The first term can be understood as $m^4$ times the {\em intra}-state correlator $\la u^2\ra ^{(\alpha)}$.
The force correlations rapidly decrease with growing $v$. Indeed, we can understand the non-analytic piece $\sim |v|$ as being due to shocks which occur with a finite density along the $v$-axis. Note the large prefactor $\sim L^{d/2}$ of $|v|$: It indicates that the product of the density and size of shocks scales as $\sim L^{d/2}$, that is, in such a way that the
system size rather than $1/m$ acts as a cutoff. Further work is in progress to clarify
the properties and the distribution of these shocks.

Let us close the discussion of shocks by sketching the analysis in the
case of continuous RSB. The typical scale in that case, $v_*=v_*^{\rm
1step} m^{-\theta/2}$, can be retrieved by the same argument as above for the one-step RSB. However, now the energy difference between two states at ultrametric distance ${\sf u}$ is
typically of order $T/{\sf u}$. This competes with the energy shift
induced by the displacement of the well:
\bea
\label{jumpFRSB}
\overline{\Delta E^2} &=& 2 \int_k g_k^{-2} |v_k|^2 [G({\sf u}_c,k)-G({\sf
u},k)]\nn\\
&&\quad\sim \frac{L^dv^2m^4}{\hat{\sf u}^{1+2/\theta}}.
\eea
The largest response is due to transitions between states differing at the
largest scales, i.e., ${\sf u}\approx {\sf u}_m\sim m^{\theta}$. The
above energy comparison at this scale immediately yields
$v_*$. However, there is no sharp selection of shocks that involve the
highest hierarchy scale. Smaller shocks contribute as well, even though
their weight decreases with increasing ${\sf u}$ characterizing the shock. 
Note that from the construction of the measure of pure states it is clear
that when a shock occurs at scale ${\sf u}={\sf u}_r$ (in a finite $K$-step RSB scheme), the
whole sub-hierarchy of state-centers $u_k^{s}$ with $K\geq s\geq r$
changes. Note that in general shocks characterized by a larger ${\sf u}$ correspond to smaller spatial rearrangements.

\section{Regime $v^2\sim N$}
\label{s:large-v}

We now compute the functional $\hat W[v]= W[j=(g^{-1} v)/T]$ defined in Section \ref{sec:observable}, and more
explicitly its second cumulant part $\hat R[v]$, in the regime $v^2 \sim N$. This is the standard regime for the
large-$N$ analysis, and in \ofrgN the associated saddle-point equation was obtained. There the focus was on the
calculation of the effective action $\Gamma[u]$, and in particular its two-replica component $R[u]$, part of which will be of use here, too, to obtain $\hat W[v]$ and $\hat R[v]$. We find, as announced in Section
\ref{sec:observable}, that the functionals $\hat R$ and $R$ are identical.
The calculations carried out here are, however,
rather different in spirit from the one in \ofrgN. We obtain $\hat R[v]$ from imposing external fields
$v^a=\pm v$ to two groups of replicas, whereas in \ofrgN  the case of $u^{ab} \neq 0$ for all $a\neq b$ was considered. It is
reassuring that the two approaches yield consistent results. The advantage of the present approach is that it
includes RSB effects, and hence it provides a complete derivation of the FRG equation on all length scales
(below {\em and} above the Larkin scale).

\subsection{General saddle point}
\label{sec:O(N)general SP}
To analyze the saddle point in the regime $v^2 \sim N$ it is convenient to introduce the
rescaled variables:
\begin{eqnarray}\label{a9}
\tilde u = \frac{u}{\sqrt{N}},\quad \tilde v = \frac{v}{\sqrt{N}}.
\end{eqnarray}
We can now rewrite the definition (\ref{defW}) and decouple the partition sum with the help of two auxiliary
fields as:
\begin{eqnarray}
 e^{\hat W[v]} &=& \overline{\prod_a Z_V\left[j^a=\frac{g^{-1} v^a}{T} \right]} = \int {\cal D} [\tu] {\cal D}[ \chi] {\cal D}[\sigma]
\rme^{ - N {\cal S} }\,, \nonumber \\
 {\cal S} &=& \frac{1}{T}\sum_a\! \int_k g_{k}^{-1} \left( \frac{1}{2} \tu^a_{-k} \cdot \tu^{a}_k
-
\tv^{a}_{-k} \cdot \tu^{a}_k \right) \qquad \label{a13} \\
&& + \int_x \left[U(\chi_{x}) + \frac{1}{2}  \sum_{ab} \sigma^{ab}_{x} \left(\chi^{ab}_{x} - \tu^{a}_{x} \cdot
\tu^{b}_{x}\right)\right]  ,~~~~~ \nn
\end{eqnarray}
where the replica matrix field $\chi^{ab}_{x}$ has been introduced through a purely imaginary Lagrange
multiplier matrix $\sigma^{ab}_{x}$.
A standard choice, as in previous sections, is $g_k^{-1}=k^2+m^2$ but our analysis is more general. We have also
included the source term shifting the center of mass. $U$ is a function of a $n \times n$ replica
matrix~\footnote{Note that $(\tu^a - \tu^b)^2=[\tu \tu]^{aa}-2[\tu \tu]^{ab}+[\tu \tu]^{bb}$. } $\tu \tu \equiv
\tu^a \cdot \tu^b$:
\BEA\label{a10}
U(\tu \tu ) &=& - \frac{1}{2T^{2}} \sum_{ab} B((\tu^a - \tu^b)^2)\nn\\
&& \quad - \frac{1}{3!T^{3}} \sum_{abc} S^{(3)}(\dots
) + ...\,,
\EEA
containing the information about the bare disorder via its cumulants, e.g., its
second cumulant being (as in (\ref{bareR0})):
\begin{equation}\label{a3}
R_{0}(u) = N B(\tilde u^{2}).
\end{equation}

Note that the rescaling (\ref{a9}) was performed in order to make explicitly appear a factor
of $N$ in front of the action ${\cal S}$, which allows for a saddle-point analysis.
One  now explicitly performs the functional integration over the field $u_{x}$ to obtain:
\begin{eqnarray}\label{lf85}
e^{\hat W[v]} &=& \int {\cal D} [\chi] {\cal D}[\sigma] \rme^{ - N S[\chi,\sigma,\tilde v] }\,, \label{largevSP}\\
 S[\chi,\sigma,\tv] &=& \frac{1}{2} \Tr \ln \left[ \frac{1}{T} (g^{-1} -
\sigma) \right] \label{a14} \\
&& + \int_x  \left[U(\chi_{x}) + \sum_{ab} \frac{1}{2 T} \sigma^{ab}_{x} \chi^{ab}_{x}\right] \nonumber \\
&& - \frac{1}{2 T} \sum_{ab}\int_{x,x'} \tv^{a}_{x}\, [(g -   g \sigma g)^{-1}]_{ xx'}^{ab} \, \tv^{b}_{x'}  ,\nn
\end{eqnarray}
where the inversions and the trace are performed in both replica and spatial-coordinate space. We use the
shorthand $g^{-1}\equiv (g^{-1})^{ab}_{xy}=g^{-1}_{xx'} \delta^{ab}$, which is diagonal in replica space.

Eq.~(\ref{largevSP})  has now the standard form for a saddle-point evaluation of the functional $\hat W[v]=: N
\tilde W[\tilde v]$ except that the saddle point is not, in general, uniform in space. At dominant order in
$1/N$, we obtain
\begin{equation}\label{W0}
 \tilde W[\tv] = \frac{1}{N} \ln \sum_{\mathrm{sp}}  \left(\rme^{- N S[\chi_\tv,\sigma_\tv,\tv] } \right)
\ .
\end{equation}
We have allowed for different saddle-points (``sp'') to contribute. To alleviate notations, we did not add an
index indicating the different saddle-points to
 $\chi_\tv$ and $\sigma_\tv$. They depend on $\tv_x$ and are the solutions
of the saddle-point equations obtained, respectively, by setting to zero the functional derivatives (at fixed
$\tv_{x}$):
\begin{eqnarray}\label{spgenWb}
 \frac{\delta S[\chi_\tv,\sigma_\tv,\tv]}
{\delta \sigma^{ab}_{x}}\bigg|_{\chi=\chi_\tv, \sigma = \sigma_\tv} &=& 0,  \\
 \frac{\delta S[\chi,\sigma,\tv]}
{\delta \chi^{ab}_{x}}\bigg |_{\chi=\chi_\tv, \sigma = \sigma_\tv}  &=&  0 \label{spgenW} \ .
\end{eqnarray}
Solving these equations, one can obtain $\hat R[v]$ from its definition (\ref{Wrepsum}).

The explicit form of the saddle-point equations is:
\begin{eqnarray}\label{a3b}
  (\chi_\tv)^{ab}_{x} &=& (G_\tv)^{ab}_{xx} + \frac{1}{T^2} (G_\tv : g^{-1} : \tv)^{a}_{x} \cdot (G_\tv : g^{-1} : \tv)^{b}_{x}   \nonumber\\
&\equiv& (G_\tv)^{ab}_{xx} +  \tu_{x}^{a} \tu_{x}^{b}\, ,\\
  (\sigma_\tv)^{ab}_x &=& - 2 T \frac{\partial }{\partial \chi^{ab}} U \!\left((\chi_\tv)_{x}\right), \\
 T G_\tv^{-1} &=&  g^{-1} -  \sigma_\tv \label{spW} \, ,
\end{eqnarray}
where $G_\tv$ is a matrix with both replica indices and spatial coordinates; thus inversion is carried out for
both. The notation for the $N$-component vector $(G:\tj)^{b}_{x} := \sum_c \int_y G^{bc}_{ x y} \tj^c_y$ is a
shorthand for a matrix product, but consistent with our above conventions, we do not write the vector indices of
$j$ explicitly. We have defined the average displacement induced by the source:
\begin{eqnarray} \label{usp}
\tilde{u}^a_x \equiv \tilde{u}^a_x(v) := \frac{1}{T} (G_\tv : g^{-1} : \tv)^a_x\,.
\end{eqnarray}
It is a function of $v$ and of the chosen saddle point ``sp'' and to simplify notations we drop in what follows the
dependence on $v$ except when an ambiguity arises. As detailed in \ofrgN, $\tilde{u}(v)$ arises in performing the
Legendre transform to obtain $\Gamma[u]$ from $W[v]$, as one may see from (\ref{legendre}) and
(\ref{changejtov}), i.e.:
\begin{eqnarray}
&& \Gamma[u] + W[v] = \frac{1}{T} u : g^{-1} : v\,, \\
&& u = T g : \frac{\delta W[v]}{\delta v}\,,
\end{eqnarray}
and using that for a given saddle point:
\begin{eqnarray}
\tilde W_{\rm sp}[\tilde v] = - S[\chi_\tv,\sigma_\tv,\tv]\,.
\end{eqnarray}
Taking the total derivative w.r.t $\tilde v$, which is equivalent to differentiating only the explicit $\tilde v$
dependence in (\ref{spgenWb}), one recovers (\ref{usp}). While in \ofrgN we had chosen sources $\tilde
v^a-\tilde v^b \neq 0$ such that $\tilde u^a-\tilde u^b \neq 0$ and therefore assumed a unique saddle point,
here we allow for spontaneous RSB and hence for many saddle points. In general the set of saddle points will include
the set of all permutations $\pi$ which leave $\tilde v$ invariant \footnote{Note that the condition is on
$\tilde v$ {\it and not} on $v$ since the saddle-point equations are expressed in terms of $\tilde v$. In other
words, the condition is $v^{\pi(a)}=v^a+o(\sqrt{N})$.}, i.e., $\tilde v^{\pi(a)}=\tilde v^a$ for all
$a=1,...,n$. Hence the $\tilde{u}^a_x(v)$ in (\ref{usp}) is identical to a partial average $\langle u^a_x
\rangle_\pi$ (corresponding to a single saddle point labelled by $\pi$) while the thermodynamic average corresponds to the full average over all equivalent saddle points,
$\langle u^a_x \rangle = \sum_\pi \langle u^a_x \rangle_\pi$. In the limit $\tilde v \to 0$ the saddle point
equations (\ref{a3b}) become identical to the saddle-point equations of the GVM, with $G_v \to G$ and $\sigma_v \to \sigma$ taking the values of the solution given by M\'ezard-Parisi (see also below). Performing the sum over equivalent saddle points ${\rm sp}$ in (\ref{W0}) one recovers the results
of Section \ref{s:small-v}.

%
%
%

\subsection{Analysis for a uniform $v$}
\label{sec:uniform v}
According to the general strategy to compute
$\hat R[v]$, as described in Section \ref{sec:evolution}, we now
solve the saddle-point equations (\ref{spgenWb}) and (\ref{spgenW}), specifying the source $\tv_x^{a} = \tv_x (1,1,\dots ,-1,\dots ,-1)$ with
$n/2$ entries $+1$, and $n/2$ entries $-1$. 
For simplicity we restrict here to a
uniform $v_x=v$, for which $\hat R[\{ v_x=v \}] = L^d \hat R(v)$. 
From the
definition (\ref{Wrepsum}) and the replica-sum expansion (\ref{expn}) one has:
\begin{eqnarray}\label{Wrepsum2}
 L^{-d} \tilde W[v] &=& \frac{m^2}{2 T } n \tilde v^2
+ \frac{1}{2 T^2} \frac{n^2}{2} \hat B(4 \tilde v^2) + O(n^3), \nonumber \\
 \hat R(v) &=& N \hat B(\tilde v^2),
\end{eqnarray}
up to a ($v$-independent) constant. The matrices $\chi_\tv$ and $\sigma$ are now independent
of $x$, and $G_{\tilde v}$ is translationally invariant.
We parametrize the $n\times n$ replica matrix $\chi_\tv$ by:
\begin{equation}\label{Ansatz}
 \chi_\tv= \left(\begin{array}{cc}
\chi_1 & \chi_{2\tv} \\
\chi_{2\tv} & \chi_1
\end{array} \right)\ ,
\end{equation}
where we anticipate that the diagonal blocks will turn out to be independent of $\tv$. We use similar notations for $\sigma$ and $G$:
\begin{equation}\label{Ansatz-2}
\sigma_\tv= \left(\begin{array}{cc}
\sigma_1 &\sigma_{2\tv} \\
\sigma_{2\tv} &\sigma_1
\end{array} \right)\ , \quad
G_\tv(k) = {\renewcommand{\arraystretch}{1.25}\left(\begin{array}{cc}
G_{1}(k)  & G_{2\tv}(k) \\
G_{2\tv}(k) & G_{1}(k)
\end{array} \right)}\,,
\end{equation}
Note that $\sigma_1$ and $G_1$ should not be confused with the variables used for the 1-step solution. The
saddle-point equations (\ref{a3b})-(\ref{spW}) now become:
\begin{eqnarray}\label{selfconsist}
\chi_1  &=& 
\tilde{u}^2\JJ + \int_{k} G_{1}(k), \\
\label{selconsist2}
\chi_{2\tv}  &=& - 
\tilde{u}^2\JJ +\int_{k} G_{2\tv}(k), \\
\label{a15} T G_\tv^{-1}(k) &=& \left (\begin{array}{cc} g^{-1}_k \JU - \sigma_1 & - \sigma_{2\tv}
\\
 - \sigma_{2\tv} & g^{-1}_k \JU - \sigma_1
\end{array} \right), \label{Ginv} \\
\sigma^{ab}_{\tv} &=& - 2 T \frac{\partial}{\partial \chi^{ab}} U (\chi_{\tv}), \label{sigmaSP}
\end{eqnarray}
and must be solved for a given value of $\tilde v$. We have introduced the notation $\JJ\equiv 1_{n/2}$ and
$\JU$ for the $\frac{n}2\times \frac n2$ matrices:
\begin{equation}
\JJ^{ab}= 1\ , \qquad \JU^{ab} = \delta^{ab}\ .
\end{equation}
In general, we expect that $\chi_{1}^{ab}$ is an ultrametric matrix while $\chi_{2\tv}^{ab}=\chi_{2\tv} \JJ^{ab}$, and
similarly for $\sigma_{1}^{ab}$ and $\sigma_{2\tv}^{ab}$. This Ansatz for the solution is motivated by the
expectation that states in different groups will be very distant in phase space (at least as distant as the farthest
equilibrium states, as described by ${\sf u}=0$), and thus their mutual overlap will not depend on the specific
replica in either of the groups. Further, we have $\tilde{u}_x^{a} =  \tilde{u}(1,1,\dots ,-1,\dots ,-1)$, and
one can show that our Ansatz implies
\begin{equation}
\tilde{u}^a_x = \tilde v^a_x \label{ueqv},
\end{equation}
which means that the average displacement is tied to the center of the harmonic well. This holds because
(\ref{usp}) implies $ T g_k \tilde{u}_k = [G^{c}_1(k) - G^{c}_{2\tv}(k)] \tilde v_k$, where $G^{c}_1(k) = \sum_{b}
G_{1,ab}(k)$ (the sum being restricted to the $n/2$ replica in the same group as $a$) and $G^{c}_{2\tv}(k) = \sum_{b}
G_{2\tv,ab}(k)$ (the sum being restricted to the $n/2$ replica in the group not containing $a$). Within our Ansatz,
as $n \to 0$, $G^{c}_1(k)=T g_k$ and $G^{c}_{2\tv}(k)$ vanishes (see (\ref{connG}) below), which establishes
(\ref{ueqv}).

To evaluate (\ref{sigmaSP}),
let us split the set of indices $a$ into two groups, $a_{+}$ for the first $n/2$ replicas, i.e.\ those
for which $\tv ^{a}=\tilde{v}$, and $a_{-}$ for the remaining $n/2$ replicas, and
consider for simplicity a bare model with gaussian disorder~\footnote{One may also consider higher bare cumulants $S^{(p)}$. While both the saddle-point equation in the regime $v\sim 1$ and the  self-consistent equation for the correlator are changed, the essential results about the connection between the FRG regime $v^2\sim N$ and the RSB regime $v\sim 1$ remain the same:
One can derive a closed self-consistency equation for the renormalized force correlator $\tilde B'$ (involving the bare cumulants $S^{(p)}$), from which the exact FRG equation can be derived; in the glass phase RSB introduces an anomaly in the selfconsistency and the FRG equation; the identity $\tilde B'(0)=-T\sigma({\sf 0})/2$ matching FRG and RSB regime holds everywhere in the phase diagram; the force correlator exhibits a cusp at the origin, if and only if the saddle-point solution is marginally stable with respect to clustering fluctuations.}:
\begin{eqnarray}\label{U}
U(\chi_{\tv}) &=&  - \frac{1}{2 T^2} \sum_{a_{+},b_{+}} B(\chi_1^{aa} + \chi_1^{bb} - \chi_1^{ab} - \chi_1^{ba}) \nonumber \\
&& - \frac{1}{2 T^2}
\sum_{a_{-},b_{-}} B(\chi_1^{aa} + \chi_1^{bb} - \chi_1^{ab} - \chi_1^{ba}) \nonumber \\
&& - \frac{1}{T^2} \sum_{a_{-},b_{+}} B(\chi_1^{aa} + \chi_1^{bb} - \chi_{2\tv}^{ab} - \chi_{2\tv}^{ba})\,.\qquad
\end{eqnarray}
One can easily see that Eqs.~(\ref{sigmaSP},\ref{U}) imply the STS property
$\sigma^c=\sum_a\sigma^{ab}_{\tv}=0$. Applying the inversion formulae (\ref{inv1}-\ref{inv3}) to (\ref{Ginv}), we find the exact relation (to all orders of $n$) $G^c=G^{c}_1+G^{c}_{2\tv}=Tg_k$ from the connected part (\ref{Ginv}). Further, to lowest order in $n$ one finds [cf., (\ref{Gc},\ref{[G]})]:
\begin{eqnarray}\label{a27}
 [G_{1}](k,{\sf u}) &=& T g_k - \frac{T} {g_k^{-1} + [\sigma_1]({\sf u})}\,, \label{a29} \\
 G_{1}(k,{\sf 0}) &=& g_k^2T \sigma_1({\sf 0}),  \label{a28} \\
 G_{2\tv} &=& g_k^2 T \sigma_{2\tv},\label{G2v}
\end{eqnarray}
and, to precision $O(n)$:
\BEA
\label{connG}
G^{ c}_1(k) &=& T g_k-(n/2)g_k^2 T \sigma_{2\tv},\\
G^{c}_{2\tv}(k) &=& (n/2)g_k^2 T \sigma_{2\tv}.\nn
\EEA
Up to negligible corrections of order $O(n)$ the saddle-point equations for the diagonal blocks are
the same as those in the regime $v=O(1)$, cf., (\ref{GabMP},\ref{sigmaMP}), independently of the external field $\tv$:
\begin{eqnarray}\label{a6}
\sigma_1^{ab} &=& - \frac{2}{T} \Bigg[ B'(\chi_1^{aa} + \chi_1^{bb} - 2 \chi_1^{ab})
\nonumber \\
&& \qquad - \delta_{ab} \sum_{c_{+}} B'(\chi_1^{aa} + \chi_1^{cc} - 2 \chi_1^{ac})\nonumber  \Bigg]\,,
\end{eqnarray}
for $a,b$ in the $+$ group. An analogous equation holds for the $-$ diagonal
block. This is not surprising since the external field $v$ does not separate the replica within a group. Since these saddle-point equations have a unique physical solution, we will identify $\sigma_1\equiv \sigma$ henceforth, and also drop the subscripts $1$ on $G_1$ and $\chi_1$. In Parisi's parametrization of ultrametric matrices, the saddle-point equation is conveniently rewritten as
\begin{eqnarray}
\sigma({\sf u}) &=& -\frac{2}{T} B'\left(2 \int_k [\tilde G(k) - G(k,{\sf u})]\right)\,, \label{a17}\\
\tilde{\sigma} &=& \int_0^1  \sigma({\sf u})\,d{\sf u}. \label{a18}
\end{eqnarray}
The saddle-point equation
for the off-diagonal part is:
\begin{eqnarray}\label{a19}
\sigma_{2\tv}^{ab} &=&-\frac{2}{T}  B'(\chi^{aa} + \chi^{bb} - 2 \chi_{2\tv}^{ab})\,, \label{a20}
\EEA
which within our Ansatz $\sigma_{2\tv}^{ab}=\sigma_{2\tv}\JJ^{ab}$ reduces to a single equation
\begin{equation}
\sigma_{2\tv}= -\frac{2}{T} B'\left(4 \tv^2 + 2 \int_k [\tilde G(k) - G_{2\tv}(k)]\right)\,. \label{sigma2first}
\end{equation}
Before analyzing these equations let us indicate how to extract $\hat R(v)$
from its solution. Let us first rewrite the part of the action at the saddle
point which depends explicitly on $v$ (in fact, on $v \cdot v$):
\begin{eqnarray}
\frac{S[\chi_\tv,\sigma_\tv,\tv]|_{\rm expl}}{L^{d}} &=& -\frac{m^4}{2 T^2} \tilde v^a G_\tv^{ab}(k=0) \tilde v^b
\\
& =& -\frac{m^4}{2 T^2} n \tilde v^2 \left[G^{c}(k=0) - G^{c}_{2\tv}(k=0)\right]\nn\\
& =& \tilde v^2 \left(-\frac{m^2}{2 T} n + \frac{\sigma_{2\tv}}{2 T} n^2\right)\,.\nn
\end{eqnarray}
Since the derivative of
$\tilde W_{\rm sp}[\tilde v] = - S[\chi_\tv,\sigma_\tv,\tv]$, w.r.t.
$\tilde v$ (in fact, w.r.t. $\tilde v \cdot \tilde v$) only involves the explicit part
we find,
and comparing with (\ref{Wrepsum2}):
\begin{eqnarray} \label{identification}
&& - \frac{2}{T} \hat B'(4 \tilde v^2) =  \sigma_{2\tv}.
\end{eqnarray}
We now discuss various cases according to whether or not the replica
symmetry is broken in the diagonal block, (the saddle point of the GVM),
indicating whether or not the system is in a glass phase.

\subsubsection{Replica-symmetric region}\label{lf1}
\label{sec:O(N) RS}
Assuming replica-symmetry, $\sigma({\sf u})=\sigma^{\text{RS}}$, we have:
\begin{eqnarray}\label{lf2}
G(k)&=& T g_k \JU + T \sigma^{\text{RS}} g_k^2 \JJ,\\
G_{2\tv}(k)&=& T \sigma^{\text{RS}}_{2\tv} g_k^2  \JJ .\label{a22}
\end{eqnarray}
This reduces the saddle-point equation (\ref{a17}) to:
\begin{equation}\label{a23}
\sigma^{\text{RS}} = -\frac{2}{T} B'\left(2 T I_1\right),
\end{equation}
independently of $\tilde v$, and Eq.~(\ref{sigma2first}) becomes:
\BEA
\sigma^{\text{RS}}_{2\tv}= -\frac{2}{T} B'\left(4 \tv^2 + 2 T\left\{I_1 + [\sigma^{\text{RS}}-\sigma^{\text{RS}}_{2\tv}]I_2\right\}\right),\label{sigma2}
\EEA
where $I_n = \int_k g_k^n$. 
Comparing with (\ref{a23}) we see that $\sigma^{\text{RS}}_{2,\tv=0}=\sigma_{\text{RS}}$, as is expected in the absence of external fields.

\subsubsection{Identification with FRG below the Larkin scale:}
\label{sec:O(N): FRGfrom RSB}
It is convenient to introduce the function $\tilde{B}'$ via the change of variables:
\begin{equation}\label{a25}
-\frac{2}{T}\tilde{B}'(4 \tilde v^2)\equiv \sigma^{\text{RS}}_{2\tv}.
\end{equation}
Plugging it into (\ref{sigma2first}), we see that it satisfies exactly the self-consistency equation (\ref{Brecursion}):
\begin{eqnarray}\label{a26}
\tilde{B}'(x)= B'\left(x + 2 T I_1 +4[\tilde{B}'(x)-\tilde{B}'(0)] I_{2}\right)\,
\end{eqnarray}
as derived
in Ref.~\onlinecite{LeDoussalWiese2003b} 
for the second cumulant function of $\tilde B(x)$ which occurs in
the effective action (defined there via $R(u) = N \tilde B(\tilde u^2)$).
Since (\ref{a26}) uniquely specifies $\tilde B$ up to a constant, a
comparison of (\ref{a25})
and (\ref{identification}) shows that
\begin{equation}\label{Bidentity}
\tilde{B}'(x) = \hat B'(x)\,,
\end{equation}
i.e., we have recovered, in the glassy regime, the general identity $\hat R=R$ announced in
(\ref{Rhat=R},\ref{relationRB}).

From
the self-consistency equation (\ref{a26}) a FRG equation for
$\hat B(x)$ can be derived, following [cf., Eqs.~(6.1)-(6.8)] in
Ref.~\onlinecite{LeDoussalWiese2003b}
\begin{eqnarray}\label{frglarkin}
 - m \partial_m \hat B'(x) &=& \hat B''(x) \Bigg[ 4 (m \partial_m I_2) [\hat B'(0) - \hat B'(x)] \nonumber \\
&&\quad \quad\quad -  \frac{2 m \partial_m T I_1}{1+4 I_2 \hat B''(0)} \Bigg]\,,
\end{eqnarray}
with initial condition $\hat B(x)=B(x)$ for $m=+\infty$.

 At this stage we only know that the FRG equation
(\ref{frglarkin}) is valid in the replica symmetric region, i.e., for $m > m_c(T)$ (see discussion in Section
\ref{sec:summary}). At the Larkin mass, $\hat B''(0)$ diverges, signaling a cusp in the force correlator, as
can be seen from the condition (\ref{instability}) and (\ref{a26}). Hence at $m=m_c^+$ the equation
(\ref{frglarkin}) is still valid, but the second term (involving the temperature explicitly) is negligibly small
there and can be dropped. It is now crucial to establish how to continue this equation {\it below} the Larkin
mass. To this end we compute $\hat B(x)$ explicitly beyond the Larkin mass, i.e., in the glassy region, where
RSB occurs within the replica groups, and derive the correct FRG equation for $m < m_c$.

\subsubsection{Self-consistency equation for the second cumulant
below the Larkin mass} \label{sec:O(N) FRG from RSB in FRSB case}

The saddle-point equation (\ref{sigma2first}) together with (\ref{identification}) yields the correct continuation of the self-consistency equation (\ref{a26}) 
below the Larkin mass:
\begin{equation} \label{sigma2_2}
\hat B'(x)= B'\left(x + 2 \int_k \tilde{G}(k) +4 \hat B'(x) I_2\right)\,.
\end{equation}
We expect that:
\begin{equation}\label{sigma2_0}
\hat B'(0) =-\frac T 2 \sigma({\sf u}={\sf 0}),
\end{equation}
remains valid when replica symmetry is broken (with of course a different and non-trivial value for $\sigma(0)$)
since it expresses the equality $\sigma({\sf u}={\sf 0})=\sigma_{2,\tv=0}$ between the self-energy
associated with distant equilibrium states, and that associated with the coupling among the two replica groups
in the limit of zero forcing, $\tilde v\to 0$. Indeed, one can check using (\ref{a28}) that (\ref{sigma2_0}) is
a solution of (\ref{sigma2_2}), given that (\ref{a17}) holds. Note that (\ref{sigma2_0}) insures that the function $\hat B(\tilde{v}^2)$ perfectly
matches the large $v^2$ limit of the $v^2 = O(1)$ results (\ref{largev},\ref{largev1}), irrespective of the scheme of RSB, as it should.

Eq.~(\ref{sigma2_2}) can be rewritten in a form similar to (\ref{a26}):
 \bea \label{frgscrsb1}
\hat B'(x)& = & B'\left(x + 2 T I_1 +4I_2[\hat{B}'(x)-\hat{B}'(0)]  + 2 \int_k A_k \right) \nn \\
& =& B'\left(\chi_0 + x +4I_2[\hat{B}'(x)-\hat{B}'(0)] \right)\,,
 \eea
where
\bea \label{defchi_0}
\chi_0:=2 \int_k\left[\tilde G(k)-G(k,{\sf 0})\right]\,,
\eea
and
\BEA
A_k &=& \tilde{G}(k) - T g_k - T \sigma(0) g_k^2 \nn\\
&=& \int_0^1 d{\sf u} [G(k,{\sf u}) - G(k,{\sf 0})]
\label{tutu}
\EEA
is an {\it anomaly} which is non-zero if and only if the replica symmetry is broken. (To get the second line in (\ref{tutu}), we used the generally valid inversion formula $G(k,{\sf 0})=T \sigma(0) g_k^2$.)

\subsubsection{Correlator and FRG equation below the Larkin mass}
\label{sec: General FRG}
From (\ref{frgscrsb1}) we easily obtain the behavior of the force correlator, $-\hat B'(x)$, at small argument.
If the RSB solution is marginally stable with respect to a clustering instability at the largest scales, i.e., if the
condition
\bea \label{marginality_at_u=0}
1=4I_2B''(\chi_0),\quad
\eea
holds, with $\chi_0$ defined in (\ref{defchi_0}), one finds a non-analytic cusp in the correlator:
\bea
-\hat B'(x)&=& \frac{T\sigma({\sf 0})}{2}-\left[\frac{-2 x}{ (4I_2)^3 B'''(\chi_0)}\right]^{\frac{1}{2}} +O(x).\quad\quad
\label{O(N)cusp}
\eea
This type of marginality is automatically ensured in the case of continuous RSB, which is marginal with respect to fluctuations on all scales (see App.~\ref{app:generalRSB}) including those corresponding to ${\sf u}_m$, which entails (\ref{marginality_at_u=0}).
 For a 1-step solution, however, it occurs naturally only on the transition line $T_c(m)$ for $m>m_*$, while in the glass phase it
requires the specific choice ${\sf u}_c={\sf u}^{\rm cp}$, cf.\ (\ref{cuspmarginality}). In the case of a non-marginal 1-step solution, the latter choice differs from the value of ${\sf
u}_c$ corresponding to thermodynamic equilibrium, and was shown to describe rare disorder configurations in Section~\ref{sec:1step_u_choice}. On the other hand, for a fully stable saddle point
with $1>4I_2B''(\chi_0)$, one finds the regular correlator:
\bea
-\hat B'(x)&=& \frac{T\sigma({\sf
0})}{2}-\frac{xB''(\chi_0)}{1-4I_2 B''(\chi_0)}+O(x^2).\label{noO(N)cusp} \quad\quad
\eea
We emphasize the difference between these
results and the generic non-analyticity (\ref{TBLRSB},\ref{cusp_O1_1step}) found in the regime $v=O(1)$ at $T=0$, occurring
irrespectively of the RSB scheme. The cusp in the regime $v=O(\sqrt{N})$ analyzed here is present even at finite
temperature, provided the RSB scheme is marginally stable in the sense of condition (\ref{marginality_at_u=0}). The non-analyticity thus reflects the criticality of the system being at the brink of an instability towards
an additional clustering of the topmost level of an ultrametric structure. In the 1-step case, the
instability is towards a two-step RSB with an additional step in the lower plateau~\cite{CrisantiSommers1992,CrisantiLeuzzi07}.

%
Taking a derivative of (\ref{sigma2_2}) with respect to $x$ and using it to simplify (\ref{frglarkin}), one finds~\cite{LeDoussalWiese2003b} repeating the steps of Eqs.~(6.4)-(6.6) in Ref.~\onlinecite{LeDoussalWiese2003b}:
\bea\label{frgrsb}
- m \partial_m \hat B'(x) &=& \hat B''(x)\times\\
&& \quad \left\{ 4 m \partial_m I_2 [\hat B'(0) - \hat B'(x)] + A(m) \right\}\,,\nn
\eea
where
\begin{eqnarray} \label{anomaly}
 A(m) &=& - 2 m \partial_m \left[ T I_1 +  \int_k A_k\right] + 4 I_2 m \partial_m \hat{B}'(0) \nonumber \\
& =& - 2 m \partial_m \int_k \tilde G(k) + 2 T \sigma({\sf 0}) m \partial_m I_2\ .
\end{eqnarray}
This equation is the general FRG equation valid in all regimes. In the non-glassy RS regime $A_k=0$ and the amplitude $A(m)$ is identical to the last term in (\ref{frglarkin}). In the glassy region Eq.~(\ref{frgrsb}) is certainly valid for all $x>0$, and it is also valid at $x=0$ if $\hat B''(0)<\infty$, i.e., if the RSB solution is not marginal in the sense of (\ref{marginality_at_u=0}). But
even if the
solution is marginal, and hence $\hat B''(0)$ is infinite (i.e., $\hat B'(x)$ exhibits a cusp), both sides of the equation have a non-trivial limit at $x=0^+$ which
yields another valid equation as can be checked using (\ref{O(N)cusp}).
Another useful expression for $A(m)$ is obtained by
rewriting the second equation in formula (\ref{anomaly}) as:
\BEA
 A(m)
 &=& - m \partial_m \chi_0 - 2 T I_2 m \partial_m \sigma(0) \label{newf}
\EEA
where we used the definition (\ref{defchi_0}) and (\ref{tutu}). The second term in (\ref{newf})
can be rewritten using the saddle-point equation (\ref{a17}) at $u=0$, and
taking a derivative w.r.t. $m$ one obtains
\begin{eqnarray}\label{amgen}
 A(m) &=& - [1 - 4 I_2 B''(\chi_0)] m \partial_m \chi_0
\end{eqnarray}
a formula valid in all cases and regimes. We already point out and will discuss further below that the amplitude $A(m)$ vanishes if the marginality condition (\ref{marginality_at_u=0}) is met.


Let us now analyze in more detail the case of continuous RSB which includes as a limiting case the marginal
1-step case occurring in $d\leq 2$ with $\gamma=\gamma_c(d)$.
In Ref.~\onlinecite{LeDoussalWiese2003b} it was found that the consistent FRG equation \footnote{For $\tilde B$, but
this should be the same as for $\hat B$.} for $m < m_c(T)$ was (\ref{frgrsb}) without the last term, i.e.,
$A(m)=0$. As we discussed above, this matches the established equation (\ref{Brecursion}) at $m=m_c^+$. The
validity of (\ref{frgrsb}) for $m < m_c$ was inferred from the study of the FRG equation in inverted variables
$x=x(\hat B')$, noting that for models I and II the FRG flow {\it completely stops} at $m=m_c^+$ (the beta
function is identically zero hence its natural continuation is to remain zero below $m_c$). In Section
VIII-D of Ref.~\onlinecite{LeDoussalWiese2003b} the analysis of the FRG flow  was extended to an arbitrary bare model (i.e., an arbitrary $B(x)$). It was found that the natural continuation is such that the function $\tilde B'$ retains a cusp for all $m \leq m_c$.
This was found to coincide with the marginality condition at ${\sf u}_m$, and
hence lead to (\ref{marginality_at_u=0},\ref{O(N)cusp}).

From the above general expression (\ref{amgen}) 
we see that the vanishing of the amplitude $A(m)=0$
is a direct consequence of the marginality of the continuous RSB at ${\sf u}={\sf u}_m$, i.e., condition (\ref{marginality_at_u=0}), and thus confirms the correctness of the assumption made in Ref.~\onlinecite{LeDoussalWiese2003b}. In addition we discover here that
the cusp can be avoided if there is a non-zero amplitude $A(m)$. This possibility was naturally not considered in Ref.~\onlinecite{LeDoussalWiese2003b}, where the analysis was based on
the self-consistency equation without anomaly and the ensuing FRG equation.
While the amplitude $A(m)$ always vanishes in the case of continuous RSB, a zero amplitude requires a specific choice of the breakpoint
${\sf u}_c={\sf u}^{\rm cp}$ in the case of 1-step RSB, as discussed in the next Section.


\subsection{Discussion of the FRG flow}

\subsubsection{General considerations}

Let us analyze the generalized self-consistency Eq.~(\ref{frgscrsb1}) for the renormalized correlator function $\hat B'(x)$. It depends only on the
(given) bare disorder function, $B'(x)$ and a number, $\chi_0=\chi_0(m,T)$,
defined in (\ref{defchi_0}). This is the only input from the saddle-point solution in the regime $v\sim O(1)$, which introduces an anomaly in the glass phase.
We emphasize again that in the case of one-step RSB, $\chi_0$
does not only depend on $T$ and $m$, but also on the choice of the breakpoint ${\sf u}_c$.

Since the force correlator (the derivative of the potential correlator)  scales like $\hat B'\sim m^{-2\theta+d+2\zeta}$, with typical values
of the argument scaling as $m^{-2\zeta}$, it is natural
to define the rescaled force correlator (as in \ofrgN):
\bea \label{scalingb}
\tilde{b}'(\hat{x})&:=&\frac{4A_d}{m^{-2\theta+d+2\zeta}}\hat{B}'\left(m^{-2\zeta}\hat{x}\right). \eea
At this stage, if we ignore all other information from the $v \sim O(1)$ regime, the two exponents $\zeta$ and $\theta$ are undetermined. Let us assume for now that they can be fixed by requiring that $\tilde{b}'(\hat{x})$ reaches a non-trivial fixed point. From (\ref{frgscrsb1}) one deduces that $\tilde b'$ satisfies the equation:
\begin{equation}
\label{SCequationrescaled}
\tilde{b}'(\hat{x}) = \frac{4A_d}{m^{2-\theta}}B'\left( \chi_0+m^{-2\zeta}\left(\hat{x} +\frac{\tilde{b}'(\hat{x})-\tilde{b}'(0)}{\epsilon}\right) \right),
\end{equation}
where we have used $I_2=A_d /(\epsilon m^{\epsilon})$ in the limit of infinite cut-off, and consequently set $\theta=d-2+2 \zeta$, as expected,
so that the last two terms in the argument of $B'$ (the bare disorder)
scale the same way. The only information needed from the $v \sim O(1)$ regime
is the value of $\chi_0$, given by the saddle-point solution. It fixes the value of $\tilde{b}'$ at the
origin:
\begin{equation}
\label{b'0}
\tilde{b}_0'(m,T):=\tilde{b}'(0) = \frac{4A_d}{m^{2-\theta}}B'( \chi_0) = -\frac{4A_d}{m^{2-\theta}}\frac{T\sigma({\sf 0})}{2}.
\end{equation}
The above two equations then completely determine the FRG flow of
$\tilde{b}'$, describing the evolution of the force correlator as the mass $m$ is decreased. The corresponding flow equation was given in (\ref{frgrsb}).
Here we directly solve (\ref{SCequationrescaled}) focusing on
models I and II, which admit simple solutions.
We distinguish the case of continuous RSB (full or marginal one-step),
which we only briefly recall since it was discussed in detail in Ref.~\onlinecite{LeDoussalWiese2003b}, and non-marginal one-step
RSB which requires a novel and thorough analysis. The difference
between the two cases can be grasped immediately. For continuous
RSB, comparing (\ref{b'0}) and (\ref{sigma0}), valid in that case for models I and II,
we immediately see that $\tilde{b}'(0)$ reaches a fixed-point value
with the same choice of exponents as in the $v \sim O(1)$ regime.
By contrast (\ref{chi01}) and (\ref{SPeq_onestep}) suggest a rapid decrease of
$\tilde{b}'(0)$ as $m \to 0$ as discussed below.

Some features are independent of the RSB scheme and worth mentioning.
Consider model I, $B(x)=e^{-x}$. If one makes the choice $\zeta=0$,
then $\tilde{b}(\hat{x})$ is the solution of
\bea
\label{fixedpointI}
\tilde{b}'(\hat{x})=\tilde{b}_0'
\exp\left(-\hat{x}-\frac{\tilde{b}'(\hat{x})-\tilde{b}_0'}{\epsilon}\right)\,.
\eea
which has clearly a fixed-point form provided $\tilde{b}_0'$,
given by
\bea
\tilde{b}_0'&=&-\frac{4A_d}{m^{\epsilon}}\exp[-\chi_0(m,T)],
\eea
reaches a fixed point as $m\to 0$. Similarly for model II,
$B'(x)=\left[1+\frac{x}{\gamma} \right]^{-\gamma}$, and (\ref{b'0})
together with the choice  $\zeta=\zeta(\gamma)=\epsilon/[2(1+\gamma)]$ leads to the
equation:
\bea
\tilde{b}_0'=
-\frac{4A_d}{m^{\epsilon\frac{\gamma}{1+\gamma}}}\left[1+\frac{\chi_0(m,T)}{\gamma} \right]^{-\gamma}\,.
\eea
Upon inversion this allows us to simplify the
equation (\ref{SCequationrescaled}) for the rescaled force correlator as:
\BEQ
\label{fixedpointII}
\tilde{b}'(\hat{x})=\tilde{b}_0'\left(1
+\left(\frac{|\tilde{b}_0'|}{4 A_d}\right)^{\frac{1}{\gamma}} \frac{1}{\gamma}
\left(\hat{x}+\frac{\tilde{b}'(\hat x)-\tilde{b}_0'}{\epsilon}\right) \right)^{-\gamma} \,,
\EEQ
which reaches a fixed point if $\tilde{b}_0'$ does. Note that these expressions, as well as (\ref{SCequationrescaled}),
are valid for all $m$, both above and below the Larkin mass.

\subsubsection{FRG flow for continuous RSB}

As discussed in Ref.~\onlinecite{LeDoussalWiese2003b} for model
I and II the flow is particularly simple. The structure of (\ref{fixedpointI}) and (\ref{fixedpointII}) imply that $\tilde b'(\hat{x})=\Phi(\hat{x}+\hat{x}_0)$ is always given by the same master function $\Phi$, shifted by an amount $\hat{x}_0$ which depends on $m$ only through $\tilde b'_0$. For model I the master function is the solution of $\Phi(x)=-\epsilon \exp[-x-1-\Phi(x)/\epsilon]$. The shift $\hat x_0$ must be positive, and $\hat x_0=0$ leads to a cusp (infinite slope) at the origin.
In the case of continuous RSB $\hat x_0$ freezes to zero as $m$ decreases below $m_c(T)$. One checks that $\tilde b'_0$ reaches its fixed point value
at $m=m_c(T)$:
\bea \label{fpfrg}
-\tilde{b}_0'&=&\epsilon,\quad \text{(I)}\\
-\tilde{b}_0'&=&\epsilon \left[\frac{4A_d}{\epsilon}\right]^{\frac{1}{1+\gamma}},\quad
\text{(II)}\nn \eea
and remains constant everywhere in the glass phase. This implies in particular that $\tilde{b}'$ reaches its fixed
point already at the phase transition $m_c(T)$ and sticks to it for all smaller $m$, a result inferred in \ofrgN
from the vanishing of the beta-function at $m_c(T)$. The FRG flow for models with continuous RSB is illustrated in Fig.~\ref{fig:flow3d}.
\begin{figure}[t]
\includegraphics[width=3.5in]{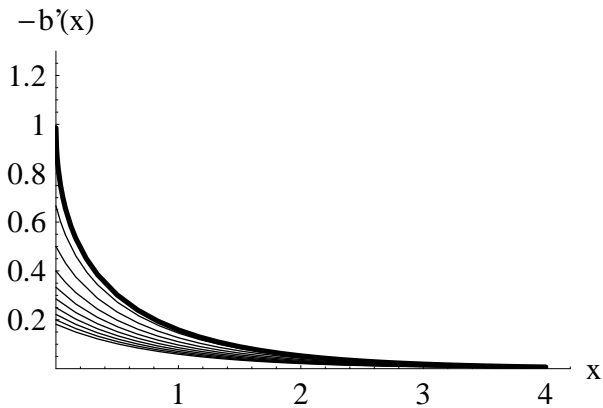}
\caption{FRG flow of the renormalized and rescaled correlator $\tilde{b}(\hat{x})$ for model I in
$d=3$. The reduced temperature was chosen to be $\hat{T}=0.5$ and the curves correspond to $1\leq
m/m_c(\hat{T})\leq 5.5$ in steps of $0.5$, from top to bottom. For $m<m_c(\hat{T})$ the force correlator sticks
to its fixed point $\tilde{b}_*'$ (thick line) as given by (\ref{fixedpointI}) with $\tilde{b}_*'(0)=\epsilon=1$. Due to the specific form of $B(x)$ in models I and II, the flow reduces to a $m$-dependent horizontal shift of an otherwise constant master function $\Phi(x)$.}
\label{fig:flow3d}
\end{figure}

As shown in Ref.~\onlinecite{LeDoussalWiese2003b} for models different from I and II, and such that there is continuous RSB for all $m < m_c(T)$,
such as linear combinations of I and II with various powers,
the flow is slightly more involved. For $m>m_c(T)$ the flow is not a simple
translation in $x$ but the shape also changes as irrelevant parts of $\hat B'(x)$
decay. At $m=m_c(T)$ a cusp appears and remains for all $m <m_c(T)$ as $\tilde b'_0$ (and the shape of the function) converges slowly towards one of the above fixed-point values (\ref{fpfrg}).

In the continuous RSB case, the existence of this family of non-analytic fixed points is related to the vanishing of the amplitude $A(m)$ in the FRG equation. In Ref.~\onlinecite{LeDoussalWiese2003b} they were derived directly from (\ref{frgrsb}) assuming $A(m)=0$. For $d>2$ and $d<2$, $\gamma \geq \gamma_c$, they lead to the same exponents $\zeta=\zeta(\gamma)$ and $\theta=\theta(\gamma)$ as the study of the $v^2=O(1)$ GVM regime, cf.\ Ref.~\oMP. Hence, all parts of the effective action scale consistently. Note that in the limit case $d<2$, $\gamma=\gamma_c=2/(2-d)$, the glass phase is described by a marginal one-step RSB solution. As discussed in details in Ref.~\onlinecite{LeDoussalWiese2003b}
(Section VI.D and Appendix F), the high-temperature phase $T>T_c$ is described by
a line of analytic fixed points which solve (\ref{SCequationrescaled}) with $\zeta=(2-d)/2=\zeta(\gamma_c)$, $\theta=\theta(\gamma_c)=0$. In that phase $\tilde b'(x)$ converges to
one of these fixed points. This line of fixed points terminates on a cuspy fixed point, towards which
$\tilde b'(x)$ converge for $T<T_c$, and which has the form (\ref{fixedpointII}, \ref{fpfrg})
with $\gamma=\gamma_c$.

By contrast, in the non-marginal one-step case, both for models I and II, the shift $\hat{x}_0$ generically vanishes only at $m_c(T)$ and becomes positive again in the glass phase, unless the choice ${\sf u}={\sf u}^{\rm cp}$ is made for the breakpoint, as we now discuss (see also Fig.~\ref{fig:shifts}).
\begin{figure}[t]
\includegraphics[width=3.5in]{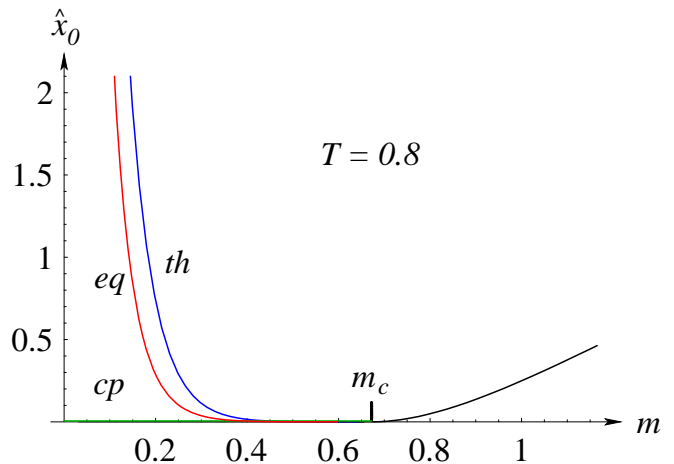}
\caption{The shifts $\hat x_0$ as a function of $m$ at constant temperature $T=0.8$ for model I in $d=1$. The minimal shift $\hat x_0=0$ is attained at the glass transition $m=m_c(T)=0.673$ and signals a cuspy correlator. In the glass phase, only the cusp condition ${\sf u}={\sf u}^{\rm cp}$ maintains $\hat x_0=0$ and hence a cusp. Equilibrium and threshold states have regular correlators ($\hat x_0>0$) in the glass phase.}
\label{fig:shifts}
\end{figure}

\subsection{FRG flow: new physics for one-step RSB}

\subsubsection{Main analysis of FRG flow for one step RSB}

The situation is radically different in the case where the saddle point is solved
by a non-marginal one-step RSB scheme. This happens for $d < 2$ and $\gamma > \gamma_c(d)=2/(2-d)$. As discussed in Section~\ref{sec:1step_u_choice}, one may consider
several possible choice for the breakpoint ${\sf u}_c$. To interpret what is actually
computed in each case, one should go back to the definition (\ref{defhatv}) of the observable
$\hat V[v]$ whose second cumulant is computed here in the regime $v^2 \sim N$.

Let us first discuss the choice ${\sf u}_c={\sf u}^{\rm eq}_c$, where one
computes the cumulant of the observable $\hat V[v] = \hat V^{\rm eq}[v]$
defined as the {\it equilibrium free energy} of a manifold (i.e., a directed polymer $d=1$, or a particle $d=0$), in an external quadratic well at position $v$ and non-zero temperature $T$, in the limit of infinite $N$. More precisely, for every $m$ this prescription
selects the metastable states of the lowest free energy density in typical disorder.
This condition is by definition equivalent to requiring the configurational entropy to remain fixed to $\Sigma=0$ as $m$ varies, and thus closely resembles the ``iso-complexity'' condition
often used to describe the rapid quench in systems undergoing one-step RSB~\cite{LopatinIoffe02}.~\footnote{In 1-step systems, the complexity usually jumps to a finite value at the transition, which is then assumed to
remain constant throughout the glass phase under conditions of a rapid quench, because the trapping threshold states become stable and do not bifurcate. In the present case the situation is similar, except that the complexity vanishes at the transition.}

The first observation
is that for $m<m_c(T)$, $A(m) >0$ and this ''anomalous'' term acts as an ''effective temperature'' in the FRG equation (\ref{frgrsb}) which avoids the occurrence of a cusp, similarly as a finite temperature does in the case of finite $N$.
Let us evaluate this amplitude at low temperature (i.e., away from the
glass phase boundary, $T < T_c$) and
$m \to 0$. One has:
\begin{eqnarray}
\chi_0 \approx \frac{4 A_d}{\epsilon (2-d)} \frac{T}{{\sf u}_c} m^{d-2},
\end{eqnarray}
and as shown in Appendix \ref{app:lowTonestep} $T/{\sf u}_c^{th} = 1/\hat {\sf u}_c^{\rm eq}$ has a finite
limit as $T \to 0$. Hence $\chi_0$ diverges as $m \to 0$ and for model I $B''(\chi_0)=-B'(\chi_0)=e^{-\chi_0}$
decays exponentially fast to zero. For model II, $4 I_2 B''(\chi_0) \sim
m^{-2 + (2-d) \gamma}$, hence in all cases where $\gamma > \gamma_c$ one finds:
\begin{eqnarray}
A(m) \approx - m \partial_m \chi_0 \approx \frac{4 A_d}{\epsilon} \frac{T}{{\sf u}_c} m^{d-2}.
\end{eqnarray}
This term in the FRG equation (\ref{frgrsb}) has precisely the same form and
scaling as a standard (one loop) temperature term \footnote{To one loop a non-zero
temperature generally produces an additional term $\sim T \nabla_u^2 R(u)$
in the FRG flow for $R(u)$. Upon proper rescaling of $R(u)$, so that
a $T=0$ fixed point exists, its coefficient generally scales as $\sim T m^\theta$
and temperature is said to be irrelevant if $\theta <0$, relevant if $\theta >0$
or marginal if $\theta=0$.} with $T \to T/{\sf u}_c$, avoiding the occurrence of a cusp.
Remarkably, this temperature takes in the limit $m \to 0$ the same value as the effective temperature $T_{\rm eff}$ which arises in the modified fluctuation dissipation relation in such systems~\cite{CugliandoloKurchan1993,CugliandoloPLD}.

Let us now discuss in more detail the solution of the self-consistent
equation (\ref{SCequationrescaled}). We start with model II. If in (\ref{scalingb}) we choose the exponent values from the region $v^2 \sim O(1)$, i.e., $\zeta=(2-d)/2$ and $\theta=d-2+2 \zeta=0$ we find that:
\begin{eqnarray}
\tilde b'_0 \sim m^{- 2 \tilde \theta} \quad , \quad \tilde \theta=1 - \frac{2-d}{2} \gamma,
\end{eqnarray}
where $\tilde \theta < 0$ for $\gamma > \gamma_c$. Hence, with this scaling the second cumulant in
the region $v^2 \sim O(N)$ tends to zero. To study this flow
for $m\to 0$ one can instead use the choice $\theta \to \tilde \theta$ in (\ref{scalingb}) and find a
fixed point for the scaled function defined in that way. Indeed, one can check easily
that with the same $\zeta=(2-d)/2$, as $m \to 0$:
\begin{eqnarray}
&& \hat B'(x) = m^{2-2 \tilde \theta} \tilde b^{* \prime}(x m^{2 \zeta}) \\
&& b^{* \prime}(\hat x)=\gamma^\gamma \left[\frac{4 A_d}{\epsilon (2-d)} \frac{T}{{\sf u}_c}+\hat x\right]^{-\gamma},\label{b'*}
\end{eqnarray}
which indicates that the force fluctuations are reduced by a factor
$m^{-2 \tilde \theta}$ in the region $v^2 \sim O(N)$ as compared to
the region $v^2 \sim O(1)$. For model I, also obtained by
taking $\gamma \to \infty$, the effect is even stronger,
the decay of correlations being exponential of the form:
\begin{eqnarray}
&& \hat B'(x) \sim e^{- \frac{1}{m^{2 \zeta}} \frac{4 A_d}{\epsilon (2-d)} \frac{T}{{\sf u}_c}}e^{-x},
\end{eqnarray}
which cannot be put in the form of a scaling function as in
(\ref{scalingb}).
The equilibrium
flow for that case in $d=1$ is shown in Fig.~\ref{fig:flow1d}. It simply amounts to the horizontal shifting of the master function $\Phi(x)$ by $\hat x_0^{\rm eq}$, which is plotted in Fig.~\ref{fig:shifts}.
While the correlator first grows with decreasing mass
in the high temperature phase (like in $d>2$), it decreases again within the glassy phase.
The reason is that as $m$ decreases below $m_c$, the stability of the one-step energy landscape increases and suppresses sample-to-sample fluctuations of the global shift $u^0_k$ of the states, which in turn sets the scale for force-fluctuations.
\begin{figure}
\includegraphics[width=3.5in]{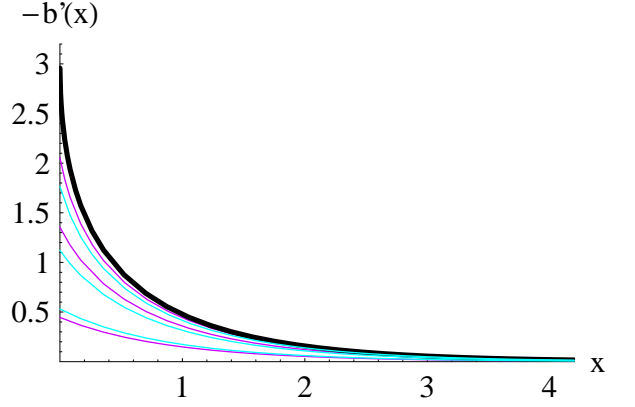}
\caption{FRG flow of the equilibrium correlator $\tilde{b}(\hat{x})$ for model I in $d=1$, at the
reduced temperature is $T=T_{\rm max}/2$. The dashed lines (light blue) correspond to
$m/m_c(T)=\{2,1.5,1.25\}$, from bottom to top, and the dotted lines (purple) to $m/m_c(T)=\{
0.4,0.3,0.2\}$ from top to bottom. Note that the correlator only exhibits a cusp exactly at the continuous
transition, $m=m_c(T)$, with $\tilde{b}'_0(m_c(T))=\epsilon=3$ (thick line). Due to the specific properties of model I the correlator always equals the same master function, displaced by a $m$-dependent horizontal shift $\hat x_0$.} \label{fig:flow1d}
\end{figure}

\subsubsection{Force correlator, shocks and the cusp}

In the limit $\gamma \to \gamma_c$ one has $\tilde \theta \to 0$, as one recovers
the marginal one step solution,~\footnote{The $x$ dependence in the limit is
delicate, since (\ref{b'*}) is only correct asymptotically as $m\to 0$ if $\gamma>\gamma_c$, but not for $\gamma=\gamma_c$. This is because, for $\gamma=\gamma_c$, one
cannot neglect the last term in the argument of $B'$ in (\ref{SCequationrescaled}), while it is negligible
for $\gamma > \gamma_c$ as $m \to 0$.} and
$\hat B'(0)$ then scales in the same way in the two regimes $v^2 \sim 1$
and $v^2 \sim N$. As noted previously the value of $\hat B'(0)$ from the $v^2 \sim N$ regime described by FRG is proportional to the contribution of the lower plateau to the two
point correlation at $k=0$. It also gives the sample to sample fluctuations of the global shift of states, $u^0_k$, in a given sample (\ref{states1}). While in the case of
continuous RSB they scale as the relative fluctuations of the states, this is not the case
for the non-marginal one step, since there $G_0(k)$ [and $\sigma_0$] is
much smaller than $G_1(k)-G_0(k)$. In the non-marginal one step case it is the
upper plateau which dominates the two point correlation (for any
$k \ll 1/m_c$).
 Upon moving the potential well by $v\sim N^{1/2}$, the global shift of
states becomes a function of $v$, $u^0_k=u^0_k(v)$, the connected
correlator of $u^0_k(0)$ and $u^0_k(v)$ being described by $G_{2v}(k)$.
In this regime the free energies are of order $N$, and one may
expect that configurations centered around $u^0_k(0)$ or $u^0_k(v)$ differ typically
by quantities of order $E(N)$ which diverge as $N\to \infty$ (at least as $N^{1/2}$).
In that case, when two levels cross as a function of $v$, the switching of
the equilibrium position takes place as a function of the scaling variable
$E(N)/T$. Hence, it turns into sharp jumps as $N\to \infty$ even at finite
$T$.

If these jumps are of order $N^{1/2}$ in the displacement $v$, they show up as a non-analyticity
in the force correlator. This is what is observed in the case of marginal one step and continuous RSB. On the other hand, stable (i.e., non marginal) one step systems
seem to have a much smoother landscape leading to an analytic response of
the system to a displacement of the well. This indicates that if there are
shocks, their size does not scale with $N^{1/2}$, i.e., the system
proceeds in small, and effectively rounded jumps in contrast to
the marginal cases.

The non-analytic response of systems with continuous RSB is tightly bound
to their marginal stability with respect to clustering. This criticality
of the system implies an anomalous response to a field ($v$) pulling apart
two replica groups, whereby the above argument suggests that this proceeds
via the occurrence of shocks of size $N^{1/2}$.
Notice that the presence of the non-analyticity only relies on the
marginality at ${\sf u}_m$ whereas it is insensitive to the marginality of
the RSB scheme at larger ${\sf u}$ (deeper down in the ultrametric
hierarchy). This indicates that the shocks are really associated with
jumps on the largest scales involving $u^0_k(v)$. That is, the whole
hierarchy of states is affected, and not only part of it, as it can happen
in the regime $v^2\sim 1$.

\subsubsection{Some subtleties of the FRG flow for one step RSB}

Although we have focused on the flow at small $m$, interesting features also
happen around the Larkin mass. At $m_c$ a cusp appears and
$A(m)$ first vanishes at $m=m_c$, but then becomes positive again. This behavior
indicates that a bifurcation in the lowest free energy state
occurs at $m_c$, where metastable states appear and the system is critical.
However, the further evolution of
the system for $m < m_c$ is smooth in the sense that the system becomes uncritical again, with no further bifurcations occurring. This is seen to translate into a response which is regular when the harmonic well is displaced by $v$ with $v^2 \sim N$.

To complete the discussion of the equilibrium FRG flow, we mention a subtlety (corresponding to the choice ${\sf u}_c={\sf u}_c^{\rm eq}$) which arises for temperatures in the range $T\in [T_c(0),T_c^{\rm max}]$.
An inspection of the phase diagram in Fig.~\ref{fig:phasediagram1step} shows that as $m\to 0$, the thermodynamic equilibrium is given by a replica symmetric saddle point in the GVM, which is reached from the glassy phase by a {\em discontinuous} transition upon decreasing $T$.
In that temperature regime, one natural continuation of the FRG flow from large to small mass is to maintain the condition $d\phi/d{\sf u}_c=0$, defining ${\sf u}_c^{\rm "eq"}$ even though it corresponds to ${\sf u}_c^{\rm "eq"}>1$ in the region where the genuine equilibrium is described by a RS solution. This reflects the fact that the selected states are thermodynamically irrelevant, in the sense that their Gibbs weight is negligible. Nevertheless, there are small corners of phase space where metastable states exist even when the mass is small, and among those states, the ones selected by ${\sf u}_c^{\rm "eq"}$  still correspond to those with the lowest available free energy density in typical disorder. We have verified that this branch of one-step solutions indeed exists down to $m=0$ for all temperatures $T<T^{\rm max}$.

We now turn to the discussion of other possible choices for the breakpoint.
If one chooses ${\sf u}_c={\sf u}^{\rm th}_c$ one
presumably computes the cumulant of an observable, $\hat V[v] = \hat V[v]^{\rm th}$,
defined as the free energy of the threshold states for the manifold in an external quadratic well at position $v$.
These threshold states being known to be
relevant for the dynamics, we may conjecture that the quantity computed using this criterion will find an interpretation as a part of the dynamical effective action.
Note that under this prescription the configurational entropy of the selected states, $\Sigma^{\rm th}(m)$ grows with decreasing mass, reflecting either the bifurcation or birth of threshold states.

Another choice of breakpoint, and the only one which leads to a cusp, is ${\sf u}_c={\sf u}^{\rm cp}$. Then the amplitude
$A(m)=0$. Hence it leads to similar fixed points as for the continuous
RSB case. In particular, it yields a fixed point of $\tilde{b}_0'$ and freezes the renormalized correlator to the cuspy shape it attains at $m_c(T)$. The exponents are
then the same, i.e., $\zeta=\zeta(\gamma)$. Note then
that $\theta=\theta(\gamma) < 0$ which means that the temperature is relevant.
That such a choice can be made was conjectured in \ofrgN for $T=0$. Here we show it to be possible at any temperature. It is not difficult to prove that one-step solutions satisfying the cusp condition
(\ref{cuspmarginality}) exist for all $T\leq T_{\rm max}$ and $m<m_c(T)$.
Note, however, that one finds ${\sf u}^{\rm cp}>1$ for $m<m_*$ and $T>\tilde T_c(m)$, where $\tilde T_c(m)$ is the (unphysical) branch of the instability line defined by (\ref{tc}), see Fig.~\ref{fig:us}.

A natural question is then: what observable would be selected by this
process? As discussed in Sec.~\ref{sec:1step_u_choice} the choice ${\sf u}^{\rm cp}$ appears to select metastable states that only exist in rare disorder configurations (the smaller $m$ the rarer). It remains
to be understood whether such an observable could be constructed by imposing some
constraint on the metastable states.

\subsection{Non-uniform $v$}

The above analysis is easily generalized to a non-uniform $v_x$, and allows
to compute the functional $\hat R[v]$ for a non-uniform $v_x$ in the regime
$v^2 \sim N$ where it takes the form $\hat R[v]= N \hat B[v^2]$
where $\hat B[v^2]:=\hat B[\{ \tilde v_x^2 \}_x]$, i.e.
it is a functional of the field $\tilde v^2_x$. We just give the final saddle-point equations
and the resulting self-consistent equation for $\hat B$, which generalize the one
obtained in Ref.~\cite{LeDoussalWiese2003b} (for $R[v]$) in the replica symmetric region (see Eq.~(3.31)
there with the correspondence in notations $\frac{\delta \tilde U_0[v\cdot v]}{\delta(v^a(x)\cdot v^b(x))}\to \sigma_{\tv}(x)$). One easily sees that the diagonal blocks (in the space of two replica groups) are again independent of $\tilde v_x$ and therefore independent of $x$, and thus simply correspond to the MP solution. Generalizing (\ref{sigma2first}), the saddle-point equation in the off-diagonal sector yields a single equation, which now involves space indices:
\BEA
\sigma_{2,\tv}(x)&=& -\frac{2}{T} B'\left(4 \tv_x^2 + 2 \left[\int_k \tilde G(k)\right] -2 G_{2,\tv}^{xx}\right),\quad \quad \label{sigma2SCnonuniform}
\end{eqnarray}
where
\bea
G_{2,\tv}^{xx}= T \int_y g^2(x-y) \sigma_{2,\tv}(y)\,,
\eea
which generalizes (\ref{G2v}). As in the uniform case $\sigma_{2,\tv=0}(x)=\sigma({\sf 0})$.
Solving for the function $\sigma_{2,\tv}(x)$ yields the desired functional since:
\bea
-\frac{2}{T} \frac{\delta \hat B[w^2]}{\delta w_x^2}\bigg|_{w^2_y=\tilde v^2_y} =\sigma_{2,\tv}(x)\,,
\eea
which is the analog of Eq.~(\ref{identification}). These equations generalize (\ref{frgscrsb1}) to a non-uniform field $\tilde v_x$. It would be interesting to see whether FRG equations can be derived also for
the non-local parts of the functional $\hat B$, but this is left for future investigation.

\section{Discussion and conclusion}
\label{s-discussion}

We have computed the renormalized disorder-correlator $R(v)$ in the large $N$ limit
for an elastic manifold in a $O(N)$ symmetric random potential in presence of an external
quadratic well. It was obtained as a physical observable describing the correlation
in free energy when the center of the quadratic well is varied by $v$. It contains
direct information about shocks, i.e., abrupt switches between two competing equilibrium positions that occur
as $v$ is varied. We have demonstrated the existence of two large-$N$ scaling regimes $v^2\sim 1$ and $v^2\sim N$ and obtain closed expressions in each regime, and their zero and low temperature limits.
Our results provide a direct connection between the GVM approach ($v=0$) using replica symmetry breaking
saddle points and previous large $N$ FRG approaches.

We found that as $v$ is increased shocks start to occur at a uniform displacement of the well of order $v \sim v_* \sim L^{-d/2}$ which decreases with system size or, equivalently for a non uniform displacement $v_x$ of order one but confined to a bounded region of volume one in space \footnote{such as a suitable norm, e.g. $\int_x v_x^2$, remains of
order one}. These shocks are rounded by temperature, but turn into a non-analytic cusp of the force correlator at $T=0$. Our results bear some similarities to shocks found in Burgers turbulence in large dimension. The shocks can be interpreted
within the RSB picture of an ultrametric phase space, which predicts that in a given disorder environment there exist several {\it states}, all centered within a single {\it big valley} whose position itself fluctuates from sample to sample.
The shock regime $v\lesssim v_*$ then describes the energy crossing between states as the harmonic well is shifted, and accordingly, it is sensitive to the replica symmetry breaking structure of the GVM approach. In particular, one finds that all GVM saddle points related by replica permutations contribute to the computation of observables. Beyond this regime, the force correlator remains constant in a large interval $N^{1/2}\gg v>v_*$ which reflects the sample-to sample fluctuations of the force density associated with the big valley, itself related to the sample-to-sample fluctuations of the global displacement $u^0$ of the valley.

This value of the force correlator matches perfectly with the value obtained from FRG calculations in the regime $v^2\sim N$. In that regime it was found that FRG recovers only the fluctuations among most distant states, which is governed by the (non-trivial) lower plateau of the self energy function, $\sigma_0$. It is now clear that the FRG in this regime captures the shock structure of
the big valley position $u^0(v)$. Indeed, imposing a position of the well on the scale $v^2\sim N$ affects the global shift $u^0$ itself, which may lead to shocks.
We have computed the decaying correlations between $u^0(0)$ and $u^0(v)$, as well as the force correlator. In the case of systems with continuous or marginal one step RSB, such as manifolds in $2\leq d\leq 4$ or directed
polymers in long range correlated disorder, we have found that their marginality towards a clustering instability implies a cusp also in this second scaling regime $v^2\sim N$.
This reflects the fact that the shift of the center of the well provokes shocks which are abrupt even at finite $T$ because they involve the crossing of energies which scale with $N$.
By establishing the persistence of the nonanalytic force correlator throughout the regime of small $m$ and $T$, we have proved that the previously obtained FRG flow had been correctly extended into the glassy regime.

A major step forward has been achieved in the case where the regime $v^2\sim 1$ displays a non-marginal one-step RSB, a problem which was left open in the previous FRG study. In particular we have found that in the regime $v^2 \sim N$ the force correlator is non-analytic only at the glass transition, but not within the glassy regime. We interpret the latter in terms of a smooth energy landscape where the manifold evolves smoothly (on the scale $N^{1/2}$) as the center of the harmonic well is varied.
In the glassy regime the disorder correlator becomes strongly suppressed with decreasing mass which is a consequence of the increasing stability of the one-step landscape and the associated smallness of sample-to-sample fluctuations of $u^0$.
Further, we have shown how the exact FRG equations in the regime $v^2\sim N$ can be extended into the glassy regime, and how replica symmetry breaking translates into an anomaly in the FRG flow equation. Moreover, the possibility of studying non-equilibrium branches of metastable states in a one-step system by tuning the break point parameter ${\sf u}_c$ was found to be equivalent to tuning the anomaly in the FRG equation.

The present study raises many questions. First, at finite $N$ the two scaling regimes in $v$ are not clearly distinct.
The main task is then to determine which aspect of either regime will persist at finite $N$. Until now the FRG flow equation in the loop expansion seemed to connect more directly to the regime $v^2 \sim N$. In particular one then expects that the
shocks of the regime $v^2 \sim  N$ should become rounded by temperature at finite $N$, as was found in FRG calculations in the loop expansion and for $N=1$. However, the present study raises the possibility of a new regime in the FRG accessible only at smaller $v$. The interpretation of the cusp at $T=0$ in terms of shocks remains valid, and offers an interesting venue for future studies going beyond the mean field picture of switching between states. These concepts also generalize to other disordered systems, e.g., to spin glasses, where the statistics and properties of shocks can be studied by similar methods. Work is in progress in this direction. This should yield valuable new insight into the structure of the phase space and elementary excitations, such as droplets.

{\em Note added in proof:}
After completion of this paper we became aware of the work of H.\ Yoshino and
T.\ Rizzo\cite{YR}, which studies related spin models and
obtains results on shocks and cusps similar to those discussed in the present work for the regime $v^2 \sim 1$.

\acknowledgments

We wish to thank Leon Balents for useful discussions at an early stage of
this work (in relation to Ref.~\oBBM) on the possibility of several scaling
regimes at large $N$ and on issues concerning the thermal boundary layer. We
also thank J.P. Bouchaud, M. M{\'e}zard, and A. Middleton for
interesting discussions.
MM was supported by grant
PA002-113151 from the Swiss National Fund for Scientific Research.
PLD and KW acknowledge support from ANR program blan05-0099-01.
We acknowledge the hospitality of the KITP where part of this work was
accomplished.

\appendix
\section{Restoring dimensions}
\label{app:dimensions}
In the main body of the text we used the natural units $r_f$ for transverse lengths, $\mu_c$ for the mass (and $1/\mu_c$ for longitudinal lengths in $d>0$), and $E_c$ for energies.
In order to recover the full dependence on the parameters $B_0$, $r_f$ and $c$, one simply has to restore the dimensionful units so as to render all quantities dimensionless:

Masses and longitudinal lengthscales:
\bea
m,k,\Lambda &\to& \frac{1}{\mu_c}\left[m,k,\Lambda\right]\nn\\
L &\to& L\mu_c,\nn
\eea
Temperature and energy densities:
\bea
T &\to& T/E_c,\nn\\
\phi,f &\to& \frac{1}{E_c\mu_c^d}\left[ \phi,f \right],\nn
\eea
Transverse fluctuations:
\bea
u,v &\to& \frac{1}{r_f}\left[u, v\right],\nn\\
G(k=0) &\to& \frac{\mu_c^d}{r_f^2} G(k=0),\nn\\
\int_k G(k),\,\chi &\to& \frac{1}{r_f^2}\left[\int_k G(k),\,\chi\right],\nn
\eea
Self-energy and other auxiliary functions:
\bea
\sigma,\Sigma_1,g^{-1}(k) &\to&\frac{r_f^2}{\mu_c^d E_c}\left[\sigma,\Sigma_1,g^{-1}(k/\mu_c)\right],\nn\\
I_n &\to& \frac{E_c}{r_f^2}\left(\frac{E_c\mu_c^d}{r_f^2}\right)^{n-1}I_n,\nn\\
\hat{B}^{(n)}(z)&\to& \frac{r_f^{2n}}{\mu_c^d E_c^2}\hat{B}^{(n)}(z/r_f^2),\nn\\
\hat{R}(v)&\to&\frac{1}{\mu_c^d E_c^2}\hat{R}(v/r_f),\nn\\
\hat{R}[v],R_n,{\cal R}_n&\to&\frac{1}{E_c^2}\left[\hat{R}[v/r_f],R_n,{\cal R}_n\right],\nn\\
\hat{W},q,Q,{\sf u},\tilde{b} &\to&  \hat{W},q,Q,{\sf u},\tilde{b}.
\eea
With these substitutions all equations turn into dimensionless identities, whereby in some cases additional factors of $c= \mu_c^{d-2}E_c r_f^2$ need to be restored. Note that amplitudes such as $A$ being dimensionlee are unchanged.

\section{Direct expansion of $\hat{W}$ to order $v^4$}
\label{app:v4}

The perturbative expansion of (\ref{start}) to second order requires
\bea
P^{(2)}_{abcd}:=\sum_\pi' G^{(\pi)}_{ab}G^{(\pi)}_{cd}.
\eea
In the case of two sets of replica with $v^a=v^{12}/2$ for $a=1,..n/2$ and $v^a=-v^{12}/2$ for $a=n/2+1,..n$, one has $G_{ab}=q_{ab}/2$ with $q_{ab}$ defined in (\ref{qdeu}).
As for the sum $P^{(1)}_{ab}:=\sum_\pi' G^{(\pi)}_{ab}=\alpha\delta_{ab}+\beta$ (cf., \ref{Gres}),
replica symmetry restricts this tensor to take the form
\bea
P^{(2)}_{abcd}&=&
A\delta_{abcd}+B(\delta_{abc}+\delta_{abd}+\delta_{acd}+\delta_{bcd})\nn\\
&&+C_1\delta_{ab}\delta_{cd}+C_2(\delta_{ac}\delta_{bd}+\delta_{ad}\delta_{bc})+D_1(\delta_{ab}+\delta_{cd})\nn\\
&&+D_2(\delta_{ac}+\delta_{ad}+\delta_{bc}+\delta_{bd})+E,
\eea
the coefficients of which can be obtained by
solving the system of linear constraints arising from the identities
\bea
\sum_a P^{(2)}_{abcd}&=&G^cP^{(1)}_{cd},\nn \\
\delta_{ab}P^{(2)}_{abcd}&=&\tilde{G} P^{(1)}_{cd},\nn\\
\sum_a \delta_{ac}\delta_{bd}P^{(2)}_{abcd}&=&\sum_aG_{ab}^2.\nn
\eea
The quantity of interest for the second
order expansion of (\ref{start}) is the second cumulant:
\BEQ
\hat W[v]^{(2)}=\frac{1}{2}\left[\sum_{abcd} v^a v^bv^cv^d
P^{(2)}_{abcd}-\left(\sum_{abcd} v^a v^b P^{(1)}_{ab} \right)^2\right].
\EEQ

For the case of two replica groups, one obtains the final result:
\bea
&&\hat W[v]^{(2)}=\frac{1}{2}v^4\left[n A+n^2(C_1+2C_2)-n^2\alpha^2\right]\\
&&\quad= \frac{v^4n^2(2-n)}{(3-n)(1-n)^2}\left[\left(\sum_{a\neq
1}G_{1a}\right)^2-(1-n)\sum_{a\neq 1}G_{1a}^2\right]\nn\\
&&\quad =\frac{2v^4}{3}\left[\int_0^1G^2({\sf u}) d{\sf u}-\left(\int_0^1G({\sf u}) d{\sf u}\right)^2\right] n^2+O(n^3).\nn
\eea
From this one derives (\ref{finalR2}) in the text, by recalling that
\bea
R_2[v]=\lim_{n\to 0} \frac{4T^2}{n^2}\hat W[v]^{(2)},
\eea
and $G({\sf u})=q({\sf u})/2$.

\section{A useful identity}
\label{app:identity}

For any function $\Phi(y)$ with derivatives decreasing strictly faster
\footnote{If the function is even, $1/|y|$ behavior is allowed.}
 than $1/|y|$ for large $|y|$, one has
\BEA
\label{Pierre}
&&\int_{-\infty}^\infty dy\, y [\Phi(y+z) + \Phi(y-z) - 2\Phi(y)] =\nn\\
&&\quad \quad \quad\quad\quad\quad\quad- z^2 [\Phi(\infty)-\Phi({-\infty})].
\EEA
Indeed,
\begin{eqnarray}\label{38-1883213723}
&&\int_{-\infty}^\infty dy\, y [\Phi(y+z) + \Phi(y-z) - 2\Phi(y)] \nn\\
&&\quad =\int_{-\infty}^\infty dy\, y \int_0^z dz' [\Phi'(y+z') - \Phi'(y-z')] \nn\\
&&\quad=
-\int_{-\infty}^\infty dy \int_0^z
dz' [\Phi(y+z') - \Phi(y-z')] \nn\\
&&\quad =-\int_{-\infty}^\infty dy \int_0^z dz' \int_{-z'}^{z'}dz'' \Phi'(y+z'')\nn\\
&&\quad =-[\Phi(\infty)-\Phi({-\infty})] \int_0^z dz' \int_{-z'}^{z'}dz'' \nn\\
&&\quad = -z^2[\Phi(\infty)-\Phi({-\infty})].\nn
\EEA
Applied to integrals over Gaussian averages, normalized s.t.\ $\left<z^2\right>_{z}=1$, Eq.~(\ref{Pierre}) implies
\BEA \label{appendixidentity}
&& \int_{-\infty}^\infty dy \,y \left[\langle \Phi( y + \sqrt{Q} z ) \rangle_z - \Phi(y)\right] =\qquad\nn\\
&&\quad\quad - \frac{Q}{2}   \left[\Phi(\infty)-\Phi(-\infty)\right]\,.
\EEA

\section{Perturbation expansion for $L^d v^2\ll 1$}
\label{app:PTfullFRG}

Here we compute the two lowest orders in the expansion
defined in Section \ref{sec:PTfullFRG}. The functions
$m_{1,2}$ satisfy, from (\ref{eqs1}):
\begin{eqnarray} \label{eqs11}
&& \dot m_1 = - \frac{1}{2} \dot q({\sf u}) \left[m_0'' + {\sf u} (m_0^2)'\right]\,, \\
&& \dot m_2 = - \frac{1}{2} \dot q({\sf u}) \left[m_1'' + {\sf u} (2 m_0 m_1)' \right]\,, \qquad\label{eqs21}
\end{eqnarray}
To lowest order, using that:
\begin{equation} \label{relm0)}
m_0'' + (m_0^2)' = 0,
\end{equation}
one has to solve:
\begin{equation} \label{C4}
 \dot m_1 = - \frac{1}{2} (1- {\sf u}) \dot q({\sf u}) m_0''\,.
\end{equation}
The solution is:
\begin{equation} \label{solum1}
m_1 = p({\sf u}) m_0'' \quad , \quad p({\sf u})=\frac{1}{2} \int_{\sf u}^{{\sf u}_c} d\tilde {\sf u} (1-\tilde{\sf u}) \dot q(\tilde {\sf u}).
\end{equation}
Integration
by parts gives $2 p({\sf u}_m)= \int_{0}^{1} d{\sf u}\, (1-{\sf u}) \dot q({\sf u}) =
-q({\sf 0})+\int_0^1 q({\sf u}) d{\sf u}$. The integration range can be extended
to the interval $[0,1]$ since $\dot q=0$ outside $[{\sf u}_m,{\sf u}_c]$.
Plugging this into (\ref{sumR}) and using $\int_0^\infty y m_0''=-\int_0^\infty m_0'= -1$,
the term proportional to $q({\sf 0})$ cancels and one obtains
(\ref{smallvFRSB1}) in the text.

The next-order correction satisfies, from (\ref{solum1}) and (\ref{eqs21}):
\begin{equation}
\dot m_2 = - \frac{1}{2} \dot q({\sf u}) p({\sf u}) \left[m_0'''' + {\sf u} (2 m_0 m_0'')' \right]\,.
\end{equation}
This gives:
\begin{equation}
m_2 = \frac{1}{2} m_0'''' \int_{{\sf u}}^{{\sf u}_c} d\tilde{\sf u}\, \dot q p + (m_0 m_0'')' \int_{{\sf u}}^{{\sf u}_c} d\tilde{\sf u}\, \tilde{\sf u} \dot q p\,.
\end{equation}
In order to calculate (\ref{sumR}), we need
\begin{eqnarray}
&& \int_{-\infty}^\infty dy\, y m_0'''' = - [m_0'']_{-\infty}^\infty = 0 ,\\
&& \int_{-\infty}^\infty dy\, y (m_0 m_0'')' = \int_{-\infty}^\infty dy\, m_0 (m_0^2)' = \frac{4}{3}\,,\qquad
\end{eqnarray}
where on the second line we used (\ref{relm0)}). Inserting into (\ref{sumR}) gives
\begin{eqnarray}
\label{smallvFRSB2}
R_2[v] = \frac{8}{3} T^2 \int_0^1 d{\sf u}\, {\sf u} \dot q({\sf u}) p({\sf u}),
\end{eqnarray}
where $p({\sf u})$ is defined in (\ref{solum1}). Substituting the definition (\ref{C4},\ref{solum1}),
${\sf u} \dot q= 2 \dot p + \dot q$, integrating by parts, using $p(1)=0$
one finds:
\begin{eqnarray}
&& \int_0^1 d{\sf u}\, {\sf u} \dot q({\sf u}) p({\sf u}) = [p^2+ q p]_0^1 - \frac{1}{2} \int_0^1 d{\sf u}\, ({\sf u}-1) q \dot q \qquad
\nonumber \\
&&=\frac{1}{4} \int_0^1 d{\sf u}\, q^2 - \frac{1}{4} q({\sf 0})^2 - p(0) [p(0)+q({\sf 0})],
\end{eqnarray}
which with $p(0)=1/2 [\int_0^1 d{\sf u}\,q - q({\sf 0})]$ yields the final result (\ref{finalR2}) in the text.

\section{Thermal boundary layer of the effective potential correlator}
\label{app:BBMrevisited}

Here we revisit and complete the calculation of the effective potential defined in Ref.~\oBBM,
in an attempt to connect with the FRG function $R(u)$. The effective potential studied there
is constructed (see Ref.~\oBBM for details) from the probability distribution for a given Fourier
mode $u=u_k$ (denoted there $\vec \phi_0$) in a given environment. This sample dependent probability is
denoted here $Z_V[u]$ (unnormalized) and there ${\cal P}_{\Omega}(\vec \phi_0)$.
The authors then introduce the potential correlator:
\BEQ
{\cal V}(u-u')=-\lim_{n\to 0}\frac{\overline{Z(u)^{n/2}Z(u')^{n/2}}-\overline{Z(u)}^{n/2}\overline{Z(u')}^{n/2}}{(n/2)^2 \beta^2}, \label{defbbm}
\EEQ
describing the second moment of the effective potential $\tilde{V}(u)=\beta^{-1}\ln[Z(u)]$.
This is the central object studied there. Although it is a very physical object, it is very different from the effective potential studied here $\hat V(v)$ - which is also a physical observable - but involves a source (implemented via a quadratic well). As a result $\tilde{V}(u)$
cannot be connected to the standard FRG \footnote{Although it is not excluded that another
version of FRG could be built from it - in a spirit close to Polchinsky RG
see discussion in Ref.~\onlinecite{LeDoussalWiese2003b}.}. Some discussion of the differences
between the two approaches was given in
Section VIII G of Ref.~\onlinecite{LeDoussalWiese2003b}. Here we discuss
more of the differences in details, based on explicit calculation.

Since this version of an effective disorder is also interesting, it is worth to push here further the
calculation of Ref.~\oBBM. To avoid confusion between $u$ (displacement field) and ${\sf u}$
(replica overlap) we replace everywhere $u \to v$, keeping in mind however that its physical
meaning is different from the one (position of well center) given in the text.
The above partition function can be evaluated via a saddle point as in the GVM:
\begin{eqnarray}
&&\overline{\prod_{a=1}^n Z(v^a)}=\sum_{\pi}\exp\left[
-\frac{\beta}{2}v^{\pi(a)}G_{ab}^{-1}(k)v^{\pi(b)}\right]\nonumber\\
&&\quad =\exp\left[-\frac{n\beta}{4}(k^2+m^2)(v^2+v'^2)\right] \times\nn\\
&&\quad\quad\quad\quad\sum_\pi
\exp\left[\frac{\beta}{2}\left(v^{\pi(a)}\sigma_{ab}v^{\pi(b)}\right)\right]\,.
\end{eqnarray}

In the same way as we derived $R[v]$, we can compute the correlator of $\tilde{{\cal V}}$ as:
\begin{eqnarray}
{\cal V}(v-v')
&=& -2T^2\int_{-\infty}^\infty dy \,y\left[ \tilde{M}({\sf 0},y)-\tanh(y)\right]\,,\nn
\end{eqnarray}
where $\tilde M$ satisfies the same flow equation as $M$, but with the ''inverse coupling'':
\bea
\tilde{q}({\sf u})=\frac{\beta}{4}(v-v')^2\sigma({\sf u}).
\eea
We can now apply our general expressions (\ref{R0FRSB},\ref{TBL}) to the above case and find for the thermal boundary layer:
\bea
-{\cal V}(v-v')&=&{\cal V}_0(v-v')+{\cal V}_1(v-v')+\dots\,,
\eea
where:
\bea
&&{\cal V}_0(v-v')=-2T^2\tilde q({\sf u}_c)= -\frac{T\sigma({\sf u}_c)}{2}(v-v')^2,\\
&&{\cal V}_1(v-v')=T^2\int_{{\sf u}_m}^{{\sf u}_c}d{\sf u}\, \left( {\sf u}\frac{d\tilde{q}}{d{\sf u}}\right)\left\langle\psi(z\sqrt{2(\tilde{q}_c-\tilde q)})\right\rangle_z\nn\\
&&\quad=(v-v')^2 {\sf u}_c\sigma_c(1+\gamma)\times\nn\\
&&\quad\int_{\frac{m^\theta}{m_c^\theta}}^1 d v\, v^{\frac{2}{\theta}-1}\left\langle\psi\left(z\left|\frac{v-v'}{T}\right|
\sqrt{\frac{T\sigma_c}{2}\left(1-v^{\frac{2}{\theta}-1}\right)}\right)\right\rangle_z\,,\nn
\eea
where $\sigma_c=\sigma({\sf u}_c)$.
These formulae have well-defined limits as $T\to 0$ (recalling that ${\sf u}_c/T= Am_c^\theta$ and $T\sigma_c=2m_c^{2-\theta}/[(2-\theta)A]$). For $m\to 0$ the last term reduces to:
\bea
{\cal V}_1(v-v')\to  \frac{(v-v')^3 {\sf u}_c\sigma_c(T\sigma_c)^{1/2}}{4}
\frac{\Gamma[1+{\theta}/(2-{\theta})]}{\Gamma[5/2+{\theta}/(2-{\theta})]}.\nn
\eea

With the notation introduced in Ref.~\oBBM (for $d\geq 2$),
\bea
1+\gamma&:=&\frac{2}{\theta}-1=\frac{4-d}{d-2},\\
g&:=& \frac{1+\gamma}{8} \beta \sigma({\sf u}_c) u_c^2(v-v')^2,
\eea
the above results can be recast into the form:
\begin{eqnarray}
{\cal V}(v-v')=-{\cal V}_0-{\cal V}_1=\frac{4}{(\beta {\sf u}_c)^2}\left[\frac{g}{1+\gamma}- {\sf u}_c^3 \Upsilon_2\left(\frac{g}{{\sf u}_c^2}\right)\right],\nn
\end{eqnarray}
where:
\begin{eqnarray}
\label{TBL3} \Upsilon_2(x)&=&\frac{x}{2} \left\langle \int_0^1 dv \,v^{1+\gamma}
\psi\left(z\sqrt{\frac{4x(1-v^{1+\gamma})}{1+\gamma}}\right)\right\rangle_z,
\end{eqnarray}
and $\psi(x)=2x\coth(x)$, as before. Using $\psi(x){\to}2|x|$ for large arguments, one further finds:
\bea
\Upsilon_2(x)\stackrel{x\to \infty}\to \sqrt{2}x^{3/2}\left[\frac{1}{(1+\gamma)^{5/2}}\frac{\Gamma(1/(1+\gamma))}{\Gamma(5/2+1/(1+\gamma)}\right]\,.\nn
\eea
The term in brackets is easily seen to tend to $1$ as $1+\gamma =\epsilon/(d-2)\to 0$. Thus, $\Upsilon^{\epsilon\to 0}_2(x\to \infty)\to \sqrt{2}x^{3/2}$.~\footnote{This result is smaller than the one given in Ref.~\oBBM by a factor of $\sqrt{2\pi}$.}

For large arguments $v$, the flow of $\tilde M$ is attracted to an intermediate fixed point $\tilde{M}({\sf u},y)\approx \tanh({\sf u}y)$. This happens for a large coupling parameter $g\gg1$ (or more explicitly for displacements such that $(v-v')^2m_c^{2+\theta}\gg1$). If in addition the mass is sufficiently small, $(v-v')^2\ll m^{-(2+\theta)}$, the  correlator ${\cal V}$ is controlled by this intermediate fixed point and can be shown to scale as
\begin{eqnarray}
{\cal V}(v-v')&\sim &g^{2\theta/(2+\theta)}\sim (v-v')^{4\theta/(2+\theta)}.
\end{eqnarray}


An important difference to the FRG correlator $R(v)$ concerns the scale on which the nonanalyticity lives in ${\cal V}(v)$. Comparing ${\cal V}_0$ and ${\cal V}_1$ we find that higher order terms become dominant for $v\geq v_{**}\sim m_c^{(2+\theta)/2}$, reflecting the fact that the effective potential correlator ${\cal V}$ is sensitive to physics on the Larkin scale, rather than to shocks occurring on the scale $m$, as is the case for the FRG correlator.

\section{Comparison with the TBL for droplets}
\label{app:droplet}

Due to the $O(N)$ symmetry, the disorder correlator is only a function of $v=|\vec v|$ (for spatially uniform $v$). Accordingly, the force correlator splits into
transverse and longitudinal parts:
\beq
- \p_{v_i}\p_{v_j}\hat R(v) = \Delta_L(v) \frac{v_i v_j}{v^2} + \Delta_T(v) \left(\delta_{ij}-\frac{v_i v_j}{v^2} \right),
\eeq
with $\Delta_L(v) = -\hat R''(v)$ and $\Delta_T(v) =  -\frac{\hat R'(v)}{v}$, derivatives being
w.r.t $v=|\vec{v}|$. Both correlators are equal at $v=0$, so that $\p_{v_i}\p_{v_j}{\cal \hat R}(v)$ is well defined
and proportional to $\delta_{ij}$ at $v=0$.
Note that at large $N$ one has $\frac1N \sum_i\partial_{v_i}^{2}{\cal R}=\Delta_T(v)+O(1/N)$,
and hence the transverse components dominate the average force correlator.

The TBL contribution to the droplet force correlator (\ref{droplet}) can be expressed in terms of longitudinal and transverse correlators:
\bea
-\Delta^{\rm drop}_L(v) &=& T\frac{m^4}{4} \left\langle y^2_1 \psi\left(\frac{m^2 y_1}{2}\hat v\right)\right\rangle_y\,,\\
-\Delta^{\rm drop}_T(v) &=& T\frac{m^4}{4} \left\langle y^2_2 \psi\left(\frac{m^2 y_1}{2}\hat v\right)\right\rangle_y\,,
\eea
where $y_1=y_\parallel$ denotes the component of $y$ parallel to $v$, while $y_2$ is an arbitrary orthogonal component.

To make contact with expressions obtained for the manifolds, we recast these expressions into the form:
\bea
-\Delta^{\rm drop}_{L,T}(v) = T\int db\, \rho^{\rm drop}_{L,T}(b)\, \psi(b |\hat v|),
\eea
with the distributions
\bea
\label{PdropL}
\rho^{\rm drop}_{L}(b)&=&\frac{2}{m^2} \int d^{N-1}y_\perp \, b^2 D\left(y_\parallel=\frac{2b}{m^2},y_\perp\right),\\
\label{PdropT}
\rho^{\rm drop}_{T}(b)&=&\frac{2}{m^2} \int d^{N-1}y_\perp \,\left(\frac{m^2y_2}{2}\right)^2 D\left(y_\parallel=\frac{2b}{m^2},y_\perp\right).\nn
\eea

In computing the force correlators (\ref{TBLforcelong}) we have used
that:
\bea
&&\p_{\hat v} \left\langle \hat v^2 \psi(\hat v x z) \right\rangle_z=
\left\langle 2\hat v \psi(\hat v x z) +\hat v z \frac{d}{dz} \psi(\hat v x z)\right\rangle_z\\
&&\quad =\left\langle \hat v(1+z^2) \psi(\hat v x z)\right\rangle_z\,,\nn\\
&&\p_{\hat v}^2 \left\langle \hat v^2 \psi(\hat v x z) \right\rangle_z=\\
&&\quad= \left\langle  (1+z^2) \psi(\hat v x z) +z(1+z^2) \frac{d}{dz} \psi(\hat v x z)\right\rangle_z\nn\\
&&\quad = \left\langle  (z^4-z^2) \psi(\hat v x z)\right\rangle_z\,.\nn
\eea

\section{1-step saddle points in the GVM and their low temperature limits}
\label{app:lowTonestep}

Here we first give for completeness the general one-step saddle-point equations for the equilibrium statics, valid for any $d$ and $\Lambda$. Equations
(\ref{G01},\ref{G01_2},\ref{tildeG},\ref{chi01},\ref{sigma01},\ref{SPeq_onestep}) form a closed set of equations, which together with  (\ref{1stepaction},\ref{equilibrium}) determines ${\sf u}_c={\sf u}_c^{{\rm eq}}$:
\bea
0 = \frac{d \phi({\sf u}_c)}{d {\sf u}_c} &=& \frac{1}{2 T} [B(\chi_0)-B(\chi_1) - (\chi_0-\chi_1) B'(\chi_0)] \nn \\
&& - \frac{T}{2 {\sf u}_c^2} \int_k \left[ \frac{\Sigma_1}{g_k^{-1}+\Sigma_1} - \ln(1 + \Sigma_1
g_k) \right]. \qquad \label{eqgen}
\eea
The easiest way to show this is to consider $\Sigma_{1}$ and ${\sf u}_{c}$ as independent variational parameters of the saddle-point equations, s.t.\ $d {\Sigma_{1}}/d {\sf u}_{c}=0$, and then use (\ref{chisp}).

Let us now analyze the limit $T\to 0^+$ of the various one-step saddle points
for $d<2$ and $\Lambda=\infty$. It is easy to see that the corresponding saddle-point equations (\ref{SPeq_onestep},\ref{chisp}) together with the three possible conditions
(\ref{equilibrium}), (\ref{replicon}) or (\ref{cuspmarginality})
admit solutions of the form ${\sf u}_c=T\hat {\sf u}_c $ with finite $\hat {\sf u}_c $ as $T\to 0$:
\bea
\chi_0&=& \frac{4A_d}{\epsilon(2-d)}\frac{1}{\hat {\sf u}_c }\left[\frac{1}{m^{2-d}}-\frac{1}{(m^2+\Sigma_1)^{1-d/2}}\right],\\
\chi_1&\sim &  T \rightarrow 0,\nn\\
\Sigma_1&=&-2\hat {\sf u}_c \left[B'(\chi_1)-B'(\chi_0)\right]\to -2\hat {\sf u}_c \left[B'(0)-B'(\chi_0)\right].\nn
\eea
The three possible values for the breakpoint are:\\
(i) existence of a  cusp in the FRG in the large-$v$ regime ($\hat {\sf u}_c  = \hat {\sf u}_c ^{\rm cp}$), i.e. condition (\ref{cuspmarginality}):
\bea
\label{cuspmarginality_lowT}
\frac{4 A_d}{\epsilon m^{\epsilon}}B''(\chi_0)=1,
\eea
(ii) the equilibrium static condition ($\hat{\sf u}_{c}= \hat {\sf u}_c ^{\rm eq}$), which from  (\ref{eqgen}) using (\ref{phionestepexpl}) reads:
\bea
&&\frac{d\phi}{d\hat {\sf u}_c }=0=\frac{1}{2}\left[B(\chi_0)-B(0)-\chi_0 B'(\chi_0)\right]\\
&&
-\frac{A_d}{\epsilon (2{-}d) \hat {\sf u}_c ^2}\left(
\frac{\Sigma_1}{(m^2{+}\Sigma_1)^{1-\frac{d}{2}}}-\frac{2}{d}\left[(m^2{+}\Sigma_1)^{\frac{d}{2}}-m^{d}\right]\right),\nn
\eea
and (iii) the condition (\ref{replicon}) for threshold states ($\hat {\sf u}_c =\hat {\sf u}_c ^{\rm th}$), which is equivalent to:
\bea
\Sigma_1=m_c^2(T)-m^2\to m_c^2(0)-m^2.
\eea
For concreteness, we give explicit results for model I in the limit $m \to 0$; an illustration for general $m$ and $d=1$ at a fixed temperature can be found  in Fig.~\ref{fig:us}.

For small $m$,  $B(\chi_{0})\sim B'(\chi_{0})\sim e^{-\chi_0} \to 0$ and $\sigma_0 \to 0$.
One finds,  from (\ref{eqgen}) {\em for any} $0<T<T_c$:
\beq
 {\sf u}_c^{{\rm eq}} e^{d (1- {\sf u}_c^{{\rm eq}})/\epsilon} = \frac{T}{T_c} \quad , \quad  \Sigma_1^{\rm eq} = 2 \frac{{\sf u}_c^{{\rm eq}}}{T} e^{-d {\sf u}_c^{{\rm eq}}/(2-d)}
\eeq
where $T_c$ is given in (\ref{Tcd}). Using condition (iii), one finds for any $0<T<T_d$:
\bea
&& {\sf u}_c^{{\rm th}} e^{2 (1- {\sf u}_c^{{\rm th}})/\epsilon} = \frac{T}{T_d} \\
&& \Sigma_1^{\rm th} = m_c^2(T)=2 \frac{{\sf u}_c^{{\rm th}}}{T} e^{-2 {\sf u}_c^{{\rm th}}/(2-d)},  \qquad
\eea
where $T_d$ 
is the dynamical transition temperature given in (\ref{Td}). In the $T \to 0$ limit one obtains:~\footnote{In Ref.~\oMP the result $\hat {\sf u}_c ^{\rm eq}=1/T_c(m=0)$ is given. This results, however, from a rather unrealistic assumption on the behavior of $B(x)$ at small $x$.}
\bea
\hat {\sf u}_c ^{\rm eq}&=&\left[\frac{2^{1+\frac{d}{2}} A_d}{d\epsilon}\right]^{\frac{2}{\epsilon}} =\frac{1}{T_c \exp(d/\epsilon)}\stackrel{d=1}{=}0.7937,\nn\\
\hat {\sf u}_c ^{\rm th}&=&\frac{m_c(0)^2}{2}=\frac{1}{2}\left(\frac{4A_d}{\epsilon}\right)^{2/\epsilon}\stackrel{d=1}{=}\frac{1}{2}.
\eea
The cusp condition (\ref{cuspmarginality_lowT}) imposes that $\chi_0\sim \log{{m}}$, and from the saddle-point
equations it follows that:
\bea
\hat {\sf u}_c ^{\rm cp} \sim m^{d-2}/\log(1/m)\to \infty.
\eea
We finally determine the $T\to 0$ limit of the configurational entropy associated to this branch of cuspy solutions,
$\Sigma(\hat {\sf u}_c ^{\rm cp})=\hat {\sf u}_c ^2 \frac{d\phi}{d\hat {\sf u}_c }$.
For $m\ll m_c$ one finds $\Sigma_1\approx 2\hat {\sf u}_c ^{\rm cp} \gg m$.
Hence, the leading term in the configurational entropy is:
\bea
\Sigma(\hat {\sf u}_c ^{\rm cp})&\approx &(\hat {\sf u}_c ^{\rm cp})^2\frac{d\phi}{d\hat {\sf u}_c }\stackrel{T\to 0,m\to 0}{\rightarrow}
-\frac{(\hat {\sf u}_c ^{\rm cp})^2}{2}\nn\\
&\approx &-8 \left[\frac{A_d m^{d-2}}{(2-d)\epsilon \log\left(4 A_d/\epsilon m^{\epsilon}\right)}\right]^2\,,
\eea
which is negative. Hence, the condition (\ref{cuspmarginality_lowT}) selects exponentially rare configurations.

\section{Thermal boundary layer in the case of one-step RSB}
\label{app:TBL1step}

Here we prove that the thermal boundary layer in the 1-step case can be written as
\begin{eqnarray}
{\cal R}_1[v]&=& 2 T^2 {\sf u}_c V\left(\sqrt{Q}\right),
\eea
where
\bea
V(\sqrt{Q})&=&\int_0^Q dQ'\langle z\sqrt{2Q'}\coth(z\sqrt{2 Q'})\rangle_z.
\eea
Since both expressions at small $Q$ are easily seen to be of order $O(Q)$, 
it is sufficient to show the equivalence for the second derivative with respect to $\sqrt{Q}$.
Taking the second derivative of ${\cal R}_1[v]$ in (\ref{1stepTBL1}), we find
\BEA
&&\frac{1}{2 T^2 {\sf u}_c}\frac{\partial^2}{\partial (\sqrt{Q})^2}{\cal R}_{1}\\
&&\quad\quad = -\int_{-\infty}^\infty dy\,
\Big\langle z^2\Phi_z''\left(\Phi_z-\Phi\right)+z^2(\Phi_z')^2\nn\\
&&\quad\quad\quad\quad\quad  -z^2\Phi_z''\left(\Phi_{z'}-\Phi\right)
-zz'\Phi_z'\Phi_{z'}' -1\Big\rangle_{z,z'}\qquad \nn
\EEA
where $\Phi=\Phi(y)=\ln\cosh(y)$, $\Phi_z=\Phi(y+z\sqrt{Q})$, and primes denote derivatives with respect to $y$.
After a partial integration (in $y$) of the terms containing $\Phi''$, the expression simplifies to
\BEA
&&\int_{-\infty}^\infty dy\,
\left \langle z (z-z') \left(1-\Phi_z'\Phi_{z'}'\right)\right\rangle_{z,z'}\nn\\
&&\quad\quad = \left \langle \frac{(z-z')^2}{2} \psi\left((z-z')\sqrt{Q}\right)\right\rangle_{z,z'}\nn\\
&&\quad\quad = \left \langle z^2 \psi(z\sqrt{2Q})\right\rangle_{z}\,,
\EEA
where $\psi(a)=2 a \coth(a)$ as in (\ref{phiw2}), and we have used the fact that $z+z'$ and $z-z'$ are independent Gaussian variables with variance $2$.

On the other hand, one finds
\bea
V'(\sqrt{Q})&=&2\sqrt{Q}\langle z\sqrt{2Q}\coth(z\sqrt{2 Q})\rangle_z,\nn\\
V''(\sqrt{Q})&=&\langle 4z\sqrt{2Q}\coth(z\sqrt{2 Q})\rangle_z\nn\\
&& +\langle 2\sqrt{2Q} z^2\p_z\coth(z\sqrt{2 Q})\rangle_z,\nn\\
&=& \langle 2\sqrt{2Q}z^3\coth(z\sqrt{2 Q})\rangle_z\nn\\
&=& \langle z^2\psi(z\sqrt{2 Q})\rangle_z\,,
\eea
where we performed a partial integration in the second but last line (with the remaining term coming form the derivative of the measure $e^{-z^{2}/2}$), completing the proof.

\section{General solution for continuous RSB}
\label{app:generalRSB}

The general continuous RSB solution has been derived in detail for the model $g_k^{-1}=k^2+m^2$ in Ref.~\onlinecite{LeDoussalWiese2003b}. Here we sketch its generalization to arbitrary $g_k^{-1}$. A continuous RSB Ansatz is always just marginally stable on all scales, as expressed in the present case by the identity
\beq
1=4B''\left(2\int_k\tilde G(k)-G(k,{\sf u})\right)\int_k\frac{1}{[g_k^{-1}+[\sigma]({\sf u})]^2}\,,
\eeq
for all
${\sf u}_m<{\sf u}<{\sf u}_c$. Using (\ref{sigmaMP}), which in continuous form reads $\sigma(u)=-\frac2TB'\left(2 \int_{k} \tilde G(k)-G(k,{\sf u})\right)$, this leads to the relation
\BEQ \label{sigmaugeneral}
\sigma({\sf u})=
-\frac{2}{T}B'\left(\left(B''\right)^{-1}\left(\frac{1}{4\int_k\left[g_k^{-1}+[\sigma]({\sf
u})\right]^{-2}}\right)\right)\,.
\EEQ
Further, the relation $1/{\sf u}=d\sigma/d[\sigma]$ allows one to solve
for $\sigma({\sf u})$.

In particular, using $[\sigma]({\sf u}_m)=0$, we find immediately the general expression for $\sigma({\sf u}_m)=\sigma({\sf
0})$: \BEA \sigma({\sf 0})=\sigma({\sf
u}_m)=-\frac{2}{T}B'\left(\left(B''\right)^{-1}\left(\frac{1}{4I_2}\right)\right). \EEA





\begin{thebibliography}{10}
\bibitem{footnoteRSB} More accurately, the integral in the0 third line can be obtained as
the limit of $(G({\sf u}_1)-G(n)) 1^n_{({\sf u}_1)} + \sum_{i=2}^{K-1}\left[G({\sf u}_i)-G({\sf u}_{i-1})\right] 1^n_{({\sf u}_i)} +
 \left[G({\sf 1})-G({\sf u}_{K-1})\right] 1^m_{({\sf u}_K)}$ (with $G({\sf 1})=G({\sf u}_K)$) of a $K$-step RSB as $K \to \infty$.
\bibitem{footnote:order_of_u's} The order $1 \leq {\sf u}_K \leq .. \leq {\sf u}_1 \leq n$ being reversed when the analytical continuation to $n<1$ is performed.
\end{thebibliography}
\end{document}